\documentclass[]{interact}

\usepackage{epstopdf}% To incorporate .eps illustrations using PDFLaTeX, etc.
\usepackage[caption=false]{subfig}% Support for small, `sub' figures and tables
\usepackage[natbibapa,nodoi]{apacite}% Citation support using apacite.sty. Commands using natbib.sty MUST be deactivated first!
\theoremstyle{plain}% Theorem-like structures provided by amsthm.sty

\theoremstyle{definition}

\theoremstyle{remark}

\usepackage{url}
\usepackage{adjustbox}

\usepackage{graphicx}
\usepackage{multirow}
\usepackage{amsmath}
\usepackage{enumitem}
\usepackage{float}
\usepackage{hhline}
\usepackage{tabu}
\usepackage{color}
\usepackage{soul}
\usepackage{lmodern}

\usepackage{algorithm,algorithmicx}
\usepackage[noend]{algpseudocode}

\newcommand\Tstrut{\rule{0pt}{2.2ex}}         % = `top' strut
\DeclareMathOperator*{\argmax}{arg\,max}
\DeclareMathOperator*{\argmin}{arg\,min}

\newtheorem{mydef}{Definition}

\graphicspath{ {figures/} }
\setlist[description]{font=\normalfont}

\begin{document}

%\articletype{ARTICLE TEMPLATE}% Specify the article type or omit as appropriate

\title{From Note-Level to Chord-Level Neural Network Models for \\Voice Separation in Symbolic Music}

\author{
%  \name{\ }
%  \affil{\ }
\name{Patrick Gray and Razvan Bunescu\thanks{CONTACT Razvan Bunescu. Email: bunescu@ohio.edu}}
\affil{School of Electrical Engineering and Computer Science\\Ohio University, Athens, Ohio 45701, USA}
}

\maketitle

\begin{abstract}
  Music is often experienced as a progression of concurrent streams of notes, or voices. The degree to which this happens depends on the position along a voice-leading continuum, ranging from monophonic, to homophonic, to polyphonic, which complicates the design of automatic voice separation models. We address this continuum by defining voice separation as the task of decomposing music into streams that exhibit both a high degree of external perceptual separation from the other streams and a high degree of internal perceptual consistency. The proposed voice separation task allows for a voice to diverge to multiple voices and also for multiple voices to converge to the same voice. Equipped with this flexible task definition, we manually annotated a corpus of popular music and used it to train neural networks that assign notes to voices either separately for each note in a chord (note-level), or jointly to all notes in a chord (chord-level). The trained neural models greedily assign notes to voices in a left to right traversal of the input chord sequence, using a diverse set of perceptually informed input features. When evaluated on the extraction of consecutive within voice note pairs, both models surpass a strong baseline based on an iterative application of an envelope extraction function, with the chord-level model consistently edging out the note-level model. The two models are also shown to outperform previous approaches on separating the voices in Bach music.
% Knowing which notes belong to the same voice can benefit a wide range of tasks in computational music analysis, such as automatic transcription or music classification and retrieval. 
%, to the maximum degree that is possible in the given musical input. 
\end{abstract}

%
% The code below should be generated by the tool at
% http://dl.acm.org/ccs.cfm
% Please copy and paste the code instead of the example below. 
%

%
% End generated code
%

%
%  Use this command to print the description
%
%\printccsdesc

% We no longer use \terms command
%\terms{Theory}


% Variables
\newcounter{fi}

\section{Introduction and Motivation}
\label{sec:introduction}

Voice separation of symbolic music refers to the processing of a musical input into sequences of musical events, such that each sequence corresponds to an auditory stream perceived by the listener. Identifying the underlying voices adds structure to music, enabling a more sophisticated analysis of music that can benefit higher level tasks such as query by humming, theme identification, or reharmonization. Existing approaches to voice separation can be classified into two major categories, depending on the type of musical events they use in their definition of voices. One set of approaches extracts voices as monophonic sequences of non-overlapping musical notes \citep{chew:cmmr04,kirlin:ismir05,madsen:icmpc06,temperley:book01,rafailidis:ismir08,jordanous:icmc08,deValk:ismir13}. A second type of approaches allows voices to contain simultaneous note events, such as chords \citep{kilian:ismir02,cambouropoulos:icmpc06,karydis:ismir07,rafailidis:smc09}. While the first set of approaches corresponds to the musicological notion of voice referenced in the Western musical tradition \citep{cambouropoulos:icmpc06,aldwell:book11}, the second type of approaches emphasizes a perceptually-founded notion of voice that corresponds more closely to independent auditory streams \citep{bregman:jep71,bregman:book90}. Orthogonal to this categorization, a relatively small number of voice separation approaches are data-driven and have parameters that are trained on symbolic music where voices have been labeled manually\citep{kirlin:ismir05,jordanous:icmc08,deValk:ismir13}.

Because an original aim was to determine whether an arbitrary musical input obeys traditional voice-leading rules, previous work on voice separation used the musicological notion of voice as a monophonic sequence of {\it non-overlapping} notes \citep{gray:ismir16}. Correspondingly, voice separation was defined as the task of {\it partitioning} music into monophonic sequences (voices) of {\it non-overlapping} notes that exhibit both a high degree of external perceptual separation from the other voices and a high degree of internal perceptual consistency, to the maximum degree that is possible in the musical input. However, subsequent annotation efforts showed this definition to be too constraining. First, there are situations where two overlapping notes that have different onsets are heard as belonging to the same voice. Second, and more interestingly, there are situations when a note is heard as belonging to multiple voices. In this paper, we present a more general definition of voice separation in symbolic music that allows a note to belong to multiple voices, reflecting more closely the auditory streams perceived by the listener. Correspondingly, multiple voices may converge into one voice or one voice may diverge into multiple voices. Equipped with this flexible task definition (Section~\ref{sec:task-definition}), we manually annotated a corpus of popular music and used it to evaluate three voice separation models that assign notes to voices in a left to right traversal of the input chord sequence (Section~\ref{sec:voice-separation}). The first model is based on an iterative application of an envelope extraction function (Section~\ref{sec:envelope-extraction}) and is meant to be used as a strong baseline. The other two voice separation models are neural networks that have been designed to allow the extraction of convergent and divergent voices. They assign notes to voices either separately for each note in a chord (Section~\ref{sec:note-level}), or jointly to all notes in a chord (Section~\ref{sec:chord-level}), using a diverse set of perceptually informed input features. When evaluated on the extraction of consecutive within voice note pairs, both models surpass the envelope extraction baseline, with the chord-level model consistently edging out the note-level model (Section~\ref{sec:evaluation}).

% In this paper, we describe a data-driven approach to voice separation that preserves the musicological notion of voice. Our aim is to obtain a segregation of music into voices that would enable a downstream system to determine whether an arbitrary musical input satisfies the known set of voice-leading rules, or conversely identify places where the input violates voice-leading rules. Such a system could form an essential component in the modeling of the so called ``musical critics'', to be used in automatic music generation systems. 

\section{Task Definition and Annotation Guidelines}
\label{sec:task-definition}

According to Huron \citeyearpar{huron:mp01}, ``the principal purpose of voice-leading is to create perceptually independent musical lines''. However, if a voice is taken to be a monophonic sequence of notes, as implied by traditional voice-leading rules \citep{aldwell:book11}, then not all music is composed of independent musical lines. In homophonic accompaniment, for example, multiple musical lines (are meant to) fuse together into one perceptual stream. As Cambouropoulos \citep{cambouropoulos:icmpc06} observes for homophonic accompaniment, ``traditional voice-leading results in perceivable musical {\it texture}, not independent musical lines''. In contrast with the traditional notion of voice used in previous voice separation approaches, Cambouropoulos redefines in \citep{cambouropoulos:icmpc06} the task of 'voice' separation as that of separating music into perceptually independent musical {\it streams}, where a stream may contain two or more synchronous notes that are perceived as fusing in the same auditory stream. This definition is used in \citep{karydis:ismir07,rafailidis:smc09} to build automatic approaches for splitting symbolic music into perceptually independent musical streams.

Since a major aim of our approach is to enable building ``musical critics'' that automatically determine whether an arbitrary musical input obeys traditional voice-leading rules, we adopt the musicological notion of voice as a {\it monophonic sequence of notes}. This definition however leads to an underspecified voice separation task: for any non-trivial musical input, there usually is a large number of possible separations into voices that satisfy the constraints that they are monophonic. Further constraining the voices to be perceptually independent would mean the definition could no longer apply to music with homophonic textures, as Cambouropoulos correctly noticed in \citep{cambouropoulos:icmpc06}. Since we intend the voice separation approach to be applicable to arbitrary musical input, we instead define voice separation as follows:
\begin{mydef}
Voice separation is the task of decomposing music into monophonic sequences of notes (voices) that exhibit both a high degree of external perceptual separation from the other voices and a high degree of internal perceptual consistency, to the maximum degree that is possible in the given musical input.
\end{mydef}
\begin{figure}[h]
\centering
\includegraphics[width=0.7\columnwidth]{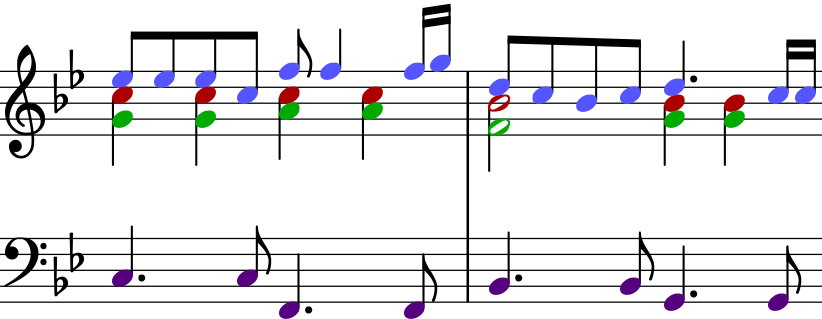}
\caption{Example voice separation from ``Earth Song'' measures 13-14.}
\label{fig:definition}
\end{figure}
Figure~\ref{fig:definition} shows a simple example of voice separation obtained using the definition above. While the soprano and bass lines can be heard as perceptually distinct voices, we cannot say the same for the tenor and alto lines shown in green and red, respectively. However, clear perceptual independence is not needed under the new task definition. The two intermediate voices exhibit a high degree of perceptual consistency: their consecutive notes satisfy to a large extent the pitch proximity and temporal continuity principles needed to evoke strong auditory streams \citep{huron:mp01}. Indeed, when heard in isolation, both the tenor and the alto are heard as continuous auditory streams, the same streams that are also heard when the two are played together. The two streams do not cross nor overlap, which helps with perceptual tracking \citep{huron:mp01}. Furthermore, out of all the streaming possibilities, they also exhibit the largest possible degree of external perceptual separation from each other and from the other voices in the given musical input.

%\section{Annotation Guidelines}
%\label{sec:annotation-guidelines}

According to the definition in Section~\ref{sec:task-definition}, voice separation requires decomposing music into monophonic sequences of notes that exhibit a high degree of perceptual salience, to the maximum extent that is possible in the given musical input. As such, an overriding principle that we followed during the manual annotation process was to always give precedence to what was heard in the music, even when this appeared to contradict formal perceptual principles, such as pitch proximity. Furthermore, whenever formal principles seemed to be violated by perceptual streams, an attempt was made to explain the apparent conflict. Providing justifications for non-trivial annotation decisions enabled refining existing formal perceptual principles and also informed the feature engineering effort.

\begin{figure}[H]
  \centering
  \includegraphics[width=0.6\columnwidth]{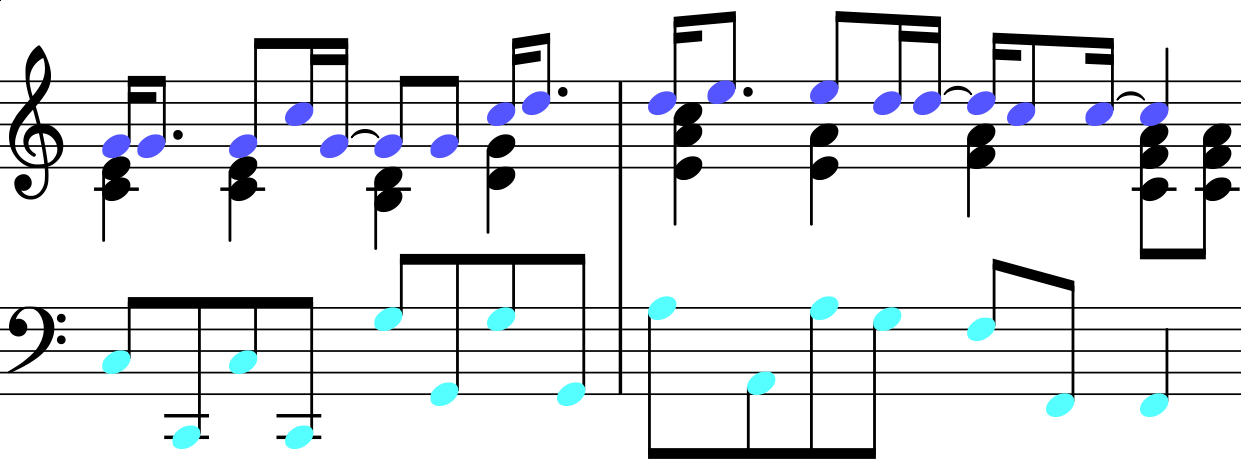}
  \caption{Annotate the most salient voices in measures 6-7 from ``Let it Be".}
  \label{fig:annotate_salient_voices}
\end{figure}

We perform the annotation in MuseScore\footnote{https://musescore.org}, by coloring notes that belong to the same voice with the same color. Listening to the original music is often not sufficient on its own for voice separation, as not all the voices contained in a given musical input can be distinctly heard. Because we give precedence to perception, we first annotate those voices that could be distinguished clearly in the music, which often means annotating first the melodic lines in the soprano and the bass, as shown in Figure~\ref{fig:annotate_salient_voices}. 

Next, we listen to the composition once again, this time focusing on hearing the less salient intermediate voices. Sometimes it is relatively easy to distinguish the intermediate voices among the surrounding soprano and bass, as shown in the first measure in Figure~\ref{fig:annotate_intermediate_voices_partial} below.

\begin{figure}[H]
  \centering
  \includegraphics[width=0.6\columnwidth]{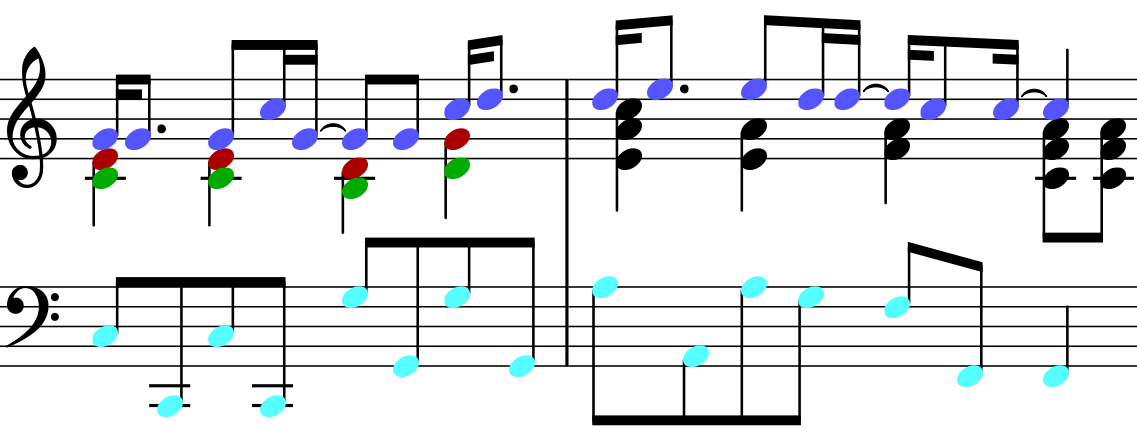}
  \caption{Partial annotation of the intermediate voices in measures 6-7 from ``Let it Be".}
  \label{fig:annotate_intermediate_voices_partial}
\end{figure}

\begin{figure}[H]
  \centering
  \includegraphics[width=0.6\columnwidth]{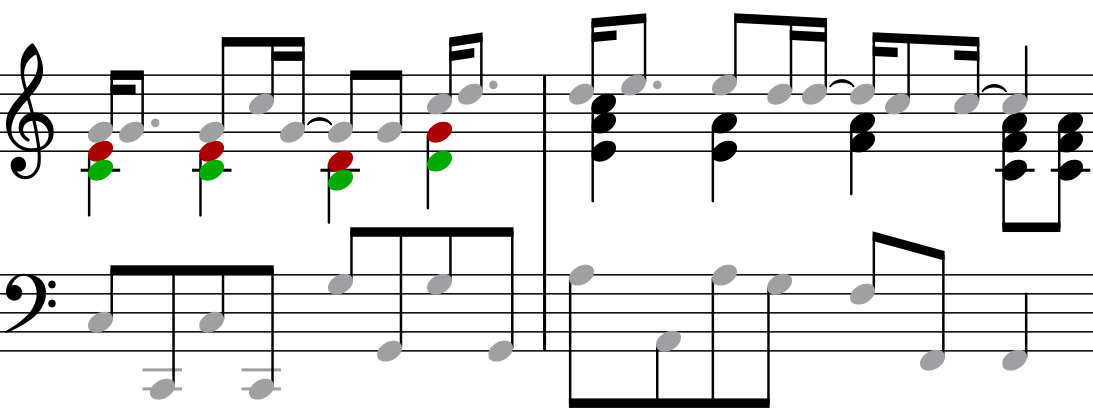}
  \caption{Listen to the intermediate voices in isolation in measures 6-7 from ``Let it Be".}
  \label{fig:annotate_intermediate_voices_isolated}
\end{figure}

However, the increase in the number of synchronous voices in the second measure can make it difficult for the listener to annotate the intermediate voices. When the intermediate voices are difficult to hear because of being masked by more salient voices (often the soprano and the bass \citep{aldwell:book11}), we temporarily mask the already annotated most prominent voices one by one and listen to the remaining combination of voices, as shown in Figure~\ref{fig:annotate_intermediate_voices_isolated}.

% solely for the purpose of annotation 

% Alternatively, when multiple conflicting voice separations were plausible, we annotated the voice that, after listening to it in isolation, was easiest to distinguish perceptually in the original music.

\begin{figure}[H]
  \centering
  \includegraphics[width=0.6\columnwidth]{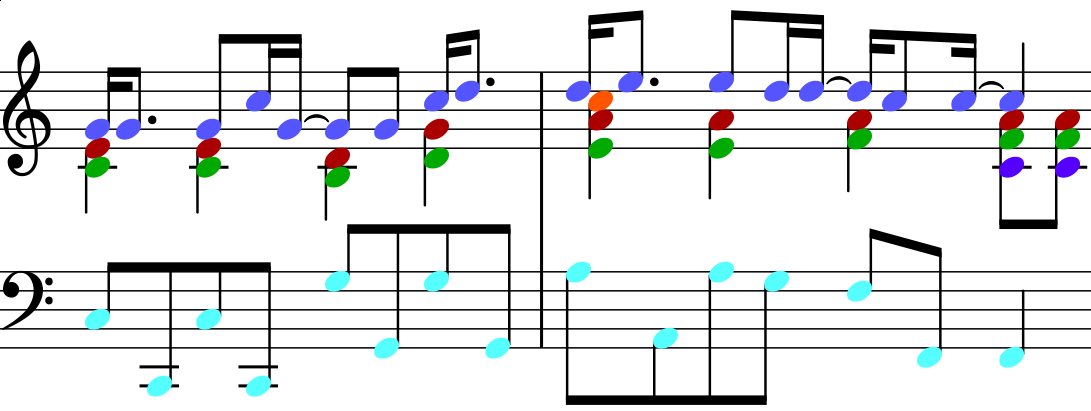}
  \caption{Annotate the intermediate voices among all voices in measures 6-7 from ``Let it Be".}
  \label{fig:annotate_intermediate_voices}
\end{figure}

While listening to the intermediate voices in isolation, we may be faced with multiple voice separations that are perceptually plausible. In such cases, we annotate the voices that are easiest to distinguish perceptually in the original, complete musical fragment. Figure~\ref{fig:annotate_intermediate_voices} shows the complete annotation of voices in measures 6-7 from ``Let it Be".

A new voice is started whenever its first note cannot be heard as continuing any of the existing active voices, as shown in Figure~\ref{fig:annotate_new_voice} below.

\begin{figure}[H]
  \centering
  \includegraphics[width=0.6\columnwidth]{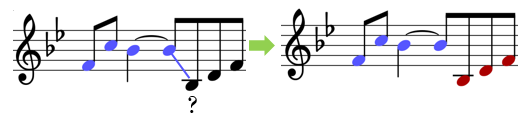}
  \caption{Annotate a new voice in measure 4 from ``See You Again".}
  \label{fig:annotate_new_voice}
\end{figure}

\subsection{Note Overlap}
\label{sec:note-overlap}

Generally, a note ends before the onset of the next note in the same voice. However, there are situations where two overlapping
notes that have different onsets are heard as belonging to the same voice, as shown in  Figure~\ref{fig:overlap} below, where $D_4$ is heard connected to the overlapping $E_4$, likely due to their close pitch proximity and the relatively long duration between their onsets.
% Formally, we say that a note $n$ overlaps another note $m$ if the onset of $n$ is stricly less than the onset of $m$ and the offset of $n$ is strictly greater than the onset of $m$. 

\begin{figure}[H]
  \centering
  \includegraphics[width=0.25\columnwidth]{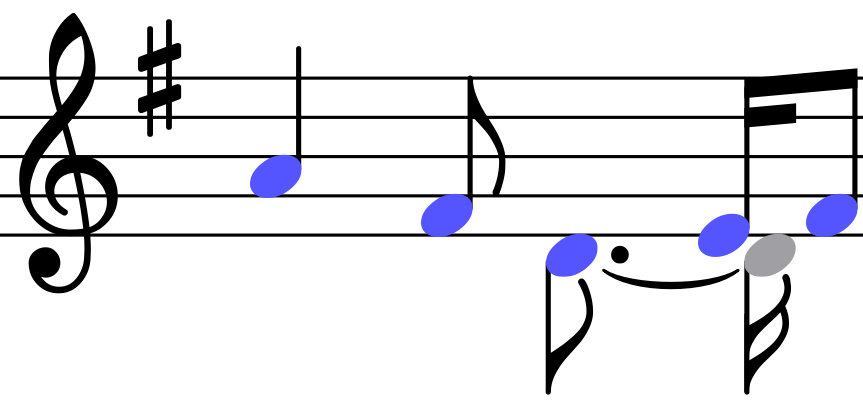}
  \caption{Overlapping voice in measure 3 from ``Greensleeves".}
  \label{fig:overlap}
\end{figure}

\subsection{Convergence and Divergence}
\label{sec:convergence-divergence}

Sometimes, a note is heard as belonging to multiple voices. Given a left-to-right traversal of the musical input, this leads to two scenarios:
\begin{enumerate}
  \item {\bf Convergence}: Two or more currently active voices move forward by using the same note at the next time step, as shown in Figure~\ref{fig:annotate_convergence}. From the model perspective, the last notes in the active voices are linked to the same note at the next time step, hence the alternative {\it many-to-one} designation. The bottom of Figure~\ref{fig:annotate_convergence} shows the voices that are implicit in the annotation.

  \begin{figure}[H]
    \centering
    \includegraphics[width=0.5\columnwidth]{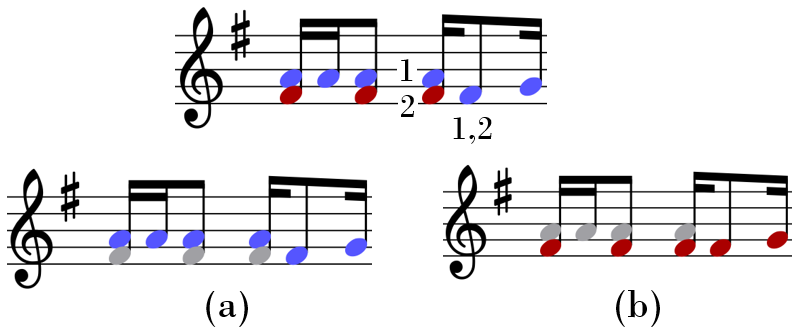}
    \caption{Annotate convergent voices in measure 7 from ``Knocking on Heaven's Door".}
    \label{fig:annotate_convergence}
    \end{figure}
  \item {\bf Divergence}: One currently active voice is continued by two or more notes, each in its own voice, as shown in Figure~\ref{fig:annotate_divergence}. From the model perspective, the last note in the active voice is linked to multiple notes at the next time step, hence the alternative {\it one-to-many} designation. The bottom of Figure~\ref{fig:annotate_divergence} shows the voices that are implicit in the annotation.

  \begin{figure}[H]
    \centering
    \includegraphics[width=0.6\columnwidth]{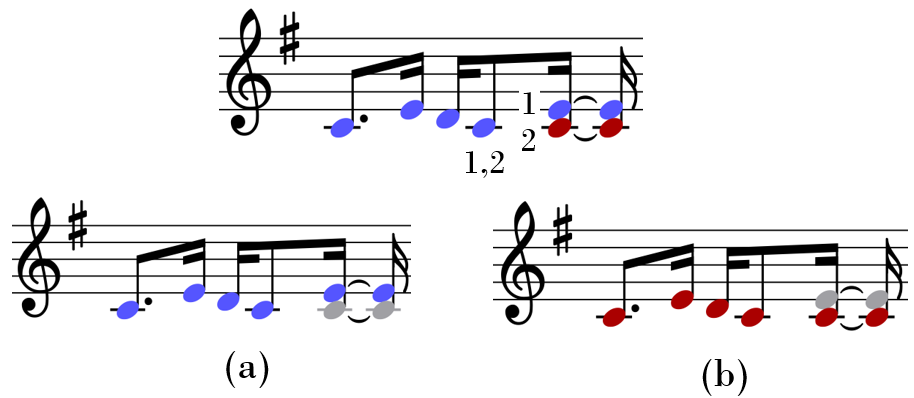}
    \caption{Annotate divergent voice in measure 8 from ``Knocking on Heaven's Door".}
    \label{fig:annotate_divergence}
  \end{figure}
\end{enumerate}
When a note belongs to only one voice, from the model perspective is forms {\it one-to-one} links with the note preceding it and the note following it in the same voice.

We annotate convergence and divergence cases using lyrics in MuseScore. Each of the converging voices is given a unique numerical identifier. The note upon which the voices converge is annotated with a comma delimited list of the converging voice identifiers. The comma delimited list always starts with the converging voice that has the most salient connection and ends with the converging voice with the weakest connection. Furthermore, the resulting voice takes on the color of the most salient converging voice. Starting from the convergence point, the converging voices share the same stream of notes. 

Analogously, the notes onto which a voice diverges are each given a unique numerical identifier. The last note in the diverging voice is then annotated with a comma delimited list of these note identifiers. The list always starts with the note that has the most salient connection with the diverging voice and ends with the note that has the weakest connection. Furthermore, the note with the most salient connection takes on the color of the diverged voice. Up to the divergence point, the diverged voices share the same stream of notes.

%It is important to note that a voice can diverge at any time regardless of whether or not it was formed from converging voices.
%\begin{figure}[H]
%  \centering
%  \includegraphics[scale=0.80]{annotation_guidelines/annotate_spontaneous_divergence.png}
%  \caption{Diverging a voice with no underlying convergent voices in measure 3-4 from ``Endless Love".}
%  \label{fig:annotate_spontaneous_divergence}
%\end{figure}

%Consider Figure~\ref{fig:annotate_spontaneous_divergence}. Voice 2 was introduced at the point of divergence. It did not exist prior to measure 4 in ``Endless Love". Thus, annotators should not feel obligated to diverge any pre-existing, underlying convergent voices. Start a new voice from an ambigous connection if it sounds new. This idea is reinforced in step 4 of the annotation guidelines.

\subsection{Notes Separated by Rests}
\label{sec:notes-rests}

\begin{figure}[h]
\centering
\includegraphics[width=0.75\columnwidth]{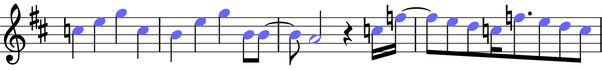}
\caption{Voice separation annotation in the treble for measures 38-41 in ``Count on Me''.}
\label{fig:rest1}
\end{figure}

Other non-trivial annotation decisions, especially in the beginning of the annotation effort, involved whether two streams separated by rests should be connected or not. Overall, we adopted the guideline that we should break the music into fewer and consequently longer voices, especially if validated perceptually. Figure~\ref{fig:rest1}, for example, shows the A$_4$ in the third measure connected to the following C$_5$. Even though the two notes are separated by a quarter rest, they are heard as belonging to the same stream, which may also be helped by the relatively long duration of A$_4$ and by the fact that the same pattern is repeated in the piece.

The repetition of motifs or patterns gives rise to a sense of familiarity that makes it easier perceptually to connect notes separated by rests. In order to avoid a proliferation of voices, we use this familiarity principle to connect notes separated by long rests, as shown in Figures~\ref{fig:block_see_you_again} and~\ref{fig:repeat_melodies} below.

\begin{figure}[H]
  \centering
  \includegraphics[width=0.6\columnwidth]{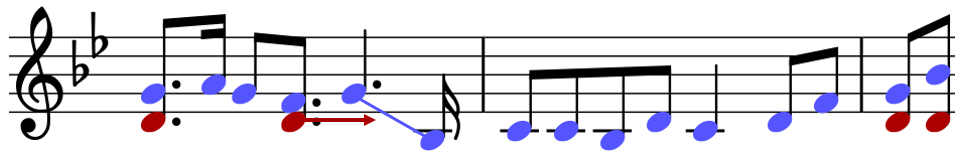}
  \caption{Pairing a blocked note in measures 5-7 ``See You Again".}
  \label{fig:block_see_you_again}
  \label{fig:see-you-block}
\end{figure}

\begin{figure}[H]
  \centering
  \includegraphics[width=0.55\columnwidth]{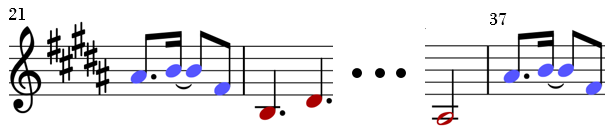}
  \caption{Voices that violate the Temporal Continuity Principle, measures 20 and 37 ``Thousand Miles".}
  \label{fig:repeat_melodies}
\end{figure}
Figure~\ref{fig:repeat_melodies} shows the continuation of the blue and red voice after a 16 measure break. We bring the blue and red voices back because measures 21 and 37 play the exact same sequence of notes. Melody repetition is common in popular music and we use it to reduce the total number of voices, by connecting repeated patterns. 

% In general, even if the connecting pair of repeated or similar sounding melodies creates dissonance, as in the Figure~\ref{fig:repeat_melodies}, logically, it makes sense to group repeated or similar sounding melodies into the same voice so as to avoid a proliferation of voice labels.

%The simplest definition of our familiarity rule is ``always annotate motifs into the same voice". Consider the following figure.
%\begin{figure}[H]
%  \centering
%  \includegraphics[scale=0.40]{annotation_guidelines/harmonic_support_voice.png}
%  \caption{Treating the harmonic support as a voice in measure 23 from ``Hymn for the Weekend".}
%  \label{fig:harmonic_support_voice}
%\end{figure}

%The bottom layer of $E_4$'s in Figure~\ref{fig:harmonic_support_voice} is not functioning as a melody but instead as a harmonic support to the soprano. However, because each $E_4$ is playing out the same musical idea of a harmonic support, we can use our familiarity rule to string together the $E_4$'s into an independent voice. In Figure~\ref{fig:repeat_melodies}, we use the same familiarity rule to join together the distant repeated motifs into the blue voice. 

\subsection{Crossing Voices}
\label{sec:crossing-voices}

Consider the annotation of voices in Figure~\ref{fig:imperceivable_cross} below. If each voice is played in a different timbre, the two crossing voices can be indeed heard as annotated.

\begin{figure}[H]
  \centering
  \includegraphics[width=0.3\columnwidth]{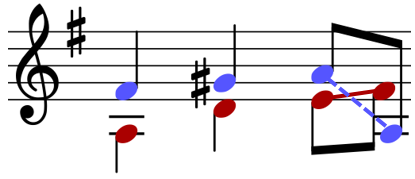}
  \caption{Voice crossing in measure 10 of ``BWV-260".}
  \label{fig:imperceivable_cross}
\end{figure}

However, when played on the piano, we perceive a bounce. As Huron observed in \citeyearpar{huron:mp01}, ``listeners are disposed to hear a bounced percept in preference to the crossing of auditory streams". Consequently, the separation shown in the figure does not reflect our perception of the voices. The $F^\#_4$ in the last chord is heard as a continuation of the blue voice whereas the last $A_3$ sounds more compatible with the red voice. 

\begin{figure}[H]
  \centering
  \includegraphics[width=0.5\columnwidth]{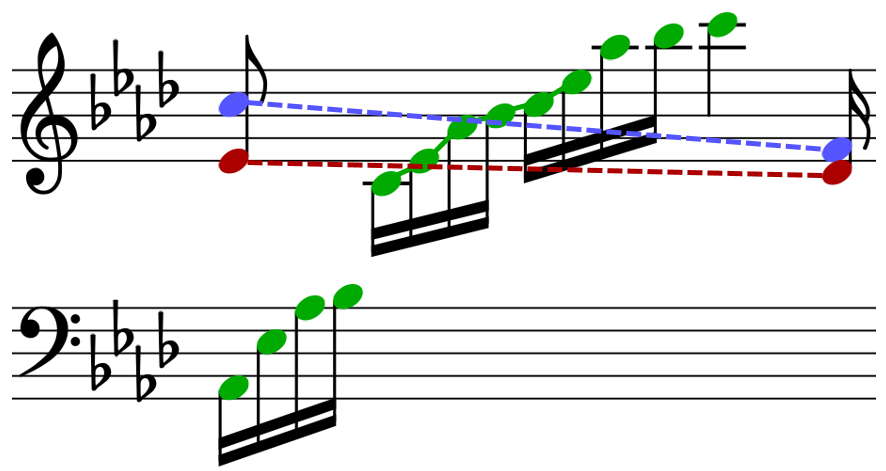}
  \caption{Perceivable voice crossing in measure 17 ``Endless Love".}
  \label{fig:perceivable_cross}
\end{figure}

Figure~\ref{fig:perceivable_cross} shows a case of voices that are indeed heard as crossing when played on the piano. Crossing is however a rare phenomenon in our datasets.

\section{Voice Separation Datasets}
\label{sec:datasets}

We compiled a corpus of piano versions of 20 popular compositions of varying complexity that are representative of many genres of music. Each song was downloaded from \textit{www.musescore.com} and converted to MusicXML. We avoided collecting piano accompaniments and gave preference to piano renditions that sounded as much as possible like the original song. Among other things, this ensured that each score contained at least one clearly defined melody. We also annotated only tonal music, as atonal music with unusual melodic structures was observed to lead to a poor perception of voices. A large set of musical fragments were annotated by both authors, in an iterative process where most differences in annotation were reconciled through further refinement of the annotation guidelines. There were however cases when the two annotators disagreed in their annotation of voices due to subjective differences in how they attended to the music material -- for example when accent-based salience conflicted with salience based on the outer vs. inner voice distinction. In such cases, the differences were documented and the annotator was instructed to follow their most salient hearing, while being consistent over the entire dataset. A first version of the dataset containing only one-to-one connections had an inter-annotator agreement (ITA) of 96.08\% \citep{gray:ismir16}. Inter-annotator differences are to be expected and are addressed by training and evaluating the system on data from just one annotator at a time (the experiments reported in this paper used the annotations coming from the first author).

Table~\ref{tab_dataset_statistics} shows a number of statistics computed over the 20 musical pieces in the dataset\footnote{The annotations will be made publicly available}, such as the total number of notes, unique onsets, voices, within-voice note pairs, within-voice one-to-one note pairs, and within-voice many-to-one (convergence) or one-to-many (divergence) note pairs under the heading many-to-many. A pair is designated many-to-many if at least one note in the pair has at least two convergent or divergent connections. The sum of one-to-one and many-to-many pairs equals the total number of pairs. 

\begin{table}[H]
  \begin{center}
    \tabcolsep=0.06cm
    \footnotesize
    \begin{tabular}{|l|r|r|r|r|r|r|}
      \hline
      Popular Music dataset & \# Notes & \# Onsets & \# Voices & \# Pairs & \# One-to-one & \# Many-to-many \\
      \hline
      21 Guns (Green Day)                   & 1969  & 666  & 25    & 2000  & 1779  & 221  \\
      Apples to the Core (Daniel Ingram)    & 923   & 397  & 8     &  924  &  894  &  30  \\
      Count On Me (Bruno Mars)              & 775   & 473  & 5     &  948  &  242  & 706  \\
      Dreams (Rogue)                        & 615   & 474  & 4     &  623  &  583  &  40  \\
      Earth Song (Michael Jackson)          & 431   & 216  & 9     &  429  &  411  &  18  \\
      Endless Love ()                       & 834   & 443  & 9     &  833  &  819  &  14  \\
      Forest (Twenty One Pilots)            & 1784  & 1090 & 21    & 1820  & 1603  & 217  \\
      Fur Elise (Ludwig van Beethoven)      & 897   & 653  & 29    &  881  &  849  &  32  \\
      Greensleeves                          & 231   & 72   & 4     &  241  &  192  &  49  \\
      How To Save a Life (The Fray)         & 438   & 291  & 11    &  433  &  415  &  18  \\
      Hymn For The Weekend (Coldplay)       & 1269  & 706  & 16    & 1284  & 1180  & 104  \\
      Knockin' on Heaven's Door (Bob Dylan) & 355   & 180  & 10    &  353  &  319  &  34  \\
      Let It Be (The Beatles)               & 563   & 251  & 10    &  562  &  522  &  40  \\
      One Call Away (Charlie Puth)          & 1369  & 602  & 38    & 1388  & 1177  & 211  \\
      See You Again (Wiz Khalifa)           & 704   & 359  & 30    &  697  &  606  &  91  \\
      Teenagers (My Chemical Romance)       & 312   & 145  &  6    &  306  &  306  &   0  \\
      Thousand Miles (Vanessa Carlton)      & 1264  & 608  & 10    & 1314  & 1064  & 250  \\
      To a Wild Rose (Edward Macdowell)     & 307   & 132  &  9    &  306  &  280  &  26  \\
      Uptown Girl (Billy Joel)              & 606   & 297  & 26    &  593  &  557  &  36  \\
      When I Look At You (Miley Cyrus)      & 1152  & 683  & 31    & 1161  & 1020  & 141  \\
      \hline
      Totals \& Averages                    & 16798 & 8738 & 15.55 & 17096 & 14818 & 2278 \\
      \hline
      \hline
      Bach Chorales dataset                 & 12282 & 4519 & 4     & 12298 & 11452 & 846  \\
      \hline
    \end{tabular}
  \end{center}
  \caption{Statistics for the Popular Music dataset and the Bach Chorales dataset.}
  \label{tab_dataset_statistics}
\end{table}

There are 16,798 labeled notes in 8,738 chords, for a total of 17,096 (within-voice) note pairs, providing a sufficient number of examples for learning the more common perceptual rules of voice separation. There are however voice separation scenarios that could benefit from more annotated data, such a pseudo-polyphony (as explained in Section~\ref{sec:evaluation}).

\begin{figure}[H]
  \centering
  \includegraphics[scale=0.75]{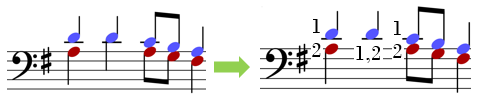}
  \caption{Replacing unison with many-to-many annotation in measure 4 of  ``BWV256".}
  \label{fig_unison_rewrite}
\end{figure}

The last line in Table~\ref{tab_dataset_statistics} shows the same type of statistics for the first 50 four-part Bach Chorales available in the Music21 corpus. We rely on the original partition of voices, where we rewrote the unisons as voices converging to one note and then diverging into multiple voices. Figure~\ref{fig_unison_rewrite} shows an example.

%The purpose of utilizing this Bach chorale dataset is to demonstrate the versatility of our assignment models. Specifically, we want to show that our assignment models are generic enough to separate any monophonic voice with a perceptually sound structure. Of course, we also want to compare the perfomance of our assignment models on the popular music dataset to a dataset with a publicly accepted ground-truth voice annotation.

\section{Voice Separation Procedure}
\label{sec:voice-separation}

\begin{figure}[H]
  \centering
  \includegraphics[width=0.8\columnwidth]{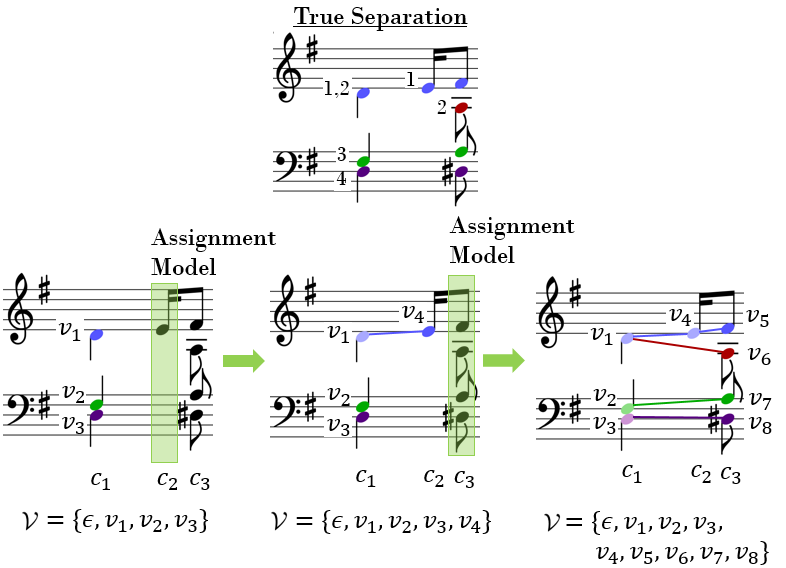}
  \caption{Three iterations of the automated voice separation procedure in measure 3 from ``Greensleeves".}
  \label{fig:separation_procedure}
\end{figure}

Manually annotating voices in a musical input is a cognitively demanding and time consuming process that cannot scale to large collections of music. In this paper, we describe a number of models that do automatic voice separation in a left-to-right processing of the musical input, as shown in the example from Figure~\ref{fig:separation_procedure}.

In a pre-processing step, all the notes are ordered based on their onsets into a sequence of chords $\mathcal{C} = \{c_1, c_2, ..., c_T\}$, where a chord is defined to be a maximal group of notes that have the same onset. Assignment of notes to voices is then performed in chronological order, from left to right, starting with the first chord $c_1$. Because voices are by definition monophonic, each note in the first chord is considered to start a separate, new voice. These first voices, together with an empty voice $\epsilon$, constitute the initial set of active voices $\mathcal{V}$. At each onset $t$, one of the 3 assignment models introduced at the end of this section is used to greedily assign each note from the current chord $c_t$ to either the empty voice or one or more of the active voices in $\mathcal{V}$. Notes assigned to the empty voice are introduced as new voices in the active set for the next timestep. 
% After finalizing all the assignments, each note in $c_t$ takes on the role of an active voice and thus are made available for pairing with the notes at onset $t + 1$.

As shown in the example from Figure~\ref{fig:separation_procedure}, every time a note $n$ is assigned to an active voice $v_i$, a new voice $v_j = v_i \leadsto n$ is created and inserted in the active set $\mathcal{V}$. Upon processing the second chord, for instance, voice $v_4$ is created by connecting the $E_4$ in the treble with the last note of $v_1$, and inserted as a new voice in the active set. Note that the voice $v_1$ is kept in the active set, even though the newly added voice $v_4$ subsumes it. This is needed in order to allow voices to diverge to notes that appear two or more chords later in the input, as is the case after processing the third chord in Figure~\ref{fig:separation_procedure}. There, the red $A_3$ note from the current chord $c_3$ is connected to voice $v_1$ from two chords earlier, effectively modeling the divergence of $v_1$ into two separate voices $v_5$ and $v_6$.

To make the formulation of voice separation algorithms easier, we distinguish between two types of active voices at each time step:
\begin{enumerate}
  \item {\bf Complete active voices}: These are voices in $\mathcal{V}$ that are not connected to any note to the right. For example, before the notes in $c_3$ are assigned to voices in Figure~\ref{fig:separation_procedure}, the complete active voices are $v_2$, $v_3$, and $v_4$. 
  \item {\bf Partial active voices}: These are voices in $\mathcal{V}$ that are connected to the right to one or more notes. For example, before the notes in $c_3$ are assigned to voices in Figure~\ref{fig:separation_procedure}, the only partial active voice is $v_1$. One step later however, the set of partial active voices grows to include $v_1$, $v_2$, $v_3$, and $v_4$.
\end{enumerate}
If a new voice symbol is added to the active set every time a note is connected to an existing voice, the number of active voices will grow linearly with the number of processed chords, which can lead to a substantial increase in the time complexity of the voice separation algorithms. In practice, two notes are rarely connected if they are separated by more than a few beats. Consequently, when processing a chord $c_t$, we eliminate from the active set all voices that end beyond a {\it beat horizon} from $c_t$, where the number of beats in the beat horizon is determined empirically from the training data.

\begin{figure}[H]
  \centering
  \includegraphics[width=0.45\columnwidth]{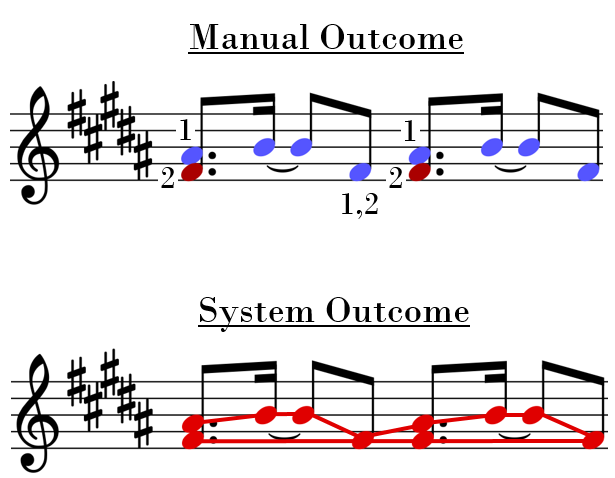}
  \caption{Outcome of the voice separation algorithm on measure 18 from ``Thousand Miles" .}
  \label{fig:separation_outcome}
\end{figure}

The voice separation algorithms assign notes $n$ from the current chord to one or more active voices $v \in \mathcal{V}$. This was shown graphically in Figure~\ref{fig:separation_procedure} by connecting a note $n$ to the last note $v.last$ of an assigned active voice $v$. To make the manual annotation process easier, notes in the same voice were marked with the same color, whereas numerical identifiers were further assigned to notes in order to capture convergence or divergence cases, as shown at the top of Figure~\ref{fig:separation_outcome}. The voice separation algorithms however do not need to use colors or numerical identifiers. As long as they link a note $n$ to the last notes $v.last$ of the correct active voices, as shown at the bottom of Figure~\ref{fig:separation_outcome}, a correct coloring and numerical identifiers can be automatically computed, due to a one-to-one correspondence between consistent colorings and numerical identifiers on one side, and note-to-note pairs on the other side.

The general voice separation procedure described above is used to formulate 3 voice separation models, as follows:
\begin{enumerate}
  \item {\bf Envelope Extraction}: This is an iterative, rule-based approach to voice-separation that repeatedly extracts the upper {\it envelope} of a given musical input, i.e. the topmost monophonic sequence of non-overlapping notes (Section~\ref{sec:envelope-extraction}).
  \item {\bf Note-Level Voice Separation}: This is a neural network model that is trained on the manual annotations to assign notes to voices separately for each note in a chord (Section~\ref{sec:note-level}).
  \item {\bf Chord-Level Voice Separation}: This is a more general neural network model that is trained on the manual annotations to assign notes to voices jointly for all the notes in a chord (Section~\ref{sec:chord-level}).
\end{enumerate}
Before we proceed to describe the three voice separation models, we will first formally define two major concepts that are used in these models: {\it voice crossing} (Section~\ref{sec:voice-crossing}) and {\it voice blocking} (Section~\ref{sec:voice-blocking}).

\subsection{Voice Crossing}
\label{sec:voice-crossing}

\begin{figure}[H]
  \centering
  \includegraphics[width=0.75\columnwidth]{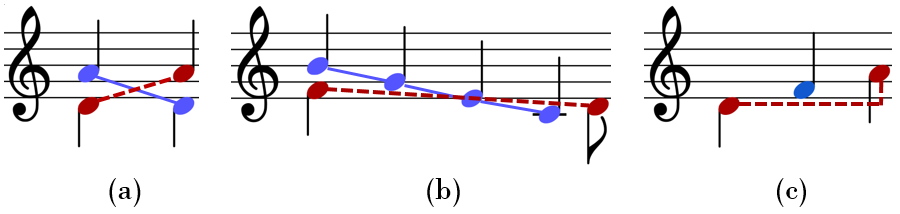}
  \caption{Definition of voice crossing.}
  \label{fig:cross}
\end{figure}

Figure~\ref{fig:cross} outlines a few examples of voice crossing. The red dashed line indicates a crossing connection.

\begin{figure}[H]
  \centering
  \includegraphics[width=0.7\columnwidth]{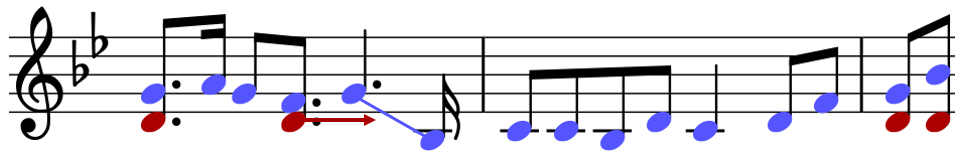}
  \caption{Pairing a blocked note in measures 5-7 ``See You Again".}
  \label{fig:block_see_you_again_repeat}
\end{figure}

\noindent Formally, we say that a voice $u$ crosses another voice $v$ if $u$ contains a pair of notes $(n_1, n_2)$ that crosses $v$. To determine whether a note pair $(n_1, n_2)$ crosses a voice $v$, we first extract the {\it maximal} subsequence of notes  $s = \langle s_1, s_2, ..., s_k \rangle$ from $v$ whose onsets are between the onsets of $n_1$ and $n_2$ as follows:
\begin{enumerate}
  \item $s_1$ has an onset greater than or equal to the onset of $n_1$ and strictly less than the onset of $n_2$, and
  \item $s_k$ has an onset strictly greater than the onset of $n_1$ and less than or equal to the onset of $n_2$. 
\end{enumerate}
Then the voice pair $(n_1, n_2)$ is said to cross a voice $v$ if the voice pair and the maximal subsequence $s$ satisfy either of the following pitch constraints:
\begin{enumerate}
  \item $s_1$ has a pitch greater than or equal to the pitch of $n_1$, and $s_k$ has a pitch less than or equal to the pitch of $n_2$, or
  \item $s_1$ has a pitch less than or equal to the pitch $n_1$, and $s_k$ has a pitch greater than or equal to the pitch of $n_2$.
\end{enumerate}
It is important for the definition of voice crossing that the subsequence $s$ is maximal. Thus, Figure~\ref{fig:block_see_you_again}, copied below as Figure~\ref{fig:block_see_you_again_repeat} for convenience, does not depict a cross from the red voice, because the crossing subset $s$ in the blue voice should include all notes from the first $F_4$ to the last $G_4$. Since the $F_4$ and $G_4$ have pitches above the $D_4$ notes in the bounding red voice pair, no cross is said to occur.

\noindent While Figure~\ref{fig:block_see_you_again_repeat} contains no voice crossing, it does show an instance of {\it blocking}, as described in the next section.

\subsection{Voice Blocking}
\label{sec:voice-blocking}

\begin{figure}[H]
  \centering
  \includegraphics[width=0.6\columnwidth]{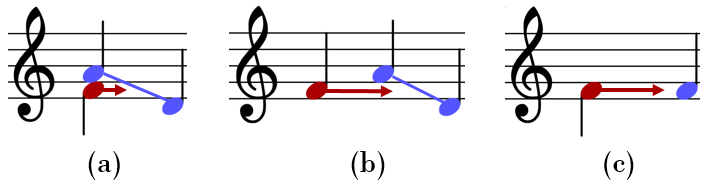}
  \caption{Examples of note blocking.}
  \label{fig:block}
\end{figure}

Figure~\ref{fig:block} shows 3 prototypical examples of a note being blocked by a voice. 

Formally, we say that a note $n$ is blocked by a voice $v$ if there is a pair of consecutive notes $(n_1, n_2)$ in $v$ such that $n$ is blocked by $(n_1, n_2)$. By definition, this can happen in either of two situations:
\begin{enumerate}
  \item The pitch of $n$ is between the pitches of $n_1$ and $n_2$, and the onset of $n$ is less than or equal to the onset of $n_1$. The first two cases in Figure~\ref{fig:block} are in this category.
  \item Note $n$ has the same pitch as $n_1$ or $n_2$, as shown in the third case in Figure~\ref{fig:block}.
\end{enumerate}

As shown in Figure~\ref{fig:block_see_you_again_repeat}, a note that is blocked by a voice can still be connected with another note ahead. Blocking will be necessary however for defining a proper ordering of active voices, as described in the next section.

\section{Envelope Extraction}
\label{sec:envelope-extraction}

\begin{figure}[H]
  \centering
  \includegraphics[width=0.5\columnwidth]{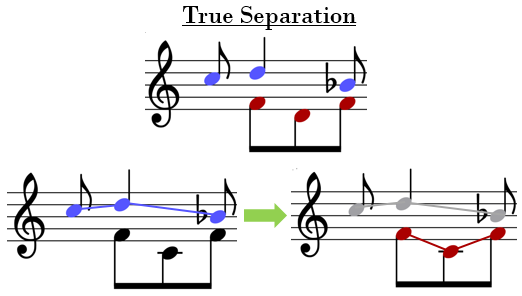}
  \caption{Voice separation as iterative envelope extraction in measure 12 of ``How to Save a Life''.}
  \label{fig:iterative_envelope}
\end{figure}

We propose a baseline system for voice-separation that iteratively extracts the upper {\it envelope} i.e. the topmost monophonic sequence of non-overlapping notes. Figure~\ref{fig:iterative_envelope} shows how the iterative envelope extraction process works on a simple measure 

The top measure represents the input, together with the manually annotated voices. The envelope extraction model starts with the raw, unannotated input and extracts its upper envelope, as shown in the bottom left measure. This envelope will become the first voice and will be removed from the input, as shown in the second measure at the bottom. We again apply the same envelope extraction process to obtain the second voice, shown in the last measure on the bottom staff. Removing this second envelope from the current input results in an empty set and correspondingly the baseline algorithm stops. In this example, the envelope extraction algorithm matches the manually annotated voices perfectly.

\subsection{Envelope Extraction as a Note to Voice Assignment Model}
\label{sec:envelope-model}

\begin{figure}[H]
  \centering
  \includegraphics[width=0.8\columnwidth]{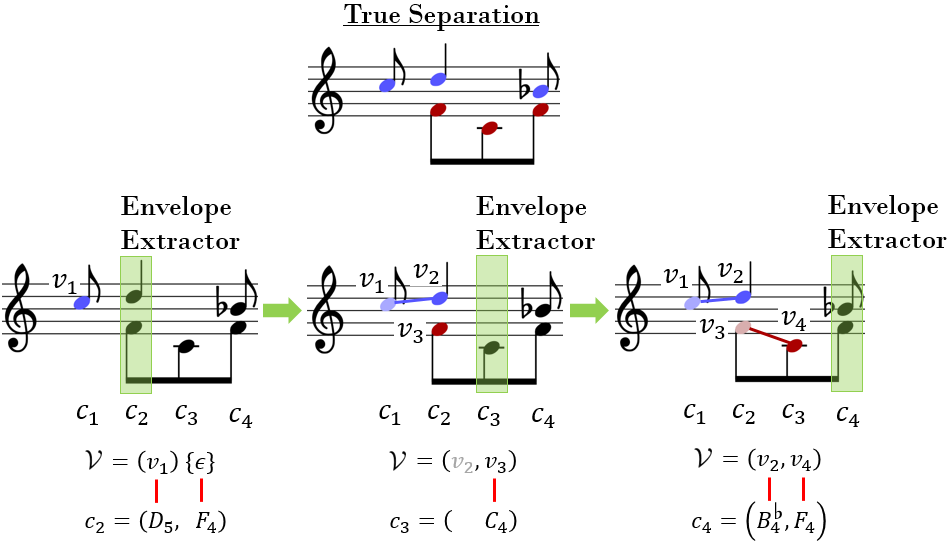}
  \caption{Iterative envelope extraction as an assignment model in measure 12 of ``How to Save a Life".}
  \label{fig:assignment_envelope}
\end{figure}

The voice separation method described above proceeds in multiple left-to-right passes over the input, extracting the top envelope in each pass and removing it from the current input, until no more notes are left in the input. The same result can be obtained in just one left-to-right pass over the chords in the input, at each step assigning notes from the current chord to voices in the active set, followed by an update of the active voice set. Whenever a note $n$ is assigned to a voice $v$, the active voice set is updated by removing $v$ and adding the new voice ending in $n$. Figure~\ref{fig:assignment_envelope} demonstrates this left-to-right note-to-voice assignment version of the envelope extractor. At each onset, we order the notes in current chord $c_t$ in descending order, based on their pitch. We also keep the active voices sorted in descending order, from ``highest'' to ``lowest''. We assign the $j^{th}$ note in $c_t$ to the $j^{th}$ {\it non-overlapping}, complete active voice in $\mathcal{V}$. An active voice $v$ is said to overlap a note $n$ if the last note in $v$ overlaps $n$.  Notes in $c_t$ not assigned to a complete active voice are assigned to the empty voice, as shown for the note F$_4$ in $c_2$ in the example. This results in a new voice being created and inserted in the set of active voices.

\noindent After processing the last chord in Figure~\ref{fig:assignment_envelope}, we see that all voices were separated correctly. One thing that still needs to be specified in the procedure above is how to keep the active voices in $\mathcal{V}$ sorted from ``highest'' to ``lowest''. Intuitively, we want voices that sound higher to be sorted above voices that sound lower, e.g. the soprano voice should be ordered before the alto voice. One possibility for sorting the active voices is to order them based on the pitch of their last note, which is consistent with how the complete active voices are ordered at each step in Figure~\ref{fig:assignment_envelope}. However, this can lead to results that contradict the original envelope extraction procedure, as shown in Figure ~\ref{fig:envelope_cross} below.

\begin{figure}[H]
  \centering
  \includegraphics[width=0.7\columnwidth]{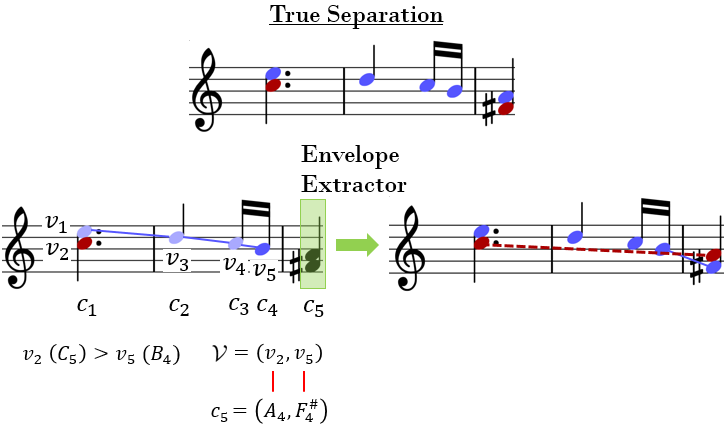}
  \caption{Ordering active voices based on the pitch value of their final notes in measures 63-65 ``Fur Elise".}
  \label{fig:envelope_cross}
\end{figure}

\noindent Upon processing chord $c_4$, the newly created complete active voice $v_5$ is placed after $v_2$, based on their last note, resulting in the ordering $\mathcal{V} = \{v_2, v_5\}$. Then the position-wise assignment of notes from chord $c_5 = \{A_4, F\#_4\}$ to voices in $\mathcal{V} = \{v_2, v_5\}$ results in two crossing voices, which would not happen in the original envelope extraction algorithm. Algorithm~\ref{alg:envelope} shown below fixes this problem by requiring complete active voices to preserve the ordering established by their partial active voice components in previous iterations. For this model, the active set $\mathcal{V}$ is constrained to contain only complete active voices. Whenever a note $n_j$ is assigned to a voice $v$, the resulting complete voice replaces the previous voice $v$ in $\mathcal{V}$ in the same position, unless the new voice crosses an existing voice.

\begin{algorithm}
\caption{{\sc EnvelopeModel}($\mathcal{C}$)}
\label{alg:envelope}
\begin{algorithmic}[1]
\Require A sequence of chords $\mathcal{C} = \{c_1, c_2, ..., c_{|\mathcal{C}|}\}$.
\Ensure A set of note-to-voice assignments $\mathcal{P}$.
\State $\mathcal{P}, \mathcal{V} \leftarrow \emptyset$
\For{{\bf each} chord $c_t = \{n_1, n_2, .., n_{l_t}\}$ in $\mathcal{C}$}
  \State let $m \leftarrow$ the number of voices in $\mathcal{V}$ that do not overlap $c_t$
  \State let $\mathcal{X} \leftarrow \emptyset$ be the set of crossing voices
  \For{$j = 1 \to \min{(m, l_t)}$}
     \Comment{Map notes to voices.}
     \State let $v \leftarrow$ the $j^{th}$ non-overlapping voice in $\mathcal{V}$
     \State create a new voice $v' \leftarrow v \leadsto n_j$
     \State update $\mathcal{P} \leftarrow \mathcal{P} \cup \{(n_j, v)\}$
     \If{$v'${\it crosses} a voice in $\mathcal{V}$}
        \State update $\mathcal{X} \leftarrow \mathcal{X} \cup \{v\}$
     \Else
        \State replace $v$ with $v'$ in $\mathcal{V}$
     \EndIf
  \EndFor
  \For{{\bf each} $v' = v \leadsto n \in \mathcal{X}$}
  \Comment{Insert the crossing voices.}
        \State update $\mathcal{V} \leftarrow \mathcal{V} - \{v\}$
        \State {\sc InsertVoice}($v', \mathcal{V}$)
  \EndFor
  \For{$j = m+1 \to l_t$}
  \Comment{Create and insert new voices.}
     \State create a new voice $v' \leftarrow \epsilon \leadsto n_j$
     \State update $\mathcal{P} \leftarrow \mathcal{P} \cup \{(n_j, \epsilon)\}$
     \State {\sc InsertVoice}($v', \mathcal{V}$)
  \EndFor
\EndFor
\State \Return $\mathcal{P}$
\end{algorithmic}
\end{algorithm}

\begin{algorithm}
\caption{{\sc InsertVoice}($v'$, $\mathcal{V}$)}
\label{alg:insert}
\begin{algorithmic}[1]
\Require A new complete voice $v'$.
\Require A sequence of complete voices $\mathcal{V} = \{v_1, v_2, ..., v_k\}$.
\Ensure The updated sequence of active voices $\mathcal{V}$.
\State let $j \leftarrow 1$
\While{$j \leq k$ and ($v_j$ is {\it blocked} by a voice in $\mathcal{V}$ or $v_j.last > v'.last$)}
   \State set $j \leftarrow j + 1$
\EndWhile
\If{$j \leq k$}
   \State insert $v'$ in $\mathcal{V}$ right before $v_j$
\Else
   \State insert $v'$ in $\mathcal{V}$ right after $v_k$
\EndIf
\State \Return $\mathcal{V}$
\end{algorithmic}
\end{algorithm}

\begin{figure}[H]
  \centering
  \includegraphics[width=0.85\columnwidth]{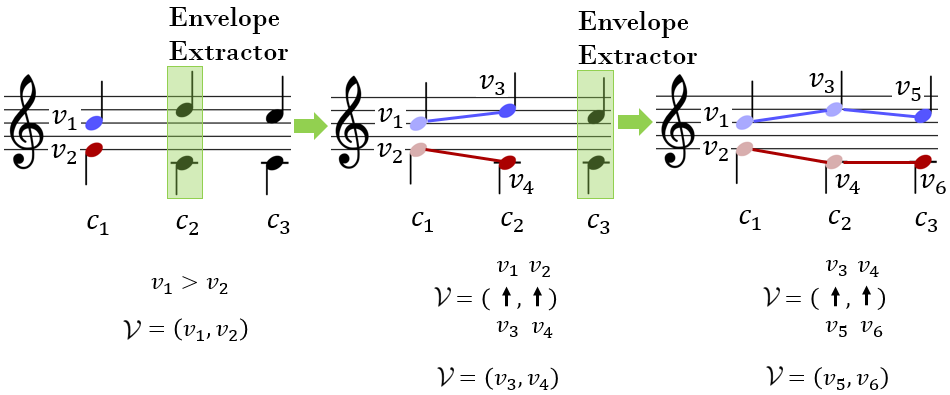}
  \caption{Active voice ordering, paired voice insertion.}
  \label{fig:active_order_pair_insert_envelope}
\end{figure}

Figure~\ref{fig:active_order_pair_insert_envelope} shows how voices are inserted in the set of active voices $\mathcal{V}$ at each step, when running Algorithm~\ref{alg:envelope} using the insertion procedure from Algorithm~\ref{alg:insert}. This is a simple example, in which no voice crossing or blocking is encountered.

\begin{figure}[H]
  \centering
  \includegraphics[width=0.7\columnwidth]{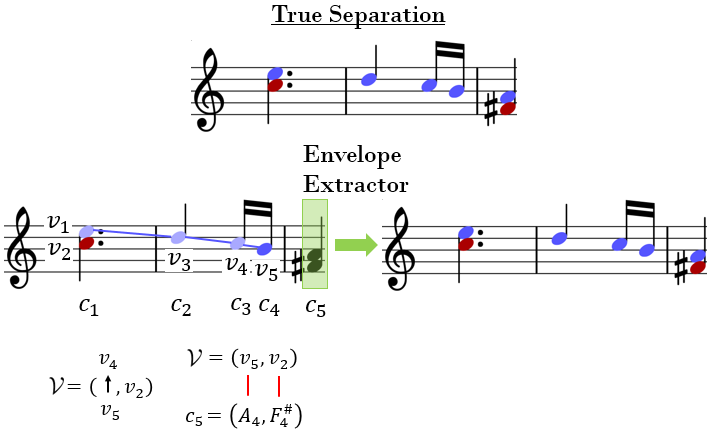}
  \caption{Active voice ordering after block in measures 63-65 ``Fur Elise".}
  \label{fig4-1g}
\end{figure}

Going back to measures 63-65 in ``Fur Elise", using Algorithm~\ref{alg:envelope} with the insertion procedure from Algorithm~\ref{alg:insert} will now result in the correct voice separation, as shown in Figure~\ref{fig4-1g}. Even though the last note in $v_5$ has a lower pitch than the last note of $v_2$, step 12 in Algorithm~\ref{alg:envelope} will replace $v_4$ with $v_5$, thus placing it before $v_2$ in the active set $\mathcal{V}$, which leads the envelope extraction algorithms to output the correct voices in this example.

\begin{figure}[H]
  \centering
  \includegraphics[width=0.4\columnwidth]{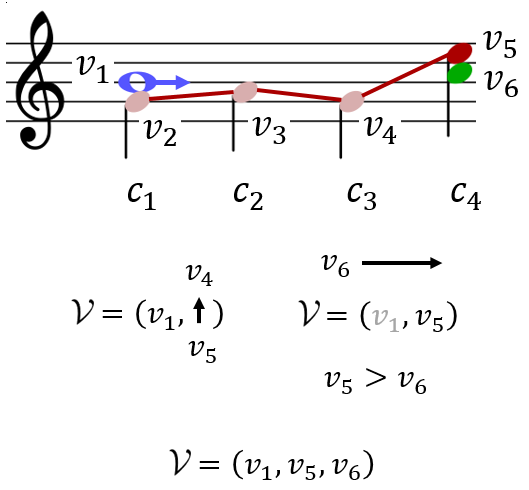}
  \caption{Active voice ordering, empty voice insertion.}
  \label{fig:active_order_empty_insert_envelope}
\end{figure}

Figure~\ref{fig:active_order_empty_insert_envelope} is provided to demonstrate the case where newly formed voices are inserted into $\mathcal{V}$. The envelope model in Algorithm~\ref{alg:envelope} constructed $v_5$ by assigning $E_5$ to the only non-overlapping active voice $v_4$ in step 6, after which it replaced $v_4$ with $v_5$ in $\mathcal{V}$ in step 12. Voice $v_6$ on the other hand was brought about by the assignment to an empty voice in step 14, after which Algorithm~\ref{alg:insert} is run to insert it into $\mathcal{V}$. At this point, the sequence of active voices is $\mathcal{V} = \{v_1, v_5\}$. Because $v_1$ is blocked by another voice in $\mathcal{V}$ (i.e. $v_5$), it is skipped in steps 2 and 3. Voice $v_5$ is skipped too, since its last note is higher than the last note of $v_6$. Therefore, in step 7 of Algorithm~\ref{alg:insert}, $v_6$ ends up being inserted at the end of the sequence of active voices.

By matching notes to voices in the order of their pitch, i.e. higher notes to higher voices, lower notes to lower voices, the envelope extraction model rarely creates crossing voices. However, because the model cannot assign an overlapping voice to a note, crossing voices may sometimes be introduced into the ordered active sequence $\mathcal{V}$. 
\begin{figure}[h]
  \centering
  \includegraphics[width=0.4\columnwidth]{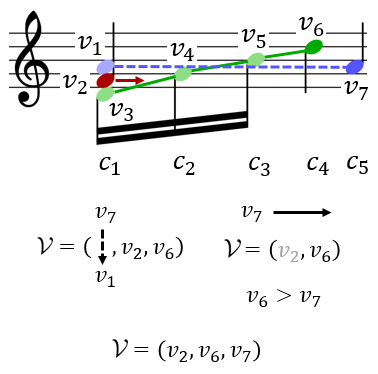}
  \caption{Active voice ordering, crossing voice insertion.}
  \label{fig:active_order_cross_insert_envelope}
\end{figure}
Figure~\ref{fig:active_order_cross_insert_envelope} demonstrates such a scenario. After processing chord $c_1$, the sequence of active voices is ordered as $\mathcal{V} = \{v_1, v_2, v_3\}$. Because $v_1$ and $v_2$ overlap with $c_2$, $c_3$, and $c_4$, the 3 notes following $c_1$ are assigned to the same voice by Algorithm~\ref{alg:envelope}, starting with $v_3$. Thus, after processing chord $c_4$, the ordered sequence of active voices will be $\mathcal{V} = \{v_1, v_2, v_6\}$. Upon reaching chord $c_5$, the last note in the highest voice $v_1$ has ended, so step 6 in the algorithm will assign it to the corresponding highest note $A_4$ from $c_5$. This results in a voice (blue $v_7$) crossing another voice (green $v_6$). Since $v_7$, which was created from $v_1$, now crosses $v_6$, voice $v_1$ will be removed from $\mathcal{V}$ in step 9, and voice $v_7$ will be inserted into the resulting $\mathcal{V} = \{v_2, v_6\}$ in step 10 by running the voice insertion Algorithm~\ref{alg:insert}. Because $v_2$ is blocked by $v_6$, it is skipped in steps 2 and 3 of this algorithm. Voice $v_6$ is skipped too, because its last note is above $v_7$. Consequently, $v_7$ is inserted after $v_6$, resulting in a sequence of active voices ordered as $\mathcal{V} = \{v_2, v_6, v_7\}$.

\section{Note-Level Voice Separation}
\label{sec:note-level}

\begin{figure}[h]
  \centering
  \includegraphics[width=0.9\columnwidth]{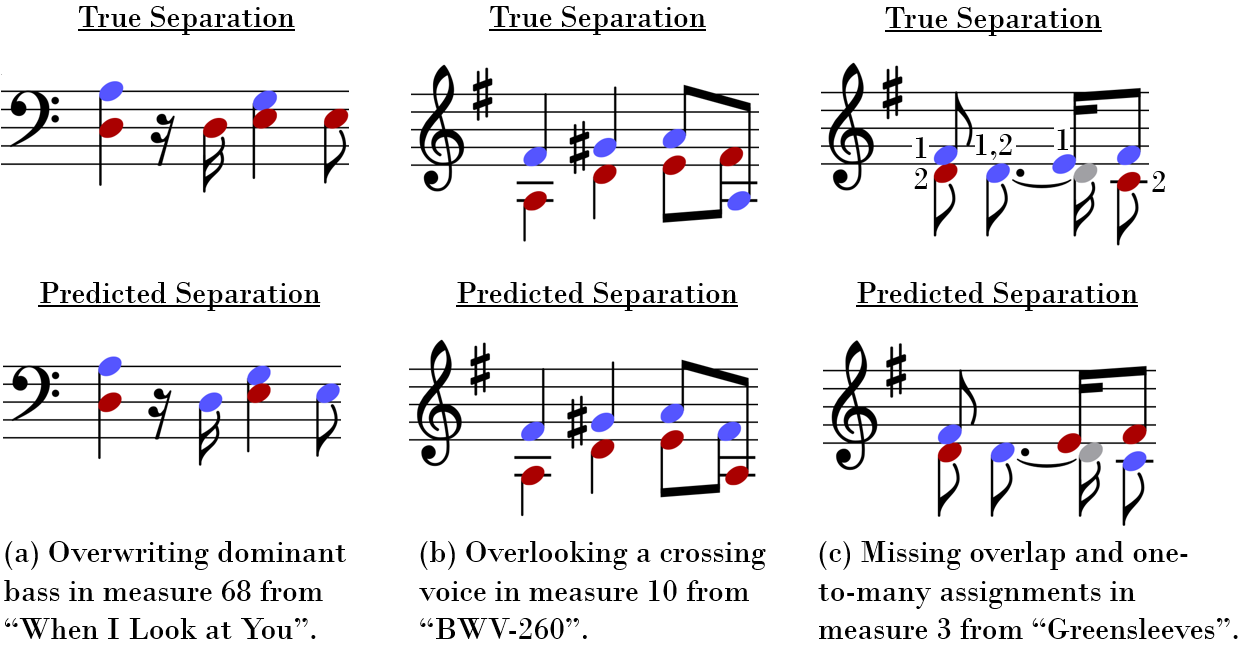}
  \caption{Notable weaknesses of the envelope extraction assignment model.}
  \label{fig:weaknesses}
\end{figure}

In the previous section, we introduced the envelope extraction assignment model, which, given an ordered chord $c_t$ and an ordered set of the active voices $\mathcal{V}$, performs a one-to-one voice separation by assigning the $i^{th}$ note in $c_t$ to the $i^{th}$ non-overlapping, complete active voice in $\mathcal{V}$. Any notes in $c_t$ that are not assigned to a complete active voice are assigned to the empty voice. While we expect a strong overall performance from this assignment model, it has a few glaring weaknesses in its design, three of which are illustrated in Figure~\ref{fig:weaknesses}. In the first example, there is a short rest separating the first two notes in the upper voice. Because the envelope extractor constructs its voices along the topmost layer of notes, when it reaches the rest it is going to wrongly assign the concurrent $D_3$ to the topmost voice. The envelope extractor will always avoid crossing voices that end on notes with the same onset, even in situations where it might be desired, as shown in the second example. Furthermore, the envelope extractor cannot handle overlapping notes in the same voice, as shown in the third example. Being by definition a one-to-one voice assignment model, the envelope extractor can handle neither convergence nor divergence.

In order to address the weaknesses of the envelope extractor, we could augment its assignment model with more perception-based separation rules. For example, to help it extract the correct top voice in the first example, we could 1) allow the assignment model to accept rests up to a certain pre-defined duration in a voice, while also 2) prohibiting the pairing of two notes separated by a pitch distance larger than some user defined threshold. Although a seemingly effective idea, there are a few caveats to consider in developing and operating such rules. Specifically, user defined parameters, such as the pitch distance threshold, are likely to be data dependent, which makes it difficult to manually tune an optimal configuration for the overall assignment model. Also, after implementing a voice separation rule, additional time and code must be invested to ensure that it does not conflict with or breaks the functionality of other system rules or features. In general, applying, updating, and maintaining the compatibility of hard-coded voice separation rules would be an overly complicated and time consuming process. Consequently, rather than construct a convoluted and rigid rule-based assignment model, we instead propose implementing a set of perceptually-informed raw features and using them as inputs to two neural network models whose parameters are automatically trained on songs annotated with voice information. The {\bf note-level} model assigns notes to voices separately for each note in a chord and is described in this section. In the next section we introduce a more general {\bf chord-level} model that trains a convolutional neural network to assign notes to voices jointly for all the notes in a chord.

At each onset $t$, the neural note-level assignment model assigns each note $n$ from the current chord $c_t$ to one or more of the voices $v$ contained in the active set $\mathcal{V}$ based on the output of a trained assignment probability $p(n, v)$. The note-level assignment procedure is described in Algorithm~\ref{alg:note-level-model} and depends on 3 hyper-parameters: a minimum threshold $\tau$ for the assignment probability $p(n, v)$, a maximum allowed number of convergent voices $\alpha$, and a maximum number of divergent voices $\beta$. When the convergence and divergence limits $\alpha$ and $\beta$ are set to 1, the the note level algorithm assigns notes to voices in a {\it one-to-one} scenario. Higher values for the assignment threshold $\tau$  leads to more new voices being create. For each chord $c_t$, of all the possible note to active-voice pairs, the maximum probability pair $(n, v)$ is selected (step 6). If the empty voice obtains the maximum score or the maximum score is less than the required threshold $\tau$ (step 7), all pairs containing $n$ are removed from further consideration (step 10) and, if the note $n$ has not been assigned any voice, it starts a new voice by being assigned to the empty voice. Otherwise, the pair $(n, v)$ is added to the assignment map and removed from further consideration (steps 12-13). If the number of voices converging to $n$ reached the maximum allowable number $\alpha$, all pairs containing note $n$ are removed from further consideration (steps 14-16). Similarly, if the number of notes into which a voice $v$ is diverging is reached the maximum allowable $\beta$, all pairs containing voice $v$ are removed from further consideration (steps 17-19). Thus, by the end of the while loop in lines 5-19, each note in the current chord $c_t$ has been assigned either to one or more active voices from $\mathcal{V}$, or to the empty voice. The newly created complete voices from $\mathcal{A}$ are then inserted in the current sequence of active voices $\mathcal{V}$ using the insertion-sort based procedure shown later in Algorithm~\ref{alg:insert-o2m}, and the newly created note to voice pairs are added to the overall set of pairs $\mathcal{P}$.

\begin{algorithm}
\caption{{\sc NoteLevelModel}($\mathcal{C}$, $\tau$, $\alpha$, $\beta$)}
\label{alg:note-level-model}
\begin{algorithmic}[1]
\Require A sequence of chords $\mathcal{C} = \{c_1, c_2, ..., c_{|\mathcal{C}|}\}$.
    \Require Assignment threshold $\tau$, convergence limit $\alpha$, divergence limit $\beta$
\Ensure A set of note-to-voice assignments $\mathcal{P}$.
\State $\mathcal{P} \leftarrow \emptyset$, $\mathcal{V} \leftarrow \{\epsilon\}$
\For{{\bf each} chord $c_t = \{n_1, n_2, .., n_{l_t}\}$ in $\mathcal{C}$}
    \State $\mathcal{A} \leftarrow \{n_1: \emptyset, n_2: \emptyset, ..., n_{l_t}: \emptyset\}$
    \Comment{notes sorted in descending order of pitch.}
    \State $Pairs \leftarrow c_t \times \mathcal{V}$
    \While{$Pairs$ not empty}
      \State let $(n,v) = \displaystyle\argmax_{(n,v) \in Pairs} p(n,v)$
      %\If{$p(n,v) < \tau$}
      % \State $v \leftarrow \epsilon$
      %\EndIf
      \If{$v = \epsilon$ {\bf or} $p(n,v) < \tau$}
         \Comment{done with note $n$?}
         \If{$\mathcal{A}[n] = \emptyset$}
         \State remove all pairs $(n, \_)$ from $Pairs$.
             \State $\mathcal{A}[n] \leftarrow \{\epsilon\}$
	 \EndIf
      \Else
         \State $\mathcal{A}[n] \leftarrow \mathcal{A}[n] \cup \{v\}$
         \State $Pairs \leftarrow  Pairs - \{(n, v)\}$
         \State let $conv \leftarrow \left|\mathcal{A}[n]\right|$
         \If{$conv = \alpha$}
            \Comment{done with note $n$?}
            \State remove all pairs $(n, \_)$ from $Pairs$
         \EndIf
         \State let $div \leftarrow$ number of pairs in $\mathcal{A}$ containing $v$
         \If{$div = \beta$}
            \Comment{done with voice $v$?}
            \State remove all pairs $(\_, v)$ from $Pairs$
         \EndIf
      \EndIf
    \EndWhile
    \State {\sc InsertVoices}($\mathcal{A}$, $\mathcal{V}$)
    \For{{\bf each} $(n, \mathcal{A}[n]) \in \mathcal{A}$}
       \For{{\bf each} $v \in \mathcal{A}[n]$}
           \State $\mathcal{P} \leftarrow \mathcal{P} \cup \{(n, v)\}$
       \EndFor
    \EndFor
    \State \Return $\mathcal{P}$
\EndFor
\end{algorithmic}
\end{algorithm}

%\begin{enumerate}%[leftmargin=1mm]
%  \item {\bf One-to-One Ranking}: Assign a note $n$ to the active voice $v$ that maximizes $p(n, v)$:
%  \begin{equation}
%    v(n) = \displaystyle\argmax_{v \in \mathcal{V}} p(n, v)
%  \end{equation}
%  \item {\bf One-to-Many Ranking}: Assign a note $n$ to all active voices $v$ whose assignment probability $p(n, v)$ exceeds some user defined threshold $\tau$:
%  \begin{equation}
%    V(n) = \{v \in \mathcal{V}\ |\ p(n, v) > \tau\}
%  \label{eq:one-to-many-ranking}
%  \end{equation}
%\end{enumerate}
%The one-to-one scenario is the simplest one and assumes that a note can belong to only a single voice. The one-to-many scenario is more general and allows for a note to belong to multiple voices. In both scenarios, notes from the current chord $c_t$ are assigned to voices in the order of their maximal score $p(n, v)$. 

To compute the assignment probability $p(n, v)$, we first define a vector $\Phi(n, v)$ of perceptually informed compatibility features (Section~\ref{sec:note-level-features}). The probability is then computed as $p(n, v | \theta) = \sigma(\mathbf{w}^Th_W(n, v))$, where $\sigma$ is the sigmoid function and $h_W(n, v)$ is the vector of activations of the neurons on the last hidden layer in a neural network with input $\Phi(n, v)$. To train the network parameters $\mathbf{\theta} = [\mathbf{w}, W]$, we maximize the likelihood of the training data:
\begin{equation}
  \displaystyle\hat{\theta} = \argmax_{\theta} \prod_{t=1}^T \prod_{n \in c_t} \prod_{v \in \mathcal{V}} p(n,\!v|\theta)^{l(n,v)}(1 - p(n,\!v|\theta))^{1\!-l(n,v)}
\label{eq:likelihood}
\end{equation}
where $l(n,v)$ is a binary label that indicates whether or not note $n$ was annotated to belong to voice $v$ in the training data. 

\begin{figure}[H]
  \centering
  \includegraphics[width=0.8\columnwidth]{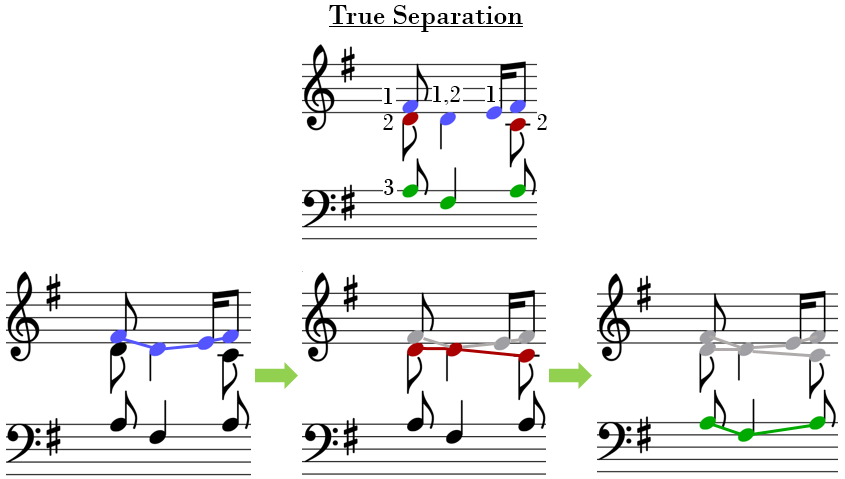}
  \caption{Voice separation as iterative envelope extraction in a one-to-many scenario.}
  \label{fig:iterative_envelope_o2m}
\end{figure}

The note-level model shown in Algorithm~\ref{alg:note-level-model} is more flexible than the envelope extraction model from Algorithm~\ref{alg:envelope}, in particular it allows for convergence and divergence. Given that voices are relatively layered and crossing is rare, it would be useful to incorporate the corresponding features of the envelope extractor into the note-level assignment model. One way to do this is by developing input features that relate the position of an active voice $v$ in an ordered sequence of active voices $\mathcal{V}$ to the position of an unassigned note $n$ in the ordered chord $c_t$. For example, a binary feature could encode whether both the active voice $v$ and the unassigned note $n$ are the highest (position 0) in the active voice sequence $\mathcal{V}$ and the chord $c_t$, respectively. In such cases, the envelope extraction model connects $n$ $v$ (bar overlapping) which perceptually makes sense, as both the active voice and the unassigned note are likely to form the soprano line. However, to use such position-based features requires the sequence of active voices $\mathcal{V}$ to be sorted. Because the note-level model allows for convergence and divergence, the active voice sequence $\mathcal{V}$ needs to contain both complete and partial active voices. Figure~\ref{fig:iterative_envelope_o2m} shows how the envelope extraction model would create the correct convergent voices if it were allowed to assign notes to partial active voices: the red $C_4$ in the fourth chord is connected back to the D$_4$ in the second chord, which is the last note in a partial voice that is followed (completed) by the E$_4$ note in the third chord.

The necessity of including partial voices in the sequence of active voices requires the design of an insertion based ordering of the active voices that is different from the one used in the envelope extraction model, where all active voices were assumed to be complete. At the same time, we need to ensure that removing the partial voices from the ordered sequence of active voices (containing both partial and complete active voices) results in the same ordered sequence of complete active voices as the one used in the envelope extraction model. This would enable the note-level model in the one-to-one setting ($\alpha$ and $\beta$ equal with 1) to replicate the envelope extraction model behavior. The insertion-based ordering of active voices is shown in Algorithm~\ref{alg:insert-o2m}.

\begin{algorithm}
  \caption{{\sc InsertVoices}($\mathcal{A}$, $\mathcal{V}$)}
  \label{alg:insert-o2m}
  \begin{algorithmic}[1]
    \Require An assignment of note-to-voice pairs $\mathcal{A}$.
    \Require A sequence of active voices $\mathcal{V} = \{v_1, v_2, ..., v_{k}\} \cup \{\epsilon\}$
    \Ensure The updated sequence of active voices $\mathcal{V}$.

    \State let $\mathcal{V}^{'} \leftarrow \emptyset$
    \For{{\bf each} $(n, \mathcal{A}[n])$ in $\mathcal{A}$}
      \State sort voices in $\mathcal{A}[n]$ based on their positions in $\mathcal{V}$
      \State create a new voice $v' \leftarrow \mathcal{A}[n] \leadsto n$

      \If{$\mathcal{A}[n] = \{\epsilon\}$ {\bf or} $v'$ crosses a voice in $\mathcal{V}$}
        \State update $\mathcal{V}^{'} \leftarrow \mathcal{V}^{'} \cup \{v'\}$
      \Else
        \For{{\bf each} $v \in \mathcal{A}[n]$ in ascending order}
          \If{$n < v.last$}
            \State insert $v'$ in $\mathcal{V}$ after the lowest voice diverging from $v$, or after $v$ if none
            \State {\bf break}
          \EndIf
        \EndFor
        \If{$v'$ was not inserted in $\mathcal{V}$ yet}
          \State insert $v'$ in $\mathcal{V}$ right before highest voice in $\mathcal{A}[n]$
        \EndIf
      \EndIf
    \EndFor

    \For{{\bf each} $v'$ in $\mathcal{V}^{'}$}
      \State let $j \leftarrow 1$
      \While{$j \leq k$ \textbf{and} ($v_j$ is {\it blocked} by a voice in $\mathcal{V}$ \textbf{or} $v'.last < v_j.last$)}
        \State set $j \leftarrow j + 1$
      \EndWhile
      \If{$j \leq k$}
        \State insert $v'$ in $\mathcal{V}$ right before $v_j$
      \Else
        \State insert $v'$ in $\mathcal{V}$ right after $v_k$
      \EndIf
    \EndFor
    \State \Return $\mathcal{V}$
  \end{algorithmic}
\end{algorithm}

\begin{figure}[H]
  \centering
  \includegraphics[width=0.8\columnwidth]{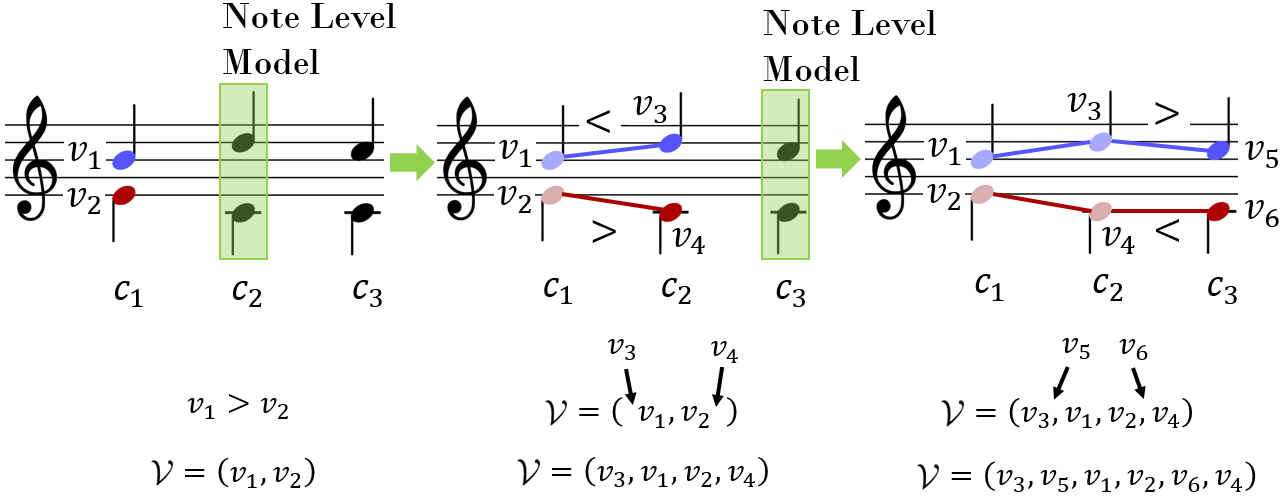}
  \caption{Active voice ordering, paired voice insertion.}
  \label{fig:active_order_pair_insert}
\end{figure}

Figure~\ref{fig:active_order_pair_insert} shows a simple one-to-one example in which the sequence of active voices is updated at every chord by Algorithm~\ref{alg:insert-o2m}, based on the assignments produced by Algorithm~\ref{alg:note-level-model} in lines 3-19. For the first chord $c_1$, these lines produce an assignment to empty voices for all notes. Lines 4-6 in Algorithm~\ref{alg:insert-o2m} then create the first voices from the notes in this first chord, and lines 14-21 sort them in $\mathcal{V}$ using insertion sort. Any new voice created from assigning notes from subsequent chords will be inserted into $\mathcal{V}$ in a position that is adjacent to its neighboring partial active voice with special treatment for divergence (lines 8-13), and for starting new voices or crossing voices (lines 5-6 and 14-21). In other words, given a new complete active voice $v$ whose last note is paired to the last note of a partial voice $u$, $v$ will be inserted into $\mathcal{V}$ in a position that is adjacent to $u$. If the last note of $v$ has a pitch that is equal to or higher than the last note of the connecting partial voice $u$, then $v$ will be placed above $u$ in $\mathcal{V}$. Conversely, if the last note of $v$ has a pitch that is strictly lower than the last note of the connecting partial voice $u$, then $v$ will be placed below $u$ in $\mathcal{V}$. This ordering procedure for $\mathcal{V}$ is presented in Figure~\ref{fig:active_order_pair_insert}. After the first iteration, $\mathcal{V}$ is comprised of the voices $v_1$ and $v_2$ sorted in descending order of pitch. At $c_2$, voice $v_3$ is inserted directly above $v_1$ because $v_1$ is the neighboring partial voice of $v_3$ and the last note of $v_3$ is higher than the last note of $v_1$. Likewise, $v_4$ is placed directly below its connecting partial voice $v_2$. At $c_3$, we see that the last note of $v_6$ is equal in pitch to its connecting note in $v_4$. Consequently, it is placed above $v_4$ in $\mathcal{V}$. 

\begin{figure}[H]
  \centering
  \includegraphics[width=0.4\columnwidth]{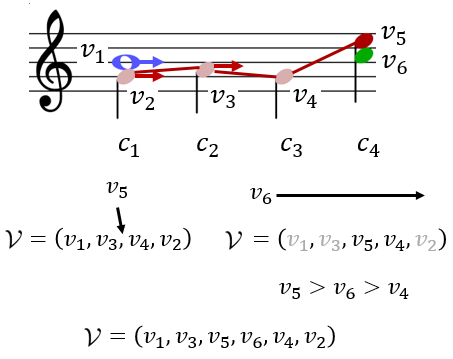}
  \caption{Active voice ordering, empty voice insertion.}
  \label{fig:active_order_empty_insert}
\end{figure}

Figure~\ref{fig:active_order_empty_insert} demonstrates how a note assigned to $\epsilon$ is inserted into $\mathcal{V}$. Assuming the two notes in $c_4$ were just assigned to the corresponding voices ($v_4$ and $\epsilon$), we now need to insert the resulting complete voices $v_5$ and $v_6$ into $\mathcal{V}$. $v_5$ will be inserted first because its last note is connected to the last note of a voice in $\mathcal{V}$. In general, complete active voices that connect to partial voices in $\mathcal{V}$ are inserted first (lines 8-13). Next, we insert the new active voices created from connection with the empty voice $\epsilon$ (the voices stored in $\mathcal{V}'$). To do so, we compare the first notes of theses active voices to the last notes of all active voices in $\mathcal{V}$ that are not blocked (line 16). Thus, we compare $v_6$ only to $v_5$ and $v_4$. $v_1$, $v_2$, and $v_3$ are ignored while inserting $v_6$ because these voices are blocked. Consequently, $v_6$ ends up positioned between $v_5$ and $v_4$ in $\mathcal{V}$.

\begin{figure}[H]
  \centering
  \includegraphics[width=0.35\columnwidth]{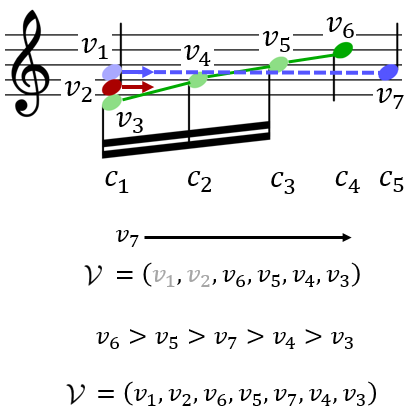}
  \caption{Active voice ordering, crossing voice insertion.}
  \label{fig:active_order_cross_insert}
\end{figure}

In situations where an active voice crosses another voice in $\mathcal{V}$, we treat the crossing voice the same as if the last note were connected to the empty voice (lines 5-6). In Figure~\ref{fig:active_order_cross_insert} shows an example, where $v_1$ created a crossing connection to $v_7$. Therefore, $v_7$ is inserted into $\mathcal{V}$ as if $v_7$ were created from the empty voice. In lines 14-21, the insertion procedure ignores the blocked voices $v_1$ and $v_2$ and only compare $v_7$ to the last notes in $v_6$, $v_5$, $v_4$, and $v_3$.

\begin{figure}[h]
  \centering
  \includegraphics[width=0.7\columnwidth]{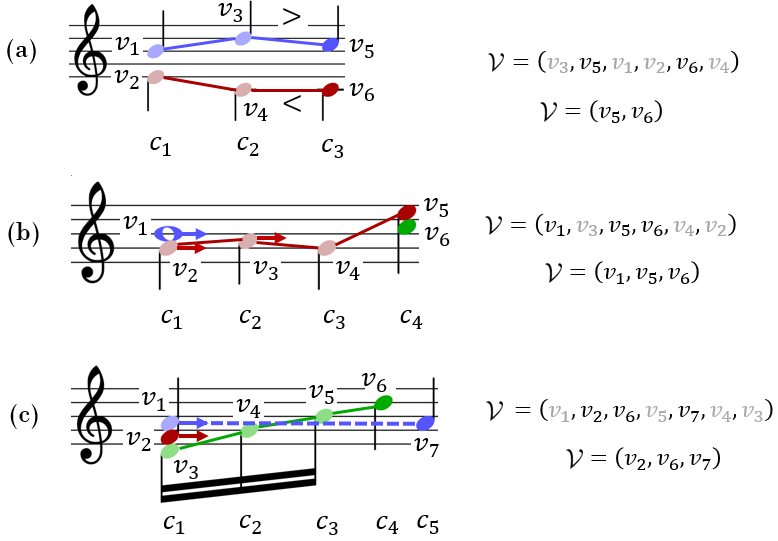}
  \caption{Active voice ordering preserves complete voice ordering.}
  \label{fig:active_order_o2o}
\end{figure}

In the examples above, we have shown only one-to-one assignments. If we wished to perform only one-to-one voice separation with the note-level model, then we could ignore all the partial active voices, as illustrated in Figure~\ref{fig:active_order_o2o}. This results in the same ordering of complete voices in $\mathcal{V}$ as the ordering produced by Algorithm~\ref{alg:insert} used by the envelope extraction model.

\begin{figure}[h]
  \centering
  \includegraphics[width=0.5\columnwidth]{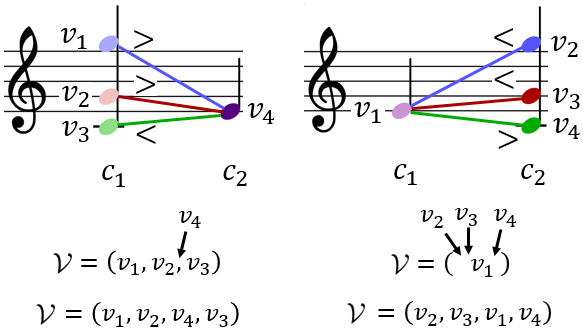}
  \caption{Active voice ordering, convergent and divergent voice insertion.}
  \label{fig:active_order_o2m_insert}
\end{figure}

Finally, Figure~\ref{fig:active_order_o2m_insert} shows how the insertion-based algorithm deals with cases of convergence and divergence. In the convergence case, $v_4$ is inserted between $v_2$ and $v_3$ as expected because the last note in the newly created $v_4$ falls between the last notes in $v_2$ and $v_3$. Similarly, in the divergence case there is nothing unusual to consider. $v_2$ and $v_3$ are placed above $v_1$ because their first notes have higher pitch than the last note of $v_1$. Furthermore, $v_2$ is positioned above $v_3$ because the first note of $v_2$ has a higher pitch than the first note of $v_3$. $v_4$ is inserted below $v_1$ because its first note is lower than the last note of $v_1$. 

\subsection{Note-Level Model: Voice Separation Features}
\label{sec:note-level-features}

The performance of the note-level voice separation model depends on the assignment probability $p(n, v)$, which is computed by a feed forward neural network based on a vector of input features $\Phi(n, v) = [\phi_1, \phi_2, ..., \phi_K]$. We define three categories of features:
\begin{enumerate}
  \item {\bf Note features:} These are features that depend only on unassigned note $n$ (Appendix~\ref{sec:note-level-note-features}).
  \item {\bf Voice features:} These are features that depend solely on the active voice $v$ (Appendix~\ref{sec:note-level-voice-features}).
  \item {\bf Pair features:} These are features that depend on both the unassigned note $n$ and the active voice $v$ (Appendix~\ref{sec:note-level-pair-features}).
\end{enumerate}
If they were to be used directly to compute $p(n, v)$, the note features and the voice features could not capture the compatibility between an unassigned note $n$ and an active voice $c$. However, when used as inputs to a neural network with one or more hidden layers, neurons on the hidden layers can learn combinations between note features and voice features that  are relevant for modeling note-voice compatibility. Thus, while pair features are manually engineered to capture known voice separation principles, the use of note and voice features gives the system the capability to learn useful combinations of note vs. voice properties that may capture different, possibly unknown data-driven voice separation principles.

All the features used in the note-level model are listed in the Appendix~\ref{sec:appendix-note-level}. There we also describe the basic concepts and notation used for defining the features (Appendix~\ref{sec:notation}), as well as the binarization scheme used for categorical values (Appendix~\ref{sec:discretization}).

\section{Chord-Level Voice Separation}
\label{sec:chord-level}

The note-level model for voice separation assigned notes to voices by computing a note-to-voice assignment probability $p(n, v)$, separately for each note in a chord and each voice in the set of active voices. This required the introduction of hyper-parameters $\tau$, $\alpha$, and $\beta$ through which the system can limit the number of voices, convergence, and divergence pairs, respectively. In the case of convergence, for example, we cannot define a feature that counts the number of active voices converging to a given note because the model assigns a note to voices only one voice at a time and thus does not know how many voices will end up being connected to that note. Hence the need for the convergence limit hyper-parameter $\alpha$ through which Algorithm~\ref{alg:note-level-model} removes a note from consideration after it has been assigned to a maximum allowed number of voices. A similar argument applies for the divergence limit hyper-parameter $\beta$. Tuning such hyper-parameters can be time consuming and can result in suboptimal assignments -- if they are not configured properly they can lead to an overabundance of convergence and divergence pairs. Furthermore, a value for $\tau$ that is too small can result in the creation of too many voices by breaking long voices into many smaller ones.

\begin{figure}[ht]
  \centering
  \includegraphics[width=0.6\textwidth]{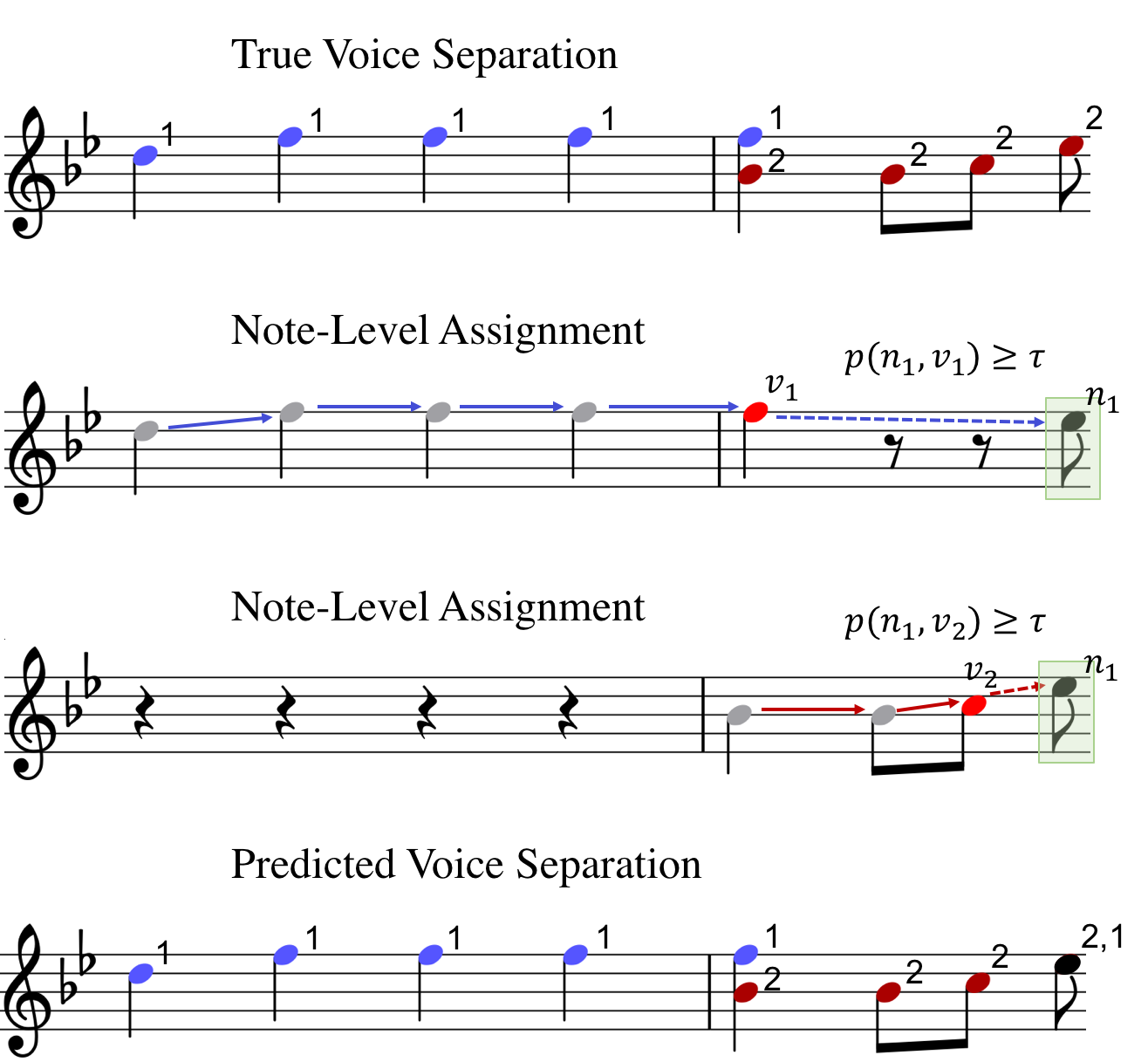}
  \caption{Expected one-to-many voice separation for measures 35-36 in ``How to Save a Life", from the note-level assignment model.}
  \label{fig:note-level-example}
\end{figure}

The most important drawback of the note-level model is its inability to utilize features that look at the entire set of note-to-voice assignments for all the notes in a chord. Through features that look at two or more candidate voice assignments, a chord-level model could use the context to determine the optimal number of voices converging to or diverging from a note. Features could also be defined to determine when crossing voices is appropriate. 
% In chapter 2, under our annotation guidelines, we present an annotated excerpt from ``21 Guns" in Figure 2.1 to help explain how the presence of one voice can influence our perception of another voice played simultaneously. 
% In the note-level assignment model, we attempt to capture a part of this influence through a few positional features, such as the voice index feature described in Section 5.1.2. Using features to capture influences between voices in a note-level assignment model is suboptimal and therefore bound to lead to incorrect note-to-voice assignments, especially when assignment model operates in a one-to-many ranking scenario. 
Figure \ref{fig:note-level-example} illustrates a type of mistake that may result from the inability of the note-level voice separation model to consider two or more voices at the same time.
Up to chord $c_{7}$, the note-level assignment model shown in Algorithm~\ref{alg:note-level-model} is assumed to have correctly assigned voices to notes using the learned assignment probability $p(n, v)$ and the assignment threshold $\tau$. In order to finalize the separation, the model must assign the $E^\flat_5$ in $c_{8}$ to either the empty voice or one or both of the active voices $v_1$ and $v_2$.
%For brevity, we only attempt to rationalize the model's likely decision to pair $E^\flat_5$ with the complete actives voices $v_1$ and $v_2$. 
In computing the assignment probability $p(n_1, v_1)$, the assignment model may penalize the pairing for the quarter rest interruption endured between the onsets of $n_1$ and $v_1$. Conversely, the close pitch proximity of $n_1$ and $v_1$ will contribute positively to $p(n_1, v_1)$. Furthermore, both $n_1$ and $v_1$ reside in the soprano, which often results in the perception of a more salient and compatible pairing. Together, these two perceptually strong melodic traits could potentially outweigh the adverse effect of the quarter rest on $p(n_1, v_1)$ to the point where $p(n_1, v_1)$ exceeds the assignment threshold $\tau$, thus resulting in a predicted pairing between $n_1$ and $v_1$. In regards to the assignment probability $p(n_1, v_2)$, given that $n_1$ and $v_2$ exhibit close pitch, positional, and temporal relationships, the assignment model will likely assign $n_1$ to $v_2$ with high confidence. Of course, with respect to the true annotation, this predicted convergence pairing would be incorrect. The voice of $v_2$ assumes the soprano starting from chord $c_6$, which drowns out the halted melody at $v_1$ and in turn makes it difficult to hear a true connection between $n_1$ and $v_1$. Because the note-level assignment model computes probabilities within a local context of a note and an active voice, without knowledge of what other voices are assigned to the same note, the model has limited ability to consider influences between voices that may or may not converge to the same note. As a result, it will be difficult to discourage the note-level assignment model from generating an excess of both predicted convergence and divergence cases.

\begin{figure}[ht]
  \centering
  \includegraphics[width=0.6\textwidth]{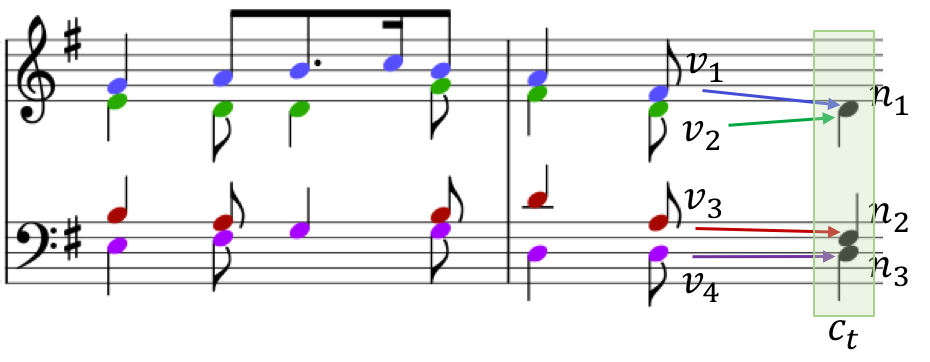}
  \caption{Convergence example in Greensleeves.}
  \label{fig:convergence}
\end{figure}

To account for interdependences among multiple voices when choosing what notes to assign to them, we propose a chord-level assignment model that simultaneously assigns notes from the current chord $c_t$ to voices in the current set of complete active voices $\mathcal{V}_t$. This is done by scoring all joint assignments of notes to voices and selecting the joint assignment that obtains the maximum score. More formally, a joint assignment is a function $a:c_t \rightarrow 2^{\mathcal{V}_t}$ that assigns each note from $c_t$ to a subset of voices from $\mathcal{V}_t$. Figure~\ref{fig:convergence} shows an example where the current chord is $c_t = \{n_1, n_2, n_3\}$ and the set of complete active voices is $\mathcal{V}_t = \{v_1, v_2, v_3, v_4, \epsilon\}$ (the empty voice $\epsilon$ is always available). A possible joint assignment is $a = \{(n_1, \{v_1, v_2\}), (n_2, \{v_3\}), (n_3, \{v_4\})\}$ in which voices $v_1$ and $v_2$ converge to note $n_1$. Let $A_t = A(c_t, \mathcal{V}_t) = \{a:c_t \rightarrow 2^{\mathcal{V}_t}\}$ be the set of all joint assignments available at position $t$ in the music, and let $f:A_t \rightarrow \mathbb{R}$ be a trainable function that maps assignments to real-valued scores. Then using the chord-level assignment model refers to finding at every position (from left to right) the joint assignment that maximizes the assignment score:
\begin{equation}
  \hat{a}_t = \argmax_{a \in \mathcal{A}_t} f(a), \;\;\; 1 \leq t \leq T
\end{equation}

The joint assignment score is computed as $f(a) = \mathbf{w}^Th(a, W)$, where $h(a, W) = h(\Phi(a), W)$ is the vector of activations of the neurons on the last hidden layer in a neural network with an input layer $\Phi(a)$ of perceptually informed compatibility features (Section~\ref{sec:chord-level-features}).
% In the chord-level assignment model, each note $n$ in the current chord $c_t$ is greedily and simultaneously assigned to one or more of the active voices $v$ by selecting the joint assignment $j$ of all notes in $c_t$ to a subset of the active voices that maximizes a trained assignment score $f(c_t, j)$, i.e. $j(c_t) = \argmax_{j \in \mathcal{J}} f(c_t, j)$, where $\mathcal{J}$ is the set of candidate joint assignments. Much like the assignment probability $p$ computed in the note-level assignment model, $f(c_t, j)$ captures the compatibility between the chord $c_t$ and the joint assignment $j$. Likewise, the joint assignment score is computed as $f(c_t, j) = \mathbf{w}^Th_W(c_t, j)$, where $h_W(c_t, j)$ is the vector of activations of the neurons on the last (hidden) layer in a neural network with an input layer $\Phi(c_t, j)$ of perceptually informed compatibility features (Section 6.1).

\begin{figure}[H]
  \centering
  \includegraphics[width=0.6\textwidth]{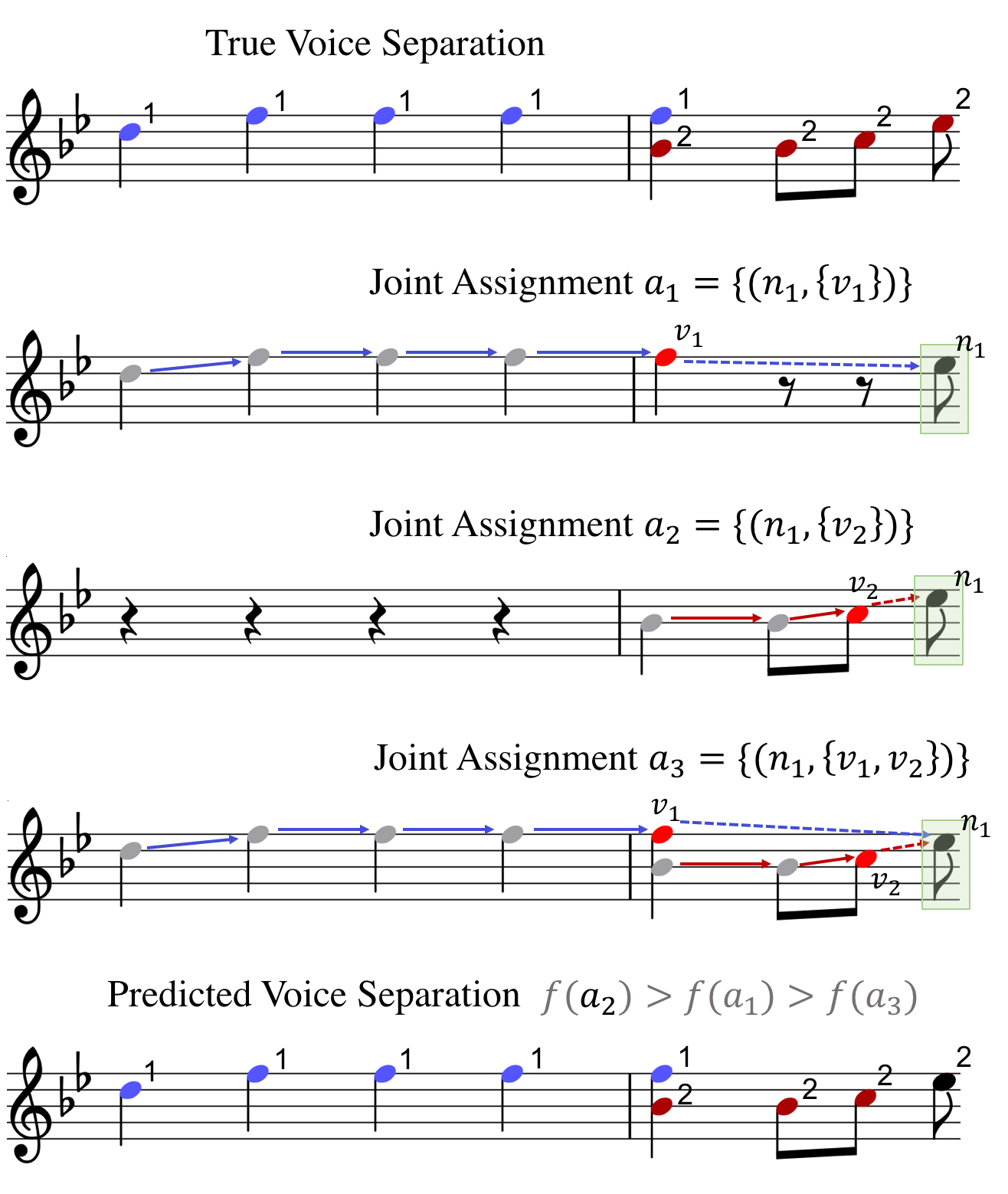}
  \caption{Expected one-to-many voice separation for measure 35-36 in ``How to Save a Life", from the chord-level assignment model.}
  \label{fig6-2}
\end{figure}

We now review in Figure \ref{fig6-2} a possible many-to-one voice separation performed by the chord-level assignment model, on the same musical segment that was used earlier in Figure~\ref{fig:note-level-example}. At the final onset, the chord-level model tests and ranks 3 different joint assignments of the singleton chord $n_1$ to the complete active voices $v_1$ and $v_2$. Again, for brevity, we do not consider the scores for the assignments that include the empty or partial active voices. Because the entire joint assignment containing the convergent voices is now scored as $f(a) = \mathbf{w}^Th(\Phi(a), W)$, we can incorporate in $\Phi(a)$ features that capture the  perceptual dominance of voice $v_1$ over voice $v_2$. As a result, we expect the model to learn to assign a lower score to this convergence case and to perform a more accurate voice separation overall.

To train the network parameters $\theta = [\mathbf{w}, W]$ used for computing the joint assignment scoring function, we minimize the following ranking criterion:

\begin{equation}
  \hat{\theta} = \argmin_{\theta} \sum_{t=1}^T max(0, 1 - f(a_t^+) + f(a_t^-))
  \label{eq:max-margin}
\end{equation}
where $a_t^+$ is the true joint assignment for chord $c_t$ and $a_t^-$ is the joint assignment different from $a_t^+$ that maximizes the assignment score:
\begin{equation}
a_t^- = \argmax_{a \in \mathcal{A}_t - \{a_t^+\}} f(a)
\label{eq:negative-sampling}
\end{equation}
The optimization objective above is based on the large-margin optimization formulation for structured and interdependent output variables introduced by Tsochantaridis et al. (\citeyear{tsochantaridis:jmlr05}), where the separation margin is defined as the minimal difference between the score of the correct label and the closest runner-up.

\begin{figure}[ht]
  \centering
  \includegraphics[width=0.75\textwidth]{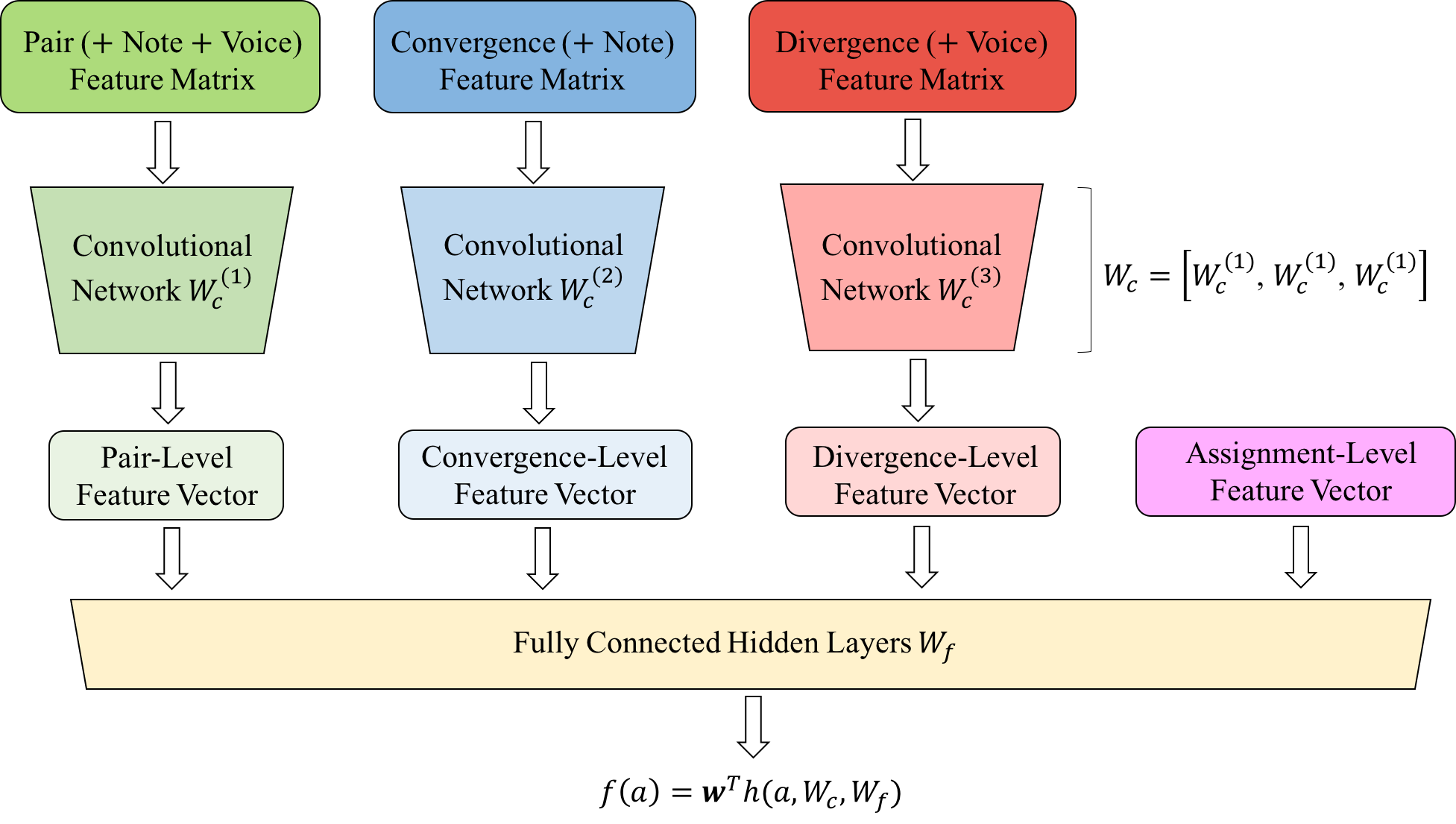}
  \caption{Complete neural network architecture of the neural chord-level assignment model.}
  \label{fig:chord-level-model}
\end{figure}

The overall neural network architecture of the chord-level assignment model is shown in Figure~\ref{fig:chord-level-model}. Four types of feature templates are used as input: pair-level, convergence-level, divergence-level, and assignment-level. The first three types of feature templates result in a variable number of features, depending on the chord-level assignment. Therefore, a special type of convolutional network parameterized by $W_c = [W_c^{(1)}, W_c^{(2)}, W_c^{(3)}]$ is used to project the variable-length feature vectors into fixed size vectors, one for each of the three feature templates. These are then concatenated with the global assignment-level feature vector and provided as input to a number of fully connected hidden layers, parameterized by $W_f$. The output of the last hidden layer is used to compute the scoring function, using a vector of parameters $\mathbf{w}$.

\begin{figure}[ht]
  \centering
  \includegraphics[width=0.7\textwidth]{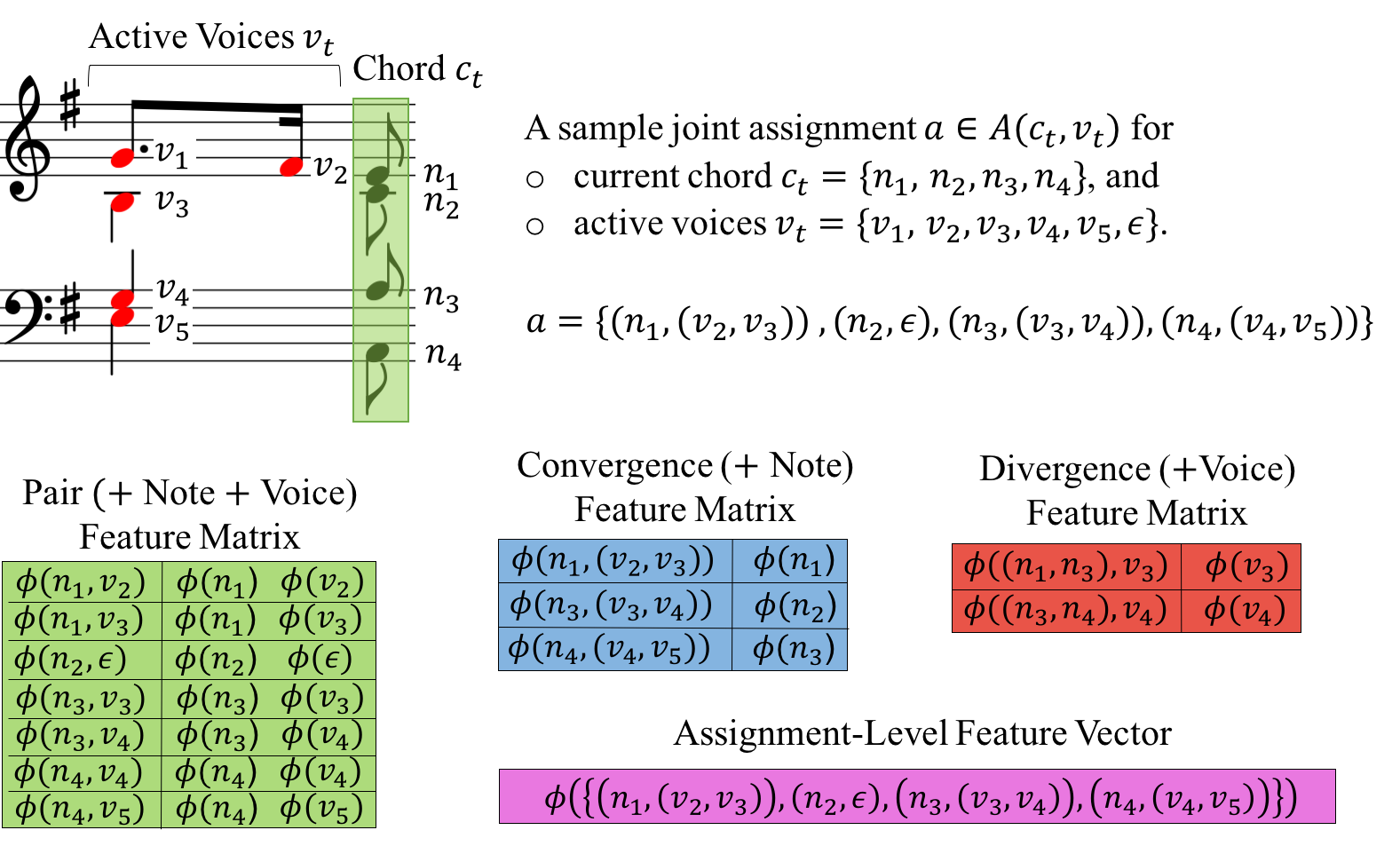}
  \caption{Feature sets comprising the input layer $\Phi(c_t, j)$ of the neural network to compute $f(c_t, j)$. The example joint assignment $j$ is drawn from chord $c_3$ in measure 7 from ``Greensleeves".}
  \label{fig:chord-level-features-example}
\end{figure}

Figure~\ref{fig:chord-level-features-example} shows a sample segment of music with a candidate chord-level assignment and the corresponding features resulting from the instantiation of feature templates for each of the four major types of features. The first type are called {\it pair features} and comprise the note (Section~\ref{sec:note-level-note-features}), voice (Section~\ref{sec:note-level-voice-features}), and pair (Section~\ref{sec:note-level-pair-features}) features that are carried over from the note-level assignment model (Section~\ref{sec:note-level}). Because the number of notes and voices in an assignment can vary, the number of {\it pair-level} feature vectors will be variable. A set of {\it convergence features} (Section~\ref{sec:convergence-features}) are defined for each convergence case contained the assignment, and concatenated with the corresponding note features. Similarly, {\it divergence features} (Section~\ref{sec:divergence-features}) are defined for each divergence case contained in the assignment, concatenated with the corresponding voice features. Likewise, the number of convergence and divergence cases varies depending on the candidate assignment, resulting in a variable number of feature vectors. Finally, a fixed-size vector of {\it assignment features} (Section~\ref{sec:assignment-features}) is defined that captures more global properties of the entire assignment.

\begin{figure}[ht]
  \centering
  \includegraphics[width=\textwidth]{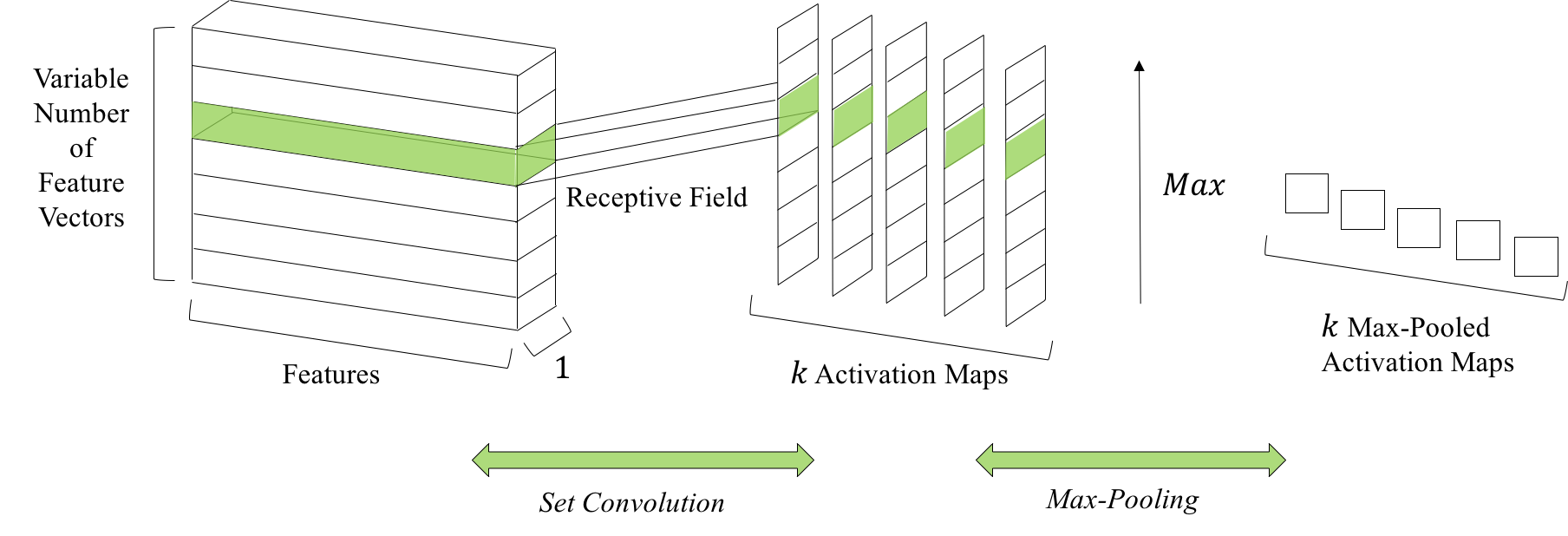}
  \caption{Convolutional network for reducing a variable number of feature vectors into a fixed-size vector.}
  \label{fig:convolution}
\end{figure}

Figure~\ref{fig:convolution} shows the convolutional network used to reduce a variable number of feature vectors into a fixed-size vector of $k$ max-pooled activations. At the convolutional layer, $k$ filters are convolved with each individual feature vector, resulting in a set of $k$ high-level feature maps. In the following layer, the $k$ activation maps are max-pooled into a set of $k$ activation values. Altogether, these $k$ values make up the fixed-size feature vector. As stated earlier, the fixed-size local input vectors obtained from the pair, convergence, and divergence features, are concatenated with the global assignment vector to form $\Phi(c_t, j)$, which is subsequently fed into the fully connected neural network used to compute $f(a)$. 

The note, voice, and pair features are carried over from the note-level assignment model whereas the convergence, divergence, and assignment level features are unique to the chord-level model and are introduced in the next section.

\subsection{Chord-Level Voice Separation Features}
\label{sec:chord-level-features}

The performance of the chord-level voice separation model depends on the scoring function $f(a)$, which is computed by the convolutional neural network from Figure~\ref{fig:chord-level-model} based on a rich set of features that are meant to capture perceptual constraints in voice separation. The first set of features is imported from the note-level model. There are also three sets of features that are specific to the chord-level model, as follows:
\begin{enumerate}
  \item {\bf Convergence features}: These features are defined for each case of two or more active voices converging on the same note in the assignment (Appendix~\ref{sec:convergence-features}). The resulting feature vector is concatenated with the corresponding note features (introduced in Appendix~\ref{sec:note-level-note-features}).
  \item {\bf Divergence features}: These features are defined for each case of one active voice diverging into two or more notes in the candidate assignment (Appendix~\ref{sec:divergence-features}). The resulting features vector is concatenated with the corresponding voice features (introduced in Appendix~\ref{sec:note-level-voice-features}).
  \item {\bf Assignment features}: These features capture global properties of the entire candidate assignment, such as whether the assignment creates crossing voices, or average pitch distances between chord notes and assigned active voices (Appendix~\ref{sec:assignment-features}).
\end{enumerate}
All the features specific to the chord-level model are listed in the Appendix~\ref{sec:appendix-chord-level}.

\section{Experimental Evaluation}
\label{sec:evaluation}

The note-level model (Section~\ref{sec:note-level}) and the chord-level model (Section~\ref{sec:chord-level}) are evaluated in a 10-fold cross validation scenario. Each dataset is partitioned into 10 folds containing an equal number of songs. The model is trained on 9 folds, tested on the remaining fold, and the process is repeated until each fold has been used for testing. The final evaluation measures are then computed by pooling the system output across the 10 folds.

The assignment probability $p(n, v| \theta)$ in the note-level model is computed using a neural network with two hidden layers consisting of 200 neurons each, with sigmoid as the activation function. The feature vector $\Phi(n, v)$ is provided as input, and an output sigmoid unit computes the probabilistic score. The network is trained to optimize an L2 regularized version of the likelihood objective shown in Equation~\ref{eq:likelihood}, using AdaDelta \citep{zeiler:corr12} and backpropagation. The regularization hyper-parameter is set to $1e\!-\!4$, while AdaDelta is run with the default setting of $\rho = 0.95$ and $\epsilon = 1e\!-\!6$. The assignment threshold in Algorithm~\ref{alg:note-level-model} is set to $\tau = 0.3$, and the model is trained for 300 epochs with a minibatch size of 20 feature vectors, i.e. 20 note-to-voice assignments made by Algorithm~\ref{alg:note-level-model}.

% {\bf Patrick: } The separation system from the ISMIR paper was trained on the one-to-one annotations of the popular and chorales datasets. The ISMIR model was given the same configurations as the multi-label note-level assignment model, however, the assignment threshold for the ISMIR model was set to 0 and only one-to-one (i.e. ranking) assignments were permitted. The ISMIR model was then tested on the one-to-many annotations of the popular and chorales datasets. 

The neural network in the chord-level model (Figure~\ref{fig:chord-level-model}) uses one hidden fully connected layer with 100 neurons, sigmoid activated. The number of kernels in the convolutional networks (Figure~\ref{fig:convolution}) is set to 50 for the pair input, 20 for the convergence input and 20 for the divergence input. The network is trained to optimize the L2 regularized version of the max-margin objective shown in Equation~\ref{eq:max-margin}, using AdaDelta and backpropagation. Similar to the note-level setting, the regularization hyper-parameter is set to $1e\!-\!4$, while AdaDelta is run with the default setting of $\rho = 0.95$ and $\epsilon = 1e\!-\!6$.
To speed up training, the maximum scoring negative assignment $a_t^-$ in the chord-level model is estimated by sampling not from the entire set $\mathcal{A}_t - \{a_t^+\}$ (Equation~\ref{eq:negative-sampling}), but from a subset of 500 negative assignments that were generated at random using an iterative procedure. First, a note $n$ is selected uniformly at random without replacement from the current chord $c_t$. Then, a voice $v$ is selected at random from the current set of active voices $\mathcal{V}$ (including the empty voice $\epsilon$) and assigned to $n$. Voice selection is repeated until $v = \epsilon$, which is when the note $n$ is removed from consideration in the next note selection iteration. This process is repeated until all notes $n$ in $c_t$ have been assigned a set of voices.

\subsection{Data-Driven Constraints}
\label{sec:constraints}

In our three voice separation models, the time complexity during training and inference increases with the number of active voices, which grows linearly with the number of chords if left unconstrained. In the case of the chord-level assignment model, the number of candidate assignments is exponential in the number of active voices. To maintain a feasible time complexity, we avoid the proliferation of voices by filtering out voices ending on notes that are not within the {\it beat horizon} of the current chord. A note $n$ is within the beat horizon of another note $m$ if the offset of $n$ is no more than a user-defined number of beats away from the onset of $m$. 

Figures~\ref{fig:see-you-block} and~\ref{fig:repeat_melodies} in Section~\ref{sec:notes-rests} showed examples of two sequences of notes being connected into the same voice, even though separated by rests or other notes. There, the repetition was strong enough to override the temporal continuity principle. Therefore, instead of using this principle to define a beat horizon, we decided to use a data-driven approach. Table~\ref{tab_histogram_onset_offset} is a histogram of the onset-offset distances between notes in every within-voice pair across the Popular and Bach Chorales datasets. For most pairs, we expect to see their two notes separated by no more than 1 beat. The counts drastically taper off for larger distances. Taking into consideration this distribution and the processing time of the chord-level assignment model, we selected a beat horizon of 4 beats. By ignoring notes separated by more than 4 beats, the models cannot cover about 200 out of 29,394 pairs in the two datasets, which is negligible.

\begin{table}[H]
  \begin{center}
  \small
    {\renewcommand{\arraystretch}{1.3}
    \begin{tabular}{c|c|c|c|c|c|c||c}
      Beat Distance & $[0, 1)$ & $[1, 2)$ & $[2, 3)$ & $[3, 4)$ & $[4, 5)$ & $[5, \infty)$ & \textbf{Total} \\
      \hline
      \# Pairs & 28236  & 698  & 197   & 50  & 33  & 180 & 29394 \\
    \end{tabular}}
  \end{center}
  \caption{Histogram of the onset-offset distance in beats between the second and first note in within-voice pairs annotated in the Popular and Bach Chorales datasets.}
  \label{tab_histogram_onset_offset}
\end{table}

Tables~\ref{tab:histogram_convergent} and~\ref{tab:histogram_divergent} show the distribution of the number of voices converging to, or diverging from, notes in the Popular and Bach Chorales datasets.
\begin{table}[h]
  \begin{center}
  \small
    {\renewcommand{\arraystretch}{1.3}
    \begin{tabular}{c|>{\centering}p{0.9cm}|>{\centering}p{0.9cm}|>{\centering}p{0.9cm}|>{\centering}p{0.9cm}|>{\centering}p{0.9cm}||c}
      Convergent Connections & 0 & 1 & 2 & 3 & 4 & \textbf{Total}\\
      \hline
      \# Notes & 488  & 27802  & 779 & 10  & 1 & 29080 \\
    \end{tabular}}
  \end{center}
  \caption{Histogram of the number of convergent voice connections to notes in the Popular and Bach Chorales datasets.}
  \label{tab:histogram_convergent}
\end{table}
\begin{table}[h]
  \begin{center}
    {\renewcommand{\arraystretch}{1.3}
    \begin{tabular}{c|>{\centering}p{0.9cm}|>{\centering}p{0.9cm}|>{\centering}p{0.9cm}|>{\centering}p{0.9cm}||c}
      Divergent Connections & 0 & 1 & 2 & 3 & \textbf{Total}\\
      \hline
      \# Notes & 496  & 27792 & 774 & 18 & 29080 \\
    \end{tabular}}
  \end{center}
  \caption{Histogram of the number of divergent voice connections from notes in the Popular and Bach Chorales datasets.}
  \label{tab:histogram_divergent}
\end{table}
According to these histograms, most notes were annotated with only one incoming and one outgoing voice connection. Only 11 notes had 3 or 4 voices converging on them, and only 18 notes had 3 voices diverging from them. Considering these numbers and the processing time of the chord-level assignment model, the assignment models were constrained to converge a maximum of 2 voices to, and to diverge a maximum of 2 voices from, any single note. As a result, the assignment models were unable to cover only 11 + 18 within-voice pairs across the two annotated. It should be noted that for the that have more than two converging or diverging voices, 2 out of those voices can still be covered by the assignment models.

\begin{figure}[H]
  \centering
  \includegraphics[width=0.3\textwidth]{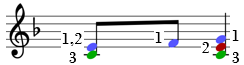}
  \caption{Voice diverging to two notes falling on different onsets in measure 82 ``21 Guns".}
  \label{fig:lookback_21_guns}
\end{figure}

Figure~\ref{fig:lookback_21_guns} shows a situation in which a voice diverges to two notes falling on different onsets. It is because of situations like this that we introduced the notion of partial active voices. In this particular example, the red voice diverging from the blue $E_4$ in the first chord ends on a note of the same pitch in the third chord. 

In general, if two consecutive notes in a diverging voice are separated by a rest, they tend to be close in pitch, as shown in the histogram from Table~\ref{tab:histogram_lookback}. The pitch distance between the first and second note in a pair is much more limited if the onset-offset beat distance separating them is greater than 0. Based on the distribution shown in the table, the note-level and chord-level assignment models were constrained to not consider assignments of a note to a partial active voice if the pitch distance to the ending note of the partial active voice is 6 half steps or larger and the onset-offset beat distance is greater than 0. 

% We would not expect to hear a second divergent connection to a note lying much further away in pitch because there is a temporal discontinuity and the sounding of other notes inbetween that inhibit our perception of the partial voice pairing. We depend on close pitch proximity to break through the surrounding distractions and bring our hearing to the attention of the partial pairing. With all that said, we can create a rule that simplifies the decision making process of the note-level and chord-level assignment models by pairing partial active voices to only those notes in the unassigned chord that lie within a restricted pitch range.

\begin{table}[H]
  \begin{center}
  \small
    {\renewcommand{\arraystretch}{1.3}
    \begin{tabular}{c||c|c|c|c|c|c|c||c}
      Beat Distance & Pitch Distance & $[0,3)$  & $[3, 6)$ & $[6, 9)$  & $[9, 12)$  & $[12, 15)$ & $[15, \infty)$ & \textbf{Total}\\
      \hline
      $0$ & \# Pairs & 532  & 341 & 124 & 43 & 244 & 193 & 1477 \\
      \hline
      \hline
      Beat Distance & Pitch Distance & $[0,3)$  & $[3, 6)$ & $[6, 9)$  & $[9, 12)$  & $[12, 15)$ & $[15, \infty)$ & \textbf{Total}\\
      \hline
      $> 0$ & \# Pairs & 103  & 15 & 5 & 1 & 1 & 0 & 125 \\
    \end{tabular}}
  \end{center}
  \caption{Histogram of the pitch distance in half steps between the first and second note in every voice pair annotated in the Popular and Bach Chorales datasets whose first note forms at least two divergent voice connections. Pair counts are further organized into separate groups based on the onset-offset beat distance between the second and first note in every pair.}
  \label{tab:histogram_lookback}
\end{table}

A synchronous cross happens when a note pair $p = (n_1, n_2)$ crosses a voice $v$ and note $n_2$ has the same onset as a note in $v$. Allowing synchronous crossings results in significantly more combinations of active voices to notes to consider, which in turn slows down training and inference considerably. However, there are only 82 cases of synchronous crossings in the Bach Chorales and 10 in the Popular music dataset, out of the close to 30,000 within-voice pairs. Therefore, to speed up training and inference, we prohibit the chord-level assignment model from crossing voices at synchronous onsets.

Pseudo-polyphony is a relatively rare phenomenon in our datasets, meaning that there is an insufficient number of examples to learn appropriate weights for the pseudo-polyphony features from Table~\ref{tab:pair-ppolphony-features}. Instead, we pre-processed the songs in the dataset and automatically identified all pseudo-polyphonic repeat chains, as explained in Section~\ref{sec:notation}. All notes that belong to the same pseudo-polyphonic repeat chain are then assigned to the same voice and kept in the same voice during training and testing. For fairness, we also apply this hard-coded rule to the envelope extraction model.

\subsection{Experimental Results}

For evaluation, we considered pairs of consecutive notes from the voices extracted by the three models and compared them with pairs of consecutive notes from the manually annotated voices. Tables~\ref{tab:popular} and~\ref{tab:chorales} show results on the Popular Music and Bach Chorales datasets in terms of the Jaccard similarity between the system pairs and the true pairs, precision, recall, and micro-averaged F-measure. Precision and recall are equivalent to the soundness and completeness measures used in \citep{kirlin:ismir05,deValk:ismir13}. The first section contains results across all within-voice pairs, while the second section reports results where pairs of notes separated by rests are ignored.

\begin{table}[ht]
 \begin{center}
  \footnotesize
  {\begin{tabular}{|c|c|c|c|c|c|c|}
    \hline
    Dataset & Evaluation & Model & Jaccard & Precision & Recall & F-measure \\
    \hline
    \hline
    \multirow{12}{*}{Popular Music} & 
    \multirow{3}{*}{\parbox{2.7cm}{\centering All within-voice pairs of notes}} & 
    Baseline    & 57.03 & 74.07 & 71.25 & 72.63 \Tstrut \\ \cline{4-7}
    & & Note-level  & 74.85 & 82.75 & 88.68 & 85.61 \Tstrut \\ \cline{4-7}
    & & Chord-level & {\bf 78.99} & 89.68 & 86.89 & {\bf 88.26} \Tstrut \\ \hhline{|~|======}
    & \multirow{3}{*}{\parbox{2.7cm}{\centering Exclude pairs separated by rests}} & 
    Baseline    & 66.69 & 81.05 & 79.03 & 80.02 \Tstrut \\ \cline{4-7}
    & & Note-level  & 82.66 & 86.06 & 95.44 & 90.51 \Tstrut \\ \cline{4-7}
    & & Chord-level &  {\bf 86.26} & 93.18 & 92.07 & {\bf 92.62} \Tstrut \\ \hhline{|~|======}
    & \multirow{3}{*}{\parbox{2.7cm}{\centering Chords Many}} & 
    Baseline    & 32.46 & 54.65 & 44.42 & 49.01 \Tstrut \\ \cline{4-7}
    & & Note-level  & {\bf 67.14} & 80.75 & 79.94 & {\bf 80.34}  \Tstrut \\ \cline{4-7}
    & & Chord-level & 61.68 & 83.10 & 70.53 & 76.30 \Tstrut \\ \hhline{|~|======}
    & \multirow{3}{*}{\parbox{2.7cm}{\centering Chords One}} & 
    Baseline    & 64.04 & 78.09 & 78.07 & 78.08 \Tstrut \\ \cline{4-7}
    & & Note-level  & 76.82 & 83.21 & 90.91 & 86.89 \Tstrut \\ \cline{4-7}
    & & Chord-level & {\bf 83.62} & 91.11 & 91.05 & {\bf 91.08} \Tstrut \\
    \hline
  \end{tabular}}
  \end{center}
  \caption{Comparative results of Baseline vs.\ Note-level Model vs.\ Chord-level Model on Popular Music.}
  \label{tab:popular}
\end{table}

\begin{table}[ht]
 \begin{center}
  \footnotesize
  {\begin{tabular}{|c|c|c|c|c|c|c|}
    \hline
    Dataset & Evaluation & Model & Jaccard & Precision & Recall & F-measure \\
    \hline
    \hline
    \multirow{12}{*}{Bach Chorales} & 
    \multirow{3}{*}{\parbox{2.7cm}{\centering All within-voice pairs of notes}} & 
    Baseline    & 86.81 & 93.77 & 92.12 & 92.94 \Tstrut \\ \cline{4-7}
    & & Note-level  & 93.95 & 96.84 & 96.91 & 96.88 \Tstrut \\ \cline{4-7}
    & & Chord-level & {\bf 94.74} & 97.55 & 97.04 & {\bf 97.30} \Tstrut \\ \hhline{|~|======}
    & \multirow{3}{*}{\parbox{2.7cm}{\centering Exclude pairs separated by rests}} & 
    Baseline    &  88.19 & 95.40 & 92.11 & 93.72 \Tstrut \\ \cline{4-7}
    & & Note-level  & 94.22 & 97.12 & 96.92 & 97.02 \Tstrut \\ \cline{4-7}
    & & Chord-level & {\bf 94.82} & 97.63 & 97.06 & {\bf 97.34} \Tstrut \\ \hhline{|~|======}
    & \multirow{3}{*}{\parbox{2.7cm}{\centering Chords Many}} & 
    Baseline    &  39.59 & 60.91 & 53.08 & 56.72 \Tstrut \\ \cline{4-7}
    & & Note-level  & 72.60 & 84.25 & 84.00 & 84.13 \Tstrut \\ \cline{4-7}
    & & Chord-level & {\bf 77.36} & 89.12 & 85.43 & {\bf 87.24} \Tstrut \\ \hhline{|~|======}
    & \multirow{3}{*}{\parbox{2.7cm}{\centering Chords One}} & 
    Baseline    & 96.73 & 98.33 & 98.34 & 98.34 \Tstrut \\ \cline{4-7}
    & & Note-level  & {\bf 97.84} & 98.84 & 98.97 & {\bf 98.91} \Tstrut \\ \cline{4-7}
    & & Chord-level &  97.76 & 98.84 & 98.90 & 98.87 \Tstrut \\
    \hline
  \end{tabular}}
  \end{center}
  \caption{Comparative results of Baseline vs.\ Note-level Model vs.\ Chord-level Model on Bach Chorales.}
  \label{tab:chorales}
\end{table}

We also run separate evaluations to determine the voice separation performance for chords that contain notes involved in convergent or divergent voices, as follows:
\begin{enumerate}
  \item {\bf Chords Many}: we consider a note $n$ to use in the evaluation if it belongs to a chord containing:
  \begin{itemize}
    \item At least one note that has either a true link back to two or more notes (i.e. convergence), or
    \item At least one note that has either a true link back to a note that diverges to two or more notes (i.e. divergence).
  \end{itemize}
  \item {\bf Chords One}: we consider a note $n$ to use in evaluation if it belongs to a chord containing:
  \begin{itemize}
    \item Only notes with at most one true link back to a note that does not diverge to two or more notes, or
  \end{itemize}
\end{enumerate}
In each of the two scenarios, we take the corresponding set of notes and consider for evaluation only the true and predicted links to those notes. Note that the definitions above imply that the set of all within-voice note pairs is partitioned between the two scenarios, which means that the performance (Jaccard or F-measure) for {\it All within-voice pairs} will always be between the performance in the {\it Chords Many} scenario and the performance in the {\it Chords One} scenario. Alternative definitions can also be obtained by replacing ``true link'' with ``predicted link'' everywhere in the definition. We have also computed performance in this alternative scenarios, but since the experimental results are similar, for brevity we only report results for the definitions above using ``true links''.

The results show that the both the note-level and chord-level models perform significantly better than the envelope extraction baseline. When pairs of notes separated by rests are excluded from evaluation, the baseline performance increases considerably, likely due to the exclusion of pseudo-polyphonic passages. Overall, the chord-level model is outperforming the note-level model, especially on the popular music dataset, thus validating empirically the importance of jointly modeling the assignment of voices to notes in a chord. One evaluation where the note-level model was significantly better than the chord-level model was in the {\it Chords Many} scenario for hte popular music dataset. Analysis of the output revealed that the note-level model created substantially more divergence and convergence pairs than the chord-level model, which for efficiency reasons was also constrained to converge from at most 2 voices and diverge to at most 2 voices (Section~\ref{sec:constraints}). This is reflected in the larger recall but lower precision for the note-level model in the {\it Chords Many} scenario. Overall though, when evaluated on all the within-voice pairs, the chord-level model is better than the note-level model.

\begin{table}[th]
 \begin{center}
 \footnotesize
 {\begin{tabular}{|c|c||c|c|c|}
  \hline
   Dataset & Model & Precision & Recall & F-measure \\
  \hline 
  \hline
   \multirow{3}{*}{10 Fugues}
      & \citep{deValk:ismir13}  & 94.07 & 93.42 & 93.74 \Tstrut \\ \cline{3-5}
      & Note-level & 94.27 & 95.27 & 94.77 \Tstrut \\ \cline{3-5}
      & Chord-level & 95.30 & 96.01 & {\bf 95.65} \Tstrut \\ \cline{3-5}
  \hline
  \hline
   \multirow{3}{*}{30 Inversions \& 48 Fugues}
      & \citep{madsen:icmpc06} & 95.94 & 70.11 & 81.01 \Tstrut \\  \cline{3-5}
      & Note-level & 95.46 & 95.91 & 95.69 \Tstrut \\ \cline{3-5}
      & Chord-level & 95.72 & 95.90 & {\bf 95.81} \Tstrut \\ \cline{3-5}
  \hline
 \end{tabular}}
\end{center}
 \caption{Comparative results on Bach datasets.}
 \label{tab:fugues}
\end{table}

Closest to our models is the data-driven approach from \citep{deValk:ismir13} for voice separation in lute tablature, which takes only the note in context as input and classifies it into a fixed number of voices. In contrast, our note-level and chord-level models use both the notes and the candidate active voices as input in order to compute a ranking probability or score, which enables them to label music with a variable number of voices. Table~\ref{tab:fugues} shows that our neural ranking models, although not specifically designed for music with a fixed number of voices, performs competitively with \citep{deValk:ismir13} when evaluated on the same datasets of 10 Fugues by Bach. The neural ranking models are also competitive with the approach from \citep{madsen:icmpc06}, which was evaluated on a different dataset containing 30 inventions and 48 fugues\footnote{In \citep{madsen:icmpc06} it is stated that soundness and completeness ``as suggested by Kirlin \citep{kirlin:ismir05}" were used for evaluation; however, the textual definitions given in \citep{madsen:icmpc06} are not consistent with \citep{kirlin:ismir05}. As was done in \citep{deValk:ismir13}, for lack of an answer to this inconsistency, we present the metrics exactly as in \citep{madsen:icmpc06}.}.

\section{Conclusion}
\label{sec:conclusion}

We introduced three approaches\footnote{The code and the dataset will be made publicly available.} for voice separation in symbolic music: an envelope extraction algorithm and two data-driven neural models that assign notes to active voices either separately for each note in a chord (note-level), or jointly for all notes in a chord (chord-level). The two neural models use a rich set of features capturing a wide diversity of perceptual factors affecting voice separation. We manually annotated a dataset of popular music with voice information using a task definition that allowed multiple voices to converge into one voice, or one voice to diverge into multiple voices. Experimental evaluations show that the two data-driven models substantially outperform the strong envelope extraction baseline, with the chord-level model edging out the note-level model. The two models are also shown to outperform previous approaches on separating the voices in Bach music. The state-of-the-art performance of the models introduced in this paper has the potential to enable a more sophisticated analysis of symbolic music that can benefit higher level tasks such as music recommendation, reharmonization, or music generation.

%\section{Acknowledgments}

\bibliographystyle{apacite}
\bibliography{jnmr18}

\pagebreak

\appendix

\section{Voice Separation Features for the Note-Level Model}
\label{sec:appendix-note-level}

\subsection{Basic Concepts and Notation}
\label{sec:notation}

Tables~\ref{tab:note-symbol-table} and~\ref{tab:voice-symbol-table} list the basic concepts and notation that are used to define the input features of the note-level assignment model. We assume that the music is traversed from left to right and segmented into a sequence of chords $\mathcal{C} = \langle c_t | 1 \leq t \leq T\rangle$, such that a chord is created every time a unique note onset or offset is encountered. A chord $c_t$ then contains all the notes whose onset is equal with $t$, sorted in descending order of their pitch. While most of the concepts introduced in Tables~\ref{tab:note-symbol-table} and~\ref{tab:voice-symbol-table} have a straightforward description, some of them require a more detailed description.

\begin{table}[ht]
  \small
  \centering
  \begin{tabular}{|c|p{11.5cm}|}
    \hline

    Symbol & \multicolumn{1}{c|}{Description} \\

    \hline
    \hline

    \multicolumn{2}{|c|}{\bf{Variables}} \\
    \hline
    \hline

    $n$ & Unassigned note in the current chord $c_t$. \\
    \hline
    $c_{m}$ & Chord of note $m$ with notes sorted in descending order of pitch. \\
    \hline
    $\mathcal{C}$ & Sequence of chords in the musical input ordered by onset. \\
    \hline
    \hline

    \multicolumn{2}{|c|}{\bf{Pitch Functions}} \\

    \hline
    \hline
    $ps(m)$ & Absolute pitch space of note $m$. Pitch values are taken from a pitch space in which C$_4$ has value 60 and a difference of 1 corresponds to one half step, e.g. C$_5$ has value 72. \\
    \hline
    \hline

    \multicolumn{2}{|c|}{\bf{Temporal Functions}} \\

    \hline
    \hline

    $bd(m)$ & Duration of note $m$ in beats.\\
    \hline
    $on(m)$ & Onset of note $m$ measured in beats relative to the first note in the score. \\
    \hline
    $\mathit{off}(m)$ & Offset of note $m$ measured in beats relative to the first note in the score. The offset is defined as the time at which a note ends. \\
    \hline
    \hline

    \multicolumn{2}{|c|}{\bf{Positional Functions}} \\

    \hline
    \hline

    $cp(m)$ & Position of note $m$ in the ordered chord $c_m$. Position is relative to the topmost note in $c_m$. \\
    \hline
    $sp(c)$ & Position of chord $c$ in the ordered sequence of chords $\mathcal{C}$. \\
    \hline
    $ab(m)$ & Neighboring note directly above note $m$ in the ordered chord $c_m$. \\
    \hline
    $bl(m)$ & Neighboring note directly below note $m$ in the ordered chord $c_m$. \\
    \hline
    $bk(m)$ & Boolean function that returns true if note $m$ is blocked. \\
    \hline
    $pp(m)$ & Boolean function that returns true if note $m$ belongs to a pseudo-polyphonic repeat chain. A pseudo-polyphonic repeat chain is a sequence of $k$ or more notes that all share the same pitch value and beat duration as $m$ and consecutive notes are separated by a distance of two times the beat duration of $m$.\\
    \hline
    $pp(x)$ & Boolean function that returns true if any note at onset $x$ belongs to a pseudo-polyphonic repeat chain. \\
    \hline
    \hline

    \multicolumn{2}{|c|}{\bf{Tonal Functions}} \\

    \hline
    \hline
    $sd(m)$ & Diatonic scale degree of note $m$. \\
    \hline
    $cs(m)$ & Diatonic chord step of note $m$. Chord step is the interval between a note $m$ and the root of its chord.\\
    \hline
    \hline

    \multicolumn{2}{|c|}{\bf{Metrical Functions}} \\

    \hline
    \hline
    $ql(m)$ & Duration of note $m$ measured relative to a quarter-note. \\
    \hline
    $bs(m)$ & Beat strength of note $m$ weighted in the range $[0, 1]$. Beat strength refers to the level of implicit accentuation each note is given in metered music. It is the weighting of strong and weak beats. \\
    \hline
  \end{tabular}

  \caption{
    Note and chord variable and function symbols used in the definitions of our note-level assignment model features.
  }

  \label{tab:note-symbol-table}
\end{table}

\begin{table}[ht]
  \small
  \centering
  \begin{tabular}{|c|p{11.5cm}|}
    \hline

    Symbol & \multicolumn{1}{c|}{Description} \\

    \hline
    \hline

    \multicolumn{2}{|c|}{\bf{Variables}} \\
    \hline
    $v$ & Active voice. $v$ may either be a complete or paritial active voice. Furthermore, in our feature definitions, $v$ is comprised only of notes that lie within the beat horizon of $l(v)$. \\
    \hline
    $l(u)$ & The last note in the active voice $u$. \\
    \hline
    $\mathcal{V}$ & The ordered sequence  of active voices at the current chord $c_t$. \\
    \hline
    \hline

    \multicolumn{2}{|c|}{\bf{Pitch Functions}} \\

    \hline
    \hline
    $cr(m, u)$ & Number of consecutive and connected notes in voice $u$, starting from $l(u)$, that share the same pitch space as note $m$. \\

    \hline
    \hline

    \multicolumn{2}{|c|}{\bf{Positional Functions}} \\

    \hline
    \hline
    $vp(u)$ & Position of active voice $u$ in the ordered active voice sequence $\mathcal{V}$. Position is relative to the topmost voice in $\mathcal{V}$. \\
    \hline
    $cvp(u)$ & Position of active voice $u$ relative to the complete active voices in the ordered active voice sequence $\mathcal{V}$. \\
    \hline
    $ab(u)$ & Neighboring complete voice directly above voice $u$ in the ordered active sequence of voices $\mathcal{V}$. \\
    \hline
    $bl(u)$ & Neighboring complete voice directly below voice $u$ in the ordered active sequence of voices $\mathcal{V}$. \\
    \hline
    $lt(m)$ & Set of notes converging to note $m$. \\
    \hline
    $rt(m)$ & Set of notes diverging from note $m$. \\
    \hline
    $top(u)$ & Linked list of the $k$ topmost notes in voice $u$ starting from $l(u)$. \\
    \hline
    $bot(u)$ & Linked list of the $k$ bottommost notes in voice $u$ starting from $l(u)$. \\
    \hline
    $dp(m)$ & Depth of the voice tree diverging from note $m$ \\
    \hline
    $div(u, w)$ & Shared divergent note between voices $u$ and $w$. \\
    \hline
    $x(m, u)$ & Boolean function that returns true if the pairing of note $m$ and voice $u$ results in a cross of another voice. \\
    \hline
  \end{tabular}

  \caption{
    Voice variable and function symbols used in the definitions of our note-level assignment model features.
  }

  \label{tab:voice-symbol-table}
\end{table}

\begin{figure}[ht]
  \centering
  \includegraphics[width=0.6\columnwidth]{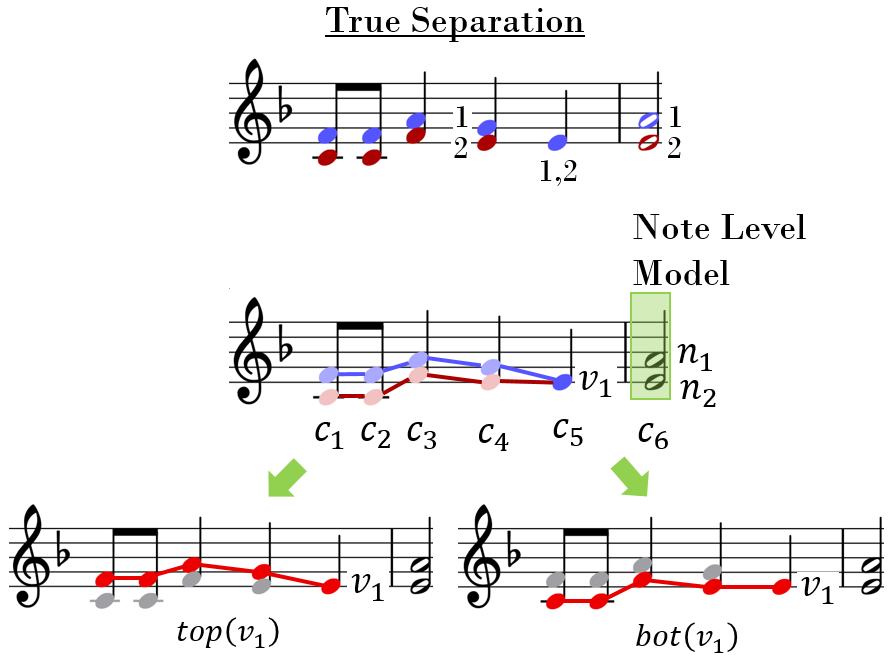}
  \caption{Demonstration of the $top(u)$ and $bot(u)$ functions in measures 31-32 from ``Apples to the Core".}
  \label{fig:top_bot_v}
\end{figure}

In our voice separation representation, multiple voices that converge to the same note result in a new voice linking that note to the converging voices. Therefore, an active voice can be traced back in time to multiple voices and thus to multiple notes sounding at the same time. In this context, we use $top(v)$ to denote the list of past notes $n$ that belong to the active voice $v$ such that $n$ is above any other note from $v$ that overlaps with it. Similarly, $bot(v)$ denotes the the list of past notes $n$ that belong to the active voice $v$ such that $n$ is below any other note from $v$ that overlaps with it. Figure~\ref{fig:top_bot_v} shows an example of the $top(v)$ and $bot(v)$ notes of a voice created by the convergence of two voices.

A note $p$ may diverge into two or more separate voices, which means that the note will be connected, directly or indirectly, to multiple notes in the future sounding at the same time, creating a note graph rooted at $p$. Given a current chord $c_t$ and a note $p$ in the past, we use $dp(p)$ to denote the note distance between $p$ and $c_t$, computed as the depth of the note graph rooted at $p$ and restricted to only notes appearing before $c_t$. Furthermore, we use $div(u, w)$ to denote the last note of the most recent voice that diverged into voices that evolved into $v$ and $w$, as shown in Figure~\ref{fig:shared_divergence} below. There, the shared divergent note between voices $v_1$ and $v_2$ has the same pitch as the sole note $n$ in chord $c_6$. A binary feature testing for this scenario would enable the assignment system to learn a preference towards converging the two voices back into a familiar sounding stream of notes.

\begin{figure}[h]
  \centering
  \includegraphics[width=0.5\columnwidth]{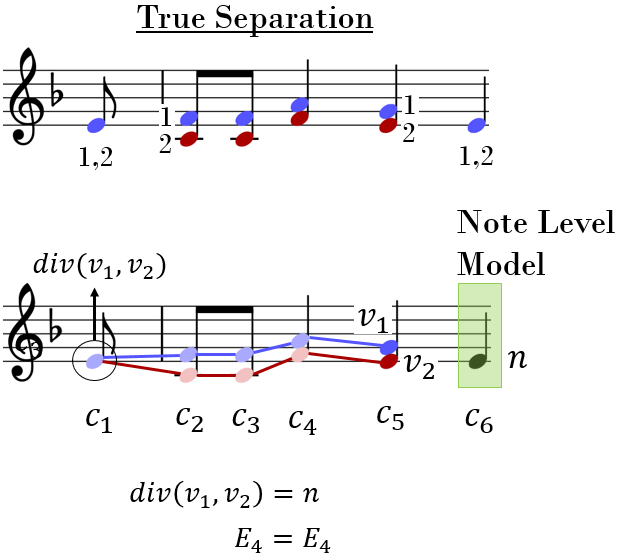}
  \caption{Shared divergent note in measures 30-31 from ``Apples to the Core".}
  \label{fig:shared_divergence}
\end{figure}

\begin{figure}[ht]
  \centering
  \includegraphics[width=0.3\columnwidth]{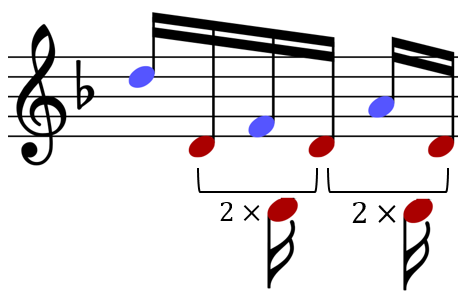}
  \caption{Pseudo-polyphonic repeat chain in measure 82 from ``Forest".}
  \label{fig:pseudopoly_repeat}
\end{figure}

The Boolean function $pp(n)$ returns true if note $n$ belongs to a {\it pseudo-polyphonic repeat chain}. In pseudo-polyphony, two perceptually independent streams are heard within a rapidly alternating, monophonic sequence of notes separated by relatively large pitch intervals. Figure~\ref{fig:pseudopoly_repeat} presents an example of pseudo-polyphony. Although the offset of each $D_4$ note is immediately followed by the onset of another note, the often large intervals and the fast tempo break the upper and lower notes into two perceptually independent streams. Another quality often seen in pseudo-polyphonic streams that reinforces the separation is the consistent repetition of a pitch in one of the two streams. We refer to this repeated stream of notes as a pseudo-polyphonic repeat chain. More formally, a pseudo-polyphonic repeat chain is a sequence of $k$ or more notes that all share the same pitch and beat duration. Furthermore, consecutive notes in the chain are separated by a distance of two times their beat durations. The red voice in Figure~\ref{fig:pseudopoly_repeat} is an example of a pseudo-polyphonic repeat chain.

The pitch repetition and consistent beat of a pseudo-polyphonic repeat chain does not formulate any salient melody but instead creates a supporting bass line that draws the listener's focus to the more freely moving notes of the second stream in a pseudo-polyphonic pair of voices. In the case of Figure~\ref{fig:pseudopoly_repeat}, the repeated and equidistant $D_4$ notes leads to the creation of monotonous tune that is easy for listeners to ignore, bringing into relatively more focus the melody of the blue voice. To help our system pick out pseudo-polyphonic voices, we feed it features that identify pseudo-polyphonic repeat chains.

\subsection{Discretization and Boundary Conditions}
\label{sec:discretization}

Categorical features, or integral feature with a small set of $K$ possible values are transformed into $K$ Boolean features, corresponding to each of the $K$ all possible values. For example, for the direction of $n$'s pitch relative to the last note $l(v)$ in voice $v$, we define three Boolean features to indicate whether $n$ is greater than, less than, or equal to $l(v)$. For feature like pitch that have e relatively large set of possible values, we discretize using 10 equal slices of the min to max range of values. For example, the maximum pitch value that is a multiple of ten is 80 and the minimum is 30, leading to 10 Boolean features corresponding to the 10 equal slices of the range $[30, 80]$. Discrete features will be identified by a top arrow in the features tables below.

Sometimes, the value computed by a function is null. For example, if $n$ is the topmost note in a chord, then $ab(n)$ will return null, since there is no other note in the chord above $n$. Correspondingly, the pitch distance between $n$ and $ab(n)$ will default to a maximum, data-dependent pitch distance to indicate that the note above is very far and thus insignificant.

\setcounter{fi}{0}

\subsection{Note-Level Model: Note Features}
\label{sec:note-level-note-features}

The note features capture properties of the unassigned note $n$ in the current chord $c_t$. By themselves, the note features are not capable of measuring any compatibility between $n$ and the active voice $v$. However, when presented alongside the voice and pair level features as input to the hidden layers of the neural network computing $p(n, v)$, they can be combined in higher level features that may encode potentially novel note-to-voice assignment rules.

\subsubsection{Pitch Features (Table~\ref{tab:note-pitch-features})}

\begin{table}[H]
  \centering
  \small

  \setcounter{fi}{0}
  {\tabulinesep=1.5mm
  \begin{tabu}{l p{7cm} | c}

    Feature & Short Description & Formula \\ 
    \hline

    \stepcounter{fi} $\vec{\phi}_{\thefi}$ & Pitch space of $n$ & $ps(n)$ \\

    \stepcounter{fi} $\phi_{\thefi}$ & Absolute pitch distance between $n$ and $ab(n)$ & $|ps(n) - ps(ab(n))|$ \\

    \stepcounter{fi} $\phi_{\thefi}$ & Absolute pitch distance between $n$ and $bl(n)$ & $|ps(n) - ps(bl(n))|$ \\

  \end{tabu}}

  \caption{
    Note-level, pitch features in $\Phi(n)$.
  }

  \label{tab:note-pitch-features}
\end{table}

According to Huron's toneness principle, the clarity of pitch perception is greatest for tones situated in the region between $F_2$ and $G_5$. $C_4$ sits in the center of this region, evoking the strongest sense of tonality. As one begins to deviate from middle $C$, the perceptual qualities of adjacent notes gradually coalesce into indiscernible tones. Consequently, the structural integrity of melodies centered around $C_4$ is more susceptible to be broken by pitch leaps and rests compared to melodies playing in the lower bass and upper soprano regions. $\phi_{1}$ captures this toneness principle, enabling the system learn what melodic intervals and overall motions are acceptable for voices lying in different pitch regions. 

Features $\phi_2$ and $\phi_3$ give the system a sense of how close surrounding notes are to $n$ and indirectly how close they are to $v$. This can help with scoring in many ways. For example, if $l(v)$ is relatively close to $n$ but even closer to $ab(n)$, then the system may be discouraged from assigning $n$ to $v$. In another scenario, if both $ab(n)$ and $bl(n)$ lie fairly close to $n$, then the system may decide to diverge $v$ if $l(v)$ is close in pitch to $n$, which of course means that $l(v)$ lies close to $ab(n)$ and $bl(n)$ also. 

\subsubsection{Temporal Features (Table~\ref{tab:note-temporal-features})}

\begin{table}[H]
  \centering
  \small

  {\tabulinesep=1.5mm
  \begin{tabu}{l l | c}

    Feature & Short Description & Formula \\ 
    \hline

    \stepcounter{fi} $\vec{\phi}_{\thefi}$ & Beat duration of $n$& $bd(n)$ \\

  \end{tabu}}

  \caption{
    Note-level, temporal features in $\Phi(n)$.
  }

  \label{tab:note-temporal-features}
\end{table}

Feature $\phi_4$ can help the system learn that large pitch leaps to a note with a long duration are more acceptable when the target note has a long duration. % The long duration somehow breaks the tension of the large pitch leap. It's like it gives you a chance to recover before continuing on with the melody. 
% A good example of this is in ``Knocking on Heavens Door" Measures 2-3 (FIGURE). Notes with shorter durations create the opposite effect.  

\subsubsection{Positional Features (Table~\ref{tab:note-positional-features})}
\begin{table}[H]
  \centering
  \small

  {\tabulinesep=1.5mm
  \begin{tabu}{l l | c}

    Feature & Short Description & Formula \\ 
    \hline

    \stepcounter{fi} $\vec{\phi}_{\thefi}$ & Chord position of $n$ & $cp(n)$ \\

    \stepcounter{fi} $\vec{\phi}_{\thefi}$ & Number of notes in $c_n$ & $|c_n|$ \\

  \end{tabu}}

  \caption{
    Note-level, positional features in $\Phi(n)$.
  }

  \label{tab:note-positional-features}
\end{table}

Notes that reside in the soprano either alone or at the top of a chord tend to be heard as the most salient. As a result, the most prominent melodic line often navigates through the topmost notes, even in situations where its active voice lies closer in pitch to the alto or tenor notes of the current chord. Notes in a low bass range that stand alone or at the bottom of a chord exhibit a similar behavior. To enable the learning model to capture this perceptual effect, we define features $\phi_5$ and $\phi_6$.

\subsubsection{Tonal Features (Table~\ref{tab:note-tonal-features})}
\begin{table}[H]
  \centering
  \small

  {\tabulinesep=1.5mm
  \begin{tabu}{l l | c}

    Feature & Short Description & Formula \\ 
    \hline

    \stepcounter{fi} $\vec{\phi}_{\thefi}$ & Scale degree of $n$ & $sd(n)$ \\

    \stepcounter{fi} $\vec{\phi}_{\thefi}$ & Chord step of $n$ & $cs(n)$ \\

  \end{tabu}}

  \caption{
    Note-level, tonal features in $\Phi(n)$.
  }

  \label{tab:note-tonal-features}
\end{table}

We use scale degrees as features in order to enable the model to learn melodic intervals that are most appropriate in a given key. For example, if a candidate active voice ends on a leading tone, then it is likely to resolve to the tonic. A similar argument can be made for the chord step feature.

\subsubsection{Metrical Features (Table~\ref{tab:note-metrical-features})}
\begin{table}[H]
  \centering
  \small

  {\tabulinesep=1.5mm
  \begin{tabu}{l l | c}

    Feature & Short Description & Formula \\ 
    \hline

    \stepcounter{fi} $\vec{\phi}_{\thefi}$ & Quarter length of $n$ & $ql(n)$ \\

    \stepcounter{fi} $\vec{\phi}_{\thefi}$ & Beat strength of $n$ & $bs(n)$ \\

  \end{tabu}}

  \caption{
    Note-level, metrical features in $\Phi(n)$.
  }

  \label{tab:note-metrical-features}
\end{table}

Feature $\phi_9$ enables the system to learn to pair notes that appear in common duration patterns, such as dotted quarter followed by an eighth.  Tonal music is organized into repeated sequences of strong and weak beats. Notes that fall on a strong beat are implicitly accentuated. Similarly, large pitch leaps tend to create more accent. When a large pitch leap contradicts the accent due to a strong beat, it can lead to a break in the corresponding voice.

\subsection{Note-Level Model: Voice Features}
\label{sec:note-level-voice-features}

Voice features capture properties of a candidate active voice $v$. By themselves, the voice features are not capable of measuring any compatibility between a note $n$ and the active voice $v$. However, when presented alongside the note and pair level features as input to the hidden layers of the neural network computing $p(n, v)$, they can be combined in higher level features that may encode potentially novel note-to-voice assignment rules.

\subsubsection{Pitch Features (Table~\ref{tab:voice-pitch-features})}
\begin{table}[!ht]
  \centering
  \small

  {\tabulinesep=1.5mm
  \begin{tabu}{l p{6cm} | c}

    Feature & Short Description & Formula \\ 
    \hline

    \stepcounter{fi} $\vec{\phi}_{\thefi}$ & Pitch space of $l(v)$ & $ps(l(v))$ \\

    \stepcounter{fi} $\vec{\phi}_{\thefi}$ & Average pitch space in $v$ & $ \frac{1}{|v|}\displaystyle\sum_{m \in v} {ps(m)}$ \\

    \stepcounter{fi} $\vec{\phi}_{\thefi}$ & Maximum pitch space in $v$ & $\displaystyle\max_{m \in v} ps(m)$  \\

    \stepcounter{fi} $\vec{\phi}_{\thefi}$ & Minimum pitch space in $v$ & $\displaystyle\min_{m \in v} ps(m)$\\

    \stepcounter{fi} $\phi_{\thefi}$ & Absolute pitch distance between $l(v)$ and $ab(l(v))$ & $|ps(l(v)) - ps(ab(l(v)))|$ \\

    \stepcounter{fi} $\phi_{\thefi}$ & Absolute pitch distance between $l(v)$ and $bl(l(v))$ & $|ps(l(v)) - ps(bl(l(v)))|$\\

    \stepcounter{fi} $\phi_{\thefi}$ & Absolute pitch distance between $l(v)$ and $l(ab(v))$ & $|ps(l(v) - ps(l(ab(v)))|$\\

    \stepcounter{fi} $\phi_{\thefi}$ & Absolute pitch distance between $l(v)$ and $l(bl(v))$ & $|ps(l(v)) - ps(l(bl(v)))|$\\

    \stepcounter{fi} $\vec{\phi}_{\thefi}$ & Direction of the pitch space of $l(v)$ relative to the pitch space of $l(ab(v))$ & $ps(l(v)) >=< ps(l(ab(v)))$\\

    \stepcounter{fi} $\vec{\phi}_{\thefi}$ & Direction of the pitch space of $l(v)$ relative to the pitch space of $l(bl(v))$ & $ps(l(v)) >=< ps(l(bl(v)))$\\

    \stepcounter{fi} $\phi_{\thefi}$ & Absolute pitch distance between notes in $top(v)$ and their neighboring notes above & $|ps(m) - ps(ab(m))|$ $\forall m \in top(v)$ \\

    \stepcounter{fi} $\phi_{\thefi}$ & Absolute pitch distance between notes in $bot(v)$ and their neighboring notes below & $|ps(m) - ps(bl(m))|$ $\forall m \in bot(v)$ \\

    \stepcounter{fi} $\phi_{\thefi}$ & Absolute pitch distance between adjacent notes in $top(v)$ & $|ps(m) - ps(lt(m))|$ $\forall m \in top(v)$ \\

    \stepcounter{fi} $\phi_{\thefi}$ & Absolute pitch distance between adjacent notes in $bot(v)$ & $|ps(m) - ps(lt(m))|$ $\forall m \in bot(v)$ \\

    \stepcounter{fi} $\vec{\phi}_{\thefi}$ & Relative directions of adjacent notes in $top(v)$ & $ps(m) >=< ps(lt(m))$ $\forall m \in top(v)$ \\

    \stepcounter{fi} $\vec{\phi}_{\thefi}$ & Relative directions of adjacent notes in $bot(v)$ & $ps(m) >=< ps(lt(m))$ $\forall m \in bot(v)$ \\

    \stepcounter{fi} $\phi_{\thefi}$ & Is $l(v)$ a blocked note & $bk(l(v))$

  \end{tabu}}

  \caption{
    Voice-level, pitch features in $\Phi(v)$.
  }

  \label{tab:voice-pitch-features}
\end{table}

Like $\phi_1$, feature $\phi_{11}$ enables the system to learn what melodic intervals and overall motions are acceptable for voices lying in different pitch regions. Features $\phi_{12}$, $\phi_{13}$, and $\phi_{14}$ complement $\phi_{11}$ and capture properties of the distribution of notes in $v$ that can help the system learn how $n$ would fit into the overall voice.  These features are computed only over the notes in $v$ that lie within the beat horizon of $l(v)$.

Features $\phi_{15}$, $\phi_{16}$ about the last note of voice $v$ mirror features $\phi_2$ and $\phi_3$  about note $n$. They give the system a sense of how close surrounding voices are to $v$ and indirectly how close they are to $v$. 

The notes directly above and below $l(v)$ in the chord $c_{l(v)}$ may not be the same as the last notes in the voices above and below, $l(ab(v))$ and $l(bl(v))$, respectively, hence the additional features $\phi_{17}$ and $\phi_{18}$. According to the definitions in Section~\ref{sec:notation}, $ab(v)$ and $bl(v)$ are the {\it complete} active voices above and below $v$, which may be different from the voices one index above and below in the order active voice sequence $\mathcal{V}$ which contains also partial active voices. We look for the voices above and below in the order active sequence whose last notes are not connected to other notes to the right up to the current chord. Most notes in the current, unassigned chord $c_t$ will pair with complete active voices, therefore it is worth defining features that use these voices. Features $\phi_{17}$ and $\phi_{18}$ share a similar function to $\phi_{15}$ and $\phi_{16}$. For example, if $v$ is distant from both $ab(v)$ and $bl(v)$ and $n$ lies close to both $ab(n)$ and $bl(n)$, we may expect to see $ab(n)$ and $bl(n)$ forgo pairing with $ab(v)$ and $bl(v)$ in favor of creating a divergent connection with $v$.

If $l(v)$ has a higher pitch than $ab(v)$ or $l(v)$ has a lower pitch than $bl(v)$, then we know that $v$ is a blocked voice. Blocked voices are limited in what notes they can connect to in the future, hence the utility of features $\phi_{19}$ and $\phi_{20}$. For example, the red voice in Figure~\ref{fig:block_see_you_again_repeat} pauses at the end of measure 5 after being blocked by the blue voice, but continues on in measure 7. The reason it is able to overcome the block is because the yellow voice just repeats the note $D_4$, in a similar motif. If the alto voice started on another note in measure 7, the connection to the read voice in measure 5 would have been weaker. 
% The reason for this has to do with the fact that the blue voice very quickly drowns out the yellow voice after the block in measure 5. 
% Overall, warning the system when a voice is blocked helps to simplify the assignment decision making process.

Features $\phi_{21}$ and $\phi_{22}$ use absolute pitch distances as a proxy for how much a voice is distinguished from surrounding voices above ($\phi_{21}$) and below ($\phi_{22}$). 
When a voice is relatively far from notes in voices immediately above and below it, it becomes more salient perceptually and thus can connect more easily to notes in the current chord. Because a voice $v$ is comprised of a varying number of notes, we restrict the number of notes $m$ for these two feature templates to the 4 most recent notes in $top(v)$ and $bot(v)$.
% As a simple example, I would suggest looking at the opening to ``How to Save a Life". Both the soprano and bass voices lie far away from each other in pitch and so there is no confusion when sequentially assigning notes ahead because we can clearly distinguish the two. Conversely, let us look at the green voice in measure 17 from ``Apples to the Core". At each $C$ onset, it is masked from above and below by voices that are pretty active in their movements and so we are more willing to drop it and forget it once we reach the end of the measure near the red $D_4$. For an example in-between, let us look at measure 39 in ``One Call Away". It would be reasonable to intially think the red voice would drop down and overtake the green voice at the second $F_4$ because the red voice is more pronounced the soprano. However, looking back, we see that many notes in the green voice are exposed in the soprano themselves and so it is easier for us to focus our hearing on the green voice as a whole and continue it onward to the exposed $F_4$ in measure 39. To summarize, these features determine to what degree a voice is masked. Masked is not good for the survival of a voice because it is drowned out by the more prominent surrounding voices. On the other hand, if a voice is more exposed, like the voices in the soprano, then the system will likely find a pairing note for it in the current, unassigned chord. 

Features $\phi_{23}$ and $\phi_{24}$ are intended to complement features $\phi_{12} to \phi_{14}$. The absolute pitch distances between consecutive notes in the top and bottom portions of a voice $v$ give a sense of how much a voice moves in the pitch space, which can help in determining if a note $n$ is a good fit for $v$. For example, in the bass of ``Let it Be", there are large pitch leaps between adjacent notes in voice $v$. By considering the pitch distances between adjacent notes, the system can learn that it is acceptable to pair up an $n$ in the bass with the $l(v)$ in the bass even if there is a large pitch distance.

Features $\phi_{25}$ and $\phi_{26}$ quantify the direction of movement in consecutive pairs of notes in the top and bottom portions of a voice, respectively. As such, they complement features $\phi_{23}$ and $\phi_{24}$.
% Again, I provide more features to complement features $\phi_{12}-\phi_{14}$ on a more fine grained level. Directions between pitches can be very important. For example, in measure 29 from ``Apples to the Core", the green voice repeats $C_4$ and so I couldn't help but perceive the last $C_4$ in the measure as a convergence pair. This features tells the system that adjacent notes in the green voice are all equal in pitch. In another related example, let us look at measure 3 from greensleeves. This feature informs us that the green voice is fairly active and moving freely, which makes it stand out in our hearing even though it is masked in the alto. Consequently, its presence is heard well enough to warrant a convergence at the quarter $D_4$. 

Feature $\phi_{27}$ enables the system to learn that blocked voices have limited mobility, i.e. are less likely to be continued.

\subsubsection{Temporal Features (Table~\ref{tab:voice-temporal-features})}

Feature $\phi_{28}$ measures the beat duration of the last note in a voice, mirroring feature $\phi_{4}$ defined similarly for a note $n$. Features $\phi_{29}$ to $\phi_{34}$ are influenced by Huron's Temporal Continuity principle, according to which strong auditory streams require continuous or recurring rather than brief or intermittent sounds. Features $\phi_{33}$ - $\phi_{36}$ inform the system on how connected or broken a voice is. A sequence of notes that is interrupted by many rests lacks the temporal continuity needed to create an auditory stream.

\begin{table}[!ht]
  \centering
  \small

  {\tabulinesep=1.5mm
  \begin{tabu}{l p{5.5cm} | c}
    Feature & Short Description & Formula \\ 
    \hline

    \stepcounter{fi} $\vec{\phi}_{\thefi}$ & Beat duration of $l(v)$ & $bd(l(v))$ \\

    \stepcounter{fi} $\phi_{\thefi}$ & Inter-onset distance between $l(v)$ and $l(ab(v))$  & $|on(l(v)) - on(l(ab(v)))|$ \\

    \stepcounter{fi} $\phi_{\thefi}$ & Inter-onset distance between $l(v)$ and $l(bl(v))$ & $|on(l(v)) - on(l(bl(v)))|$ \\

    \stepcounter{fi} $\vec{\phi}_{\thefi}$ & Direction of the onset of $l(v)$ relative to the onset of $l(ab(v))$ & $on(l(v)) >=< on(l(ab(v)))$ \\

    \stepcounter{fi} $\vec{\phi}_{\thefi}$ & Direction of the onset of $l(v)$ relative to the onset of $l(bl(v))$ & $on(l(v)) >=< on(l(bl(v)))$ \\

    \stepcounter{fi} $\phi_{\thefi}$ & Onset-offset distance between adjacent notes in $top(v)$ & $\displaystyle\max(on(m) - \mathit{off}(lt(m)), 0)$ $\forall m \in top(v)$ \\

    \stepcounter{fi} $\phi_{\thefi}$ & Onset-offset distance between adjacent notes in $bot(v)$ & $\displaystyle\max(on(m)- \mathit{off}(lt(m)), 0)$ $\forall m \in bot(v)$ \\

    \stepcounter{fi} $\vec{\phi}_{\thefi}$ & Number of rests in $top(v)$ & $\displaystyle\sum_{m \in top(v)} on(m) > \mathit{off}(lt(m))$ \\

    \stepcounter{fi} $\vec{\phi}_{\thefi}$ & Number of rests in $bot(v)$ & $\displaystyle\sum_{m \in bot(v)} on(m) > \mathit{off}(lt(m))$ \\

  \end{tabu}}

  \caption{
    Voice-level, temporal features in $\Phi(v)$.
  }

  \label{tab:voice-temporal-features}
\end{table}

% To explain further, let us look at the purple voice in measures 14-16 in ``Uptown Girl". It is broken up by a lot of rests and thus lacks a coherent melody. Without a melody, it fades into a supporting role for the other voices and thus the purple voice is likely not to be involved in any exciting convergence or divergence cases. On the other hand, the red voice above is broken by only one rest in measure 16, indicating that this is a pretty fluid and prominent voice and thus most likely will be assigned to notes ahead.

\subsubsection{Positional Features (Table~\ref{tab:voice-positional-features})}

Features $\phi_{37}$ to $\phi_{39}$ consider the position of ab active voice in the chord of its last note or in the current sequence of active voices. As such, they mirror the note feature $\phi_{5}$ and help the system learn the layering of voices that is at the core of the envelope extraction model. Features $\phi_{40}$ - $\phi_{42}$ measure the number of concurrent notes in the current chord or the number of concurrent voices in the current sequence of active voices, mirroring the note feature $\phi_{6}$. Features $\phi_{43}$ - $\phi_{45}$ look only at notes within the beat horizon and give a sense of the note density of a voice. 
% A voice comprised of many notes within the beat horizon of $l(v)$) is likely to be perceived as more active and thus more likely to be continued. 
Feature $\phi_{46}$ is meant to help the system learn that a partial voice that is far from the current chord is unlikely to be used by the system for pairing with a note in the current chord.
% As a simple example, look at measure 2 from ``Hymn for the Weekend". Already, when assigning notes to a current chord in measure 2, the partial voices in measure one are drowned out by quite a few notes ahead. In a specific example, the number of notes separating the partial voice ending at $A_4$ at the first onset is 12 notes deep in the complete active red voice ending at $G_4$ in the first onset of measure 2. The system from this point on can just ignore the very first $A_4$. On the other hand, looking at measure 21, at the last green $E_4$, it is a partial voice that is only 1 note away from the complete active voice ending at $B^b_4$ in the same green voice when working with the unassigned, current chord at the first onset of measure 22. This 1 note difference is not as big of a deal as the 12 note difference case, and thus we hope the system will consider it for pairing still.
Features $\phi_{47} - \phi_{50}$ determine whether there are notes above and below a voice $v$ and complement the pitch distance features features $\phi_{21}$ and $\phi_{22}$.
Feature $\phi_{51}$ enables the system learn an upper bound on the divergence limit. For example, if 3 notes are diverging from $l(v)$, then $v$ it is unlikely to be connect to another note because the data shows it is unlikely to have a voice diverge into more than 3 voices.

\begin{table}[!ht]
  \centering
  \small

  {\tabulinesep=1.5mm
  \begin{tabu}{l p{6cm} | c}
    Feature & Short Description & Formula \\ 
    \hline

    \stepcounter{fi} $\vec{\phi}_{\thefi}$ & Chord position of $l(v)$ & $cp(l(v))$ \\

    \stepcounter{fi} $\vec{\phi}_{\thefi}$ & Active voice position of $v$ & $vp(v)$ \\

    \stepcounter{fi} $\vec{\phi}_{\thefi}$ & Complete active voice position of $v$ & $cvp(v)$ \\

    \stepcounter{fi} $\vec{\phi}_{\thefi}$ & Number of notes in $c_{l(v)}$ & $|c_{l(v)}|$ \\

    \stepcounter{fi} $\vec{\phi}_{\thefi}$ & Number of active voices in $\mathcal{V}$ & $|\mathcal{V}|$ \\

    \stepcounter{fi} $\vec{\phi}_{\thefi}$ & Number of complete active voices in $\mathcal{V}$ & $\displaystyle\sum_{v \in \mathcal{V}} dp(l(v)) = 0$ \\

    \stepcounter{fi} $\vec{\phi}_{\thefi}$ & Number of notes in $v$ & $|v|$ \\

    \stepcounter{fi} $\vec{\phi}_{\thefi}$ & Number of notes in $top(v)$ & $|top(v)|$ \\

    \stepcounter{fi} $\vec{\phi}_{\thefi}$ & Number of notes in $bot(v)$ & $|bot(v)|$ \\

    \stepcounter{fi} $\vec{\phi}_{\thefi}$ & Depth of the voice tree diverging from $l(v)$ & $dp(l(v))$ \\

    \stepcounter{fi} $\phi_{\thefi}$ & Do notes in $top(v)$ have a neighboring note above & $cp(m) \neq 1$ $\forall m \in top(v)$ \\

    \stepcounter{fi} $\phi_{\thefi}$ & Do notes in $bot(v)$ have a neighboring note below & $cp(m) \neq |c_m|$ $\forall m \in bot(v)$ \\

    \stepcounter{fi} $\vec{\phi}_{\thefi}$ & Number of notes in $top(v)$ that have a neighboring note above & $\displaystyle\sum_{m \in top(v)} cp(m) \neq 1$ \\

    \stepcounter{fi} $\vec{\phi}_{\thefi}$ & Number of notes in $bot(v)$ that have a neighboring note below & $\displaystyle\sum_{m \in bot(v)} cp(m) \neq |c_m|$ \\

    \stepcounter{fi} $\vec{\phi}_{\thefi}$ & Number of notes diverging from $l(v)$ & $|rt(l(v))|$ \\

  \end{tabu}}

  \caption{
    Voice-level, positional features in $\Phi(v)$.
  }

  \label{tab:voice-positional-features}
\end{table}

\subsubsection{Tonal Features (Table~\ref{tab:voice-tonal-features})}

These tonal voice-level features mirror their note-level counterparts.

\begin{table}[h]
  \centering
  \small

  {\tabulinesep=1.5mm
  \begin{tabu}{l l | c}
    Feature & Short Description & Formula \\ 
    \hline

    \stepcounter{fi} $\vec{\phi}_{\thefi}$ & Scale degree of $l(v)$ & $sd(l(v))$ \\

    \stepcounter{fi} $\vec{\phi}_{\thefi}$ & Chord step of $l(v)$ & $cs(l(v))$ \\
  \end{tabu}}

  \caption{
    Voice-level, tonal features in $\Phi(v)$.
  }

  \label{tab:voice-tonal-features}
\end{table}

\subsubsection{Metrical Features (Table~\ref{tab:voice-metrical-features})}

These metrical voice-level features mirror their note-level counterparts.

\begin{table}[h]
  \centering
  \small

  {\tabulinesep=1.5mm
  \begin{tabu}{l l | c}
    Feature & Short Description & Formula \\ 
    \hline

    \stepcounter{fi} $\vec{\phi}_{\thefi}$ & Quarter length of $l(v)$ & $ql(l(v))$ \\

    \stepcounter{fi} $\vec{\phi}_{\thefi}$ & Beat strength of $l(v)$ & $bs(l(v)$ \\
  \end{tabu}}

  \caption{
    Voice-level, metrical features in $\Phi(v)$.
  }

  \label{tab:voice-metrical-features}
\end{table}

\subsection{Note-Level Model: Pair Features}
\label{sec:note-level-pair-features}

Pair features depend on both the unassigned note $n$ and the active voice $v$, as such are meant to more directly guide the system towards modeling the likelihood $p(n, v)$ of assigning a note to an active voice.

\subsubsection{Pitch Features (Table~\ref{tab:pair-pitch-features})}

Features $\phi_{56}$ to $\phi_{66}$ are the most significant in this category and are meant to capture Huron's pitch proximity principle according to which "the coherence of an auditory stream is maintained by close pitch proximity in successive tones within the stream".
% JUSTIFY $\phi_{67}$ - $\phi_{70}$. Look at measures 14 and 15 in ``Thousand Miles".

\begin{table}[!ht]
  \centering
  \small
  {\tabulinesep=1.5mm
  \begin{tabu}{l p{6cm} | c}
    Feature & Short Description & Formula \\ 
    \hline

    \stepcounter{fi} $\phi_{\thefi}$ & Absolute pitch distance between $n$ and $l(v)$ & $|ps(n) - ps(l(v))|$ \\

    \stepcounter{fi} $\vec{\phi}_{\thefi}$ & Direction of the pitch space of $n$ relative to the pitch space of $l(v)$ & $ps(n) >=< ps(l(v))$ \\

    \stepcounter{fi} $\phi_{\thefi}$ & Absolute pitch distance between $n$ and the average pitch in $v$ & $ |ps(n) - \frac{1}{|v|}\displaystyle\sum_{m \in v} {ps(m)}|$ \\

    \stepcounter{fi} $\vec{\phi}_{\thefi}$ & Direction of the pitch space of $n$ relative to the average pitch in $v$ & $ ps(n) >=< \frac{1}{|v|}\displaystyle\sum_{m \in v} {ps(m)}$ \\

    \stepcounter{fi} $\phi_{\thefi}$ & Absolute pitch distance between $n$ and the maximum pitch in $v$ & $|ps(n) - \displaystyle\max_{m \in v} ps(m)|$ \\

    \stepcounter{fi} $\vec{\phi}_{\thefi}$ & Direction of the pitch space of $n$ relative to the maximum pitch in $v$ & $ps(n) >=< \displaystyle\max_{m \in v} ps(m)$ \\

    \stepcounter{fi} $\phi_{\thefi}$ & Absolute pitch distance between $n$ and the minimum pitch in $v$ & $|ps(n) - \displaystyle\min_{m \in v} ps(m)|$ \\

    \stepcounter{fi} $\vec{\phi}_{\thefi}$ & Direction of the pitch space of $n$ relative to the minimum pitch in $v$ & $ps(n) >=< \displaystyle\min_{m \in v} ps(m)$ \\

    \stepcounter{fi} $\phi_{\thefi}$ & Absolute distance between $\phi_{58}$ and the standard deviation of the pitches in $v$ & $\Bigl|\phi_{58} - \sqrt{\frac{1}{|v|}\displaystyle\sum_{m \in v} {(ps(m) - \frac{1}{|v|}\displaystyle\sum_{m \in v} ps(m))^2}}\Bigr|$ \\

    \stepcounter{fi} $\phi_{\thefi}$ & Is $\phi_{58}$ less than the standard deviation of the pitches in $v$ & $\phi_{58} < \sqrt{\frac{1}{|v|}\displaystyle\sum_{m \in v} {(ps(m) - \frac{1}{|v|}\displaystyle\sum_{m \in v} ps(m))^2}}$ \\

    \stepcounter{fi} $\vec{\phi}_{\thefi}$ & Consecutive repetition of the pitch space of $n$ in $v$ & $cr(n, v)$ \\

    \stepcounter{fi} $\phi_{\thefi}$ & Absolute pitch distance between $n$ and the diverging note connecting $v$ and $ab(v)$  & $|ps(n) - ps(div(v, ab(v)))|$ \\

    \stepcounter{fi} $\vec{\phi}_{\thefi}$ & Direction of the pitch space of $n$ relative to the pitch space of the diverging note connecting $v$ and $ab(v)$ & $ps(n) >=< ps(div(v, ab(v)))$ \\

    \stepcounter{fi} $\phi_{\thefi}$ & Absolute pitch distance between $n$ and the diverging note connecting $v$ and $bl(v)$ & $|ps(n) - ps(div(v, bl(v)))|$ \\

    \stepcounter{fi} $\vec{\phi}_{\thefi}$ & Direction of the pitch space of $n$ relative to the pitch space of the diverging note connecting $v$ and $bl(v)$ & $ps(n) >=< ps(div(v, bl(v)))$ \\

  \end{tabu}}
  \caption{
    Pair-level, pitch features in $\Phi(n, v)$.
  }
  \label{tab:pair-pitch-features}
\end{table}

\subsubsection{Temporal Features (Table~\ref{tab:pair-temporal-features})}

The temporal features in Table~\ref{tab:pair-temporal-features} are meant to model Huron's Temporal Continuity principle.

\begin{table}[!ht]
  \centering
  \small
  {\tabulinesep=1.5mm
  \begin{tabu}{l p{6cm} | c}
    Feature & Short Description & Formula \\ 
    \hline

    \stepcounter{fi} $\phi_{\thefi}$ & Inter-onset distance between $n$ and $l(v)$ & $|on(n) - on(l(v))|$ \\

    \stepcounter{fi} $\phi_{\thefi}$ & Onset-offset distance between $n$ and $l(v)$ & $\displaystyle\max(on(n) - \mathit{off}(l(v)), 0)$ \\

    \stepcounter{fi} $\vec{\phi}_{\thefi}$ & Direction of the onset of $n$ relative to the offset of $l(v)$ & $on(n) >=< \mathit{off}(l(v))$ \\

    \stepcounter{fi} $\phi_{\thefi}$ & Absolute difference in beat duration between $n$ and $l(v)$ & $|bd(n) - bd(l(v))|$ \\

    \stepcounter{fi} $\vec{\phi}_{\thefi}$ & Length of the beat duration of $n$ in relation to the beat duration of $l(v)$ & $bd(n) >=< bd(l(v))$ \\

  \end{tabu}}
  \caption{
    Pair-level, temporal features in $\Phi(n, v)$.
  }
  \label{tab:pair-temporal-features}
\end{table}

\subsubsection{Positional Features (Table~\ref{tab:pair-positional-features})}

Features $\phi_{76}$ - $\phi_{83}$ in Table~\ref{tab:pair-positional-features} are meant to enable the note-level assignment model to learn the core behavior of the envelope extraction model, which assigns notes to voices based on their position in the current chord and sequence of complete active voices, respectively.
Feature $\phi_{84}$ is meant to enable the system learn that the strength of the perceptual connection between $n$ and $l(v)$ decreases with the number of onsets between the two.

\begin{table}[H]
  \centering
  \small
  {\tabulinesep=1.5mm
  \begin{tabu}{l p{6cm} | c}
    Feature & Short Description & Formula \\ 
    \hline

    \stepcounter{fi} $\phi_{\thefi}$ & Absolute difference between the chord positions of $n$ and $l(v)$ & $|cp(n) - cp(l(v))|$ \\

    \stepcounter{fi} $\vec{\phi}_{\thefi}$ & Chord positions of $n$  relative to the chord position of $l(v)$ & $cp(n) >=< cp(l(v))$ \\

    \stepcounter{fi} $\phi_{\thefi}$ & Absolute difference between the chord position of $n$ and the voice position of $v$ & $|cp(n) - vp(v)|$ \\

    \stepcounter{fi} $\vec{\phi}_{\thefi}$ & Chord positions of $n$  relative to the voice position of $v$ & $cp(n) >=< vp(v)$ \\

    \stepcounter{fi} $\phi_{\thefi}$ & Absolute difference between the chord position of $n$ and the complete active voice position of $v$ & $|cp(n) - cvp(v)|$ \\

    \stepcounter{fi} $\vec{\phi}_{\thefi}$ & Chord positions of $n$  relative to the complete active voice position of $v$ & $cp(n) >=< cvp(v)$ \\

    \stepcounter{fi} $\phi_{\thefi}$ & Absolute difference between the length of $c_n$ and the length of $c_{l(v)}$ & $\left| |c_n| - |c_{l(v)}| \right|$ \\

    \stepcounter{fi} $\phi_{\thefi}$ & Absolute difference between the length of $c_n$ and the length of $\mathcal{V}$ & $\left| |c_n| - |V|\right|$ \\

    \stepcounter{fi} $\vec{\phi}_{\thefi}$ & Number of unique onsets between $c_n$ and $c_{l(v)}$ & $sp(c_n) - sp(c_{l(v)})$ \\

    \stepcounter{fi} $\vec{\phi}_{\thefi}$ & Does the pairing of $n$ and $v$ result in a cross & $x(n, v)$ \\
  \end{tabu}}
  \caption{
    Pair-level, positional features in $\Phi(n, v)$.
  }
  \label{tab:pair-positional-features}
\end{table}

\subsubsection{Tonal Features (Table~\ref{tab:pair-tonal-features})}

The features in Table~\ref{tab:pair-tonal-features} mirror the note-level and voice-level tonal features.

\begin{table}[H]
  \centering
  \small
  {\tabulinesep=1.5mm
  \begin{tabu}{l p{6cm} | c}
    Feature & Short Description & Formula \\ 
    \hline

    \stepcounter{fi} $\phi_{\thefi}$ & Absolute distance between the scale degrees of $n$ and $l(v)$ & $|sd(n) - sd(l(v))|$ \\

    \stepcounter{fi} $\vec{\phi}_{\thefi}$ & Direction of the scale degree of $n$ relative to the scale degree of $l(v)$ & $sd(n) >=< sd(l(v))$ \\

    \stepcounter{fi} $\phi_{\thefi}$ & Absolute distance between the chord steps of $n$ and $l(v)$ & $|cs(n) - cs(l(v))|$ \\

    \stepcounter{fi} $\vec{\phi}_{\thefi}$ & Direction of the chord step of $n$ relative to the chord step of $l(v)$ & $cs(n) >=< cs(l(v))$ \\

  \end{tabu}}
  \caption{
    Pair-level, tonal features in $\Phi(n, v)$.
  }
  \label{tab:pair-tonal-features}
\end{table}

\subsubsection{Metrical Features(Table~\ref{tab:pair-metrical-features}) }

The features in Table~\ref{tab:pair-metrical-features} mirror the note-level and voice-level metrical features.

\begin{table}[H]
  \centering
  \small
  {\tabulinesep=1.5mm
  \begin{tabu}{l p{6cm} | c}
    Feature & Short Description & Formula \\ 
    \hline

    \stepcounter{fi} $\phi_{\thefi}$ & Absolute distance between the quarter lengths of $n$ and $l(v)$ & $|ql(n) - ql(l(v))|$ \\

    \stepcounter{fi} $\vec{\phi}_{\thefi}$ & Direction of the quarter length of $n$ relative to the quarter length of $l(v)$ & $ql(n) >=< ql(l(v))$ \\

    \stepcounter{fi} $\phi_{\thefi}$ & Absolute difference between the beat strengths of $n$ and $l(v)$ & $|bs(n) - bs(l(v))|$ \\

    \stepcounter{fi} $\vec{\phi}_{\thefi}$ & Direction of the beat strength of $n$ relative to the beat strength of $l(v)$ & $bs(n) >=< bs(l(v))$ \\
  \end{tabu}}
  \caption{
    Pair-level, metrical features in $\Phi(n, v)$.
  }
  \label{tab:pair-metrical-features}
\end{table}

\subsubsection{Pseudo-Polyphony Features (Table~\ref{tab:pair-ppolphony-features})}

The features in Table~\ref{tab:pair-ppolphony-features} are meant to determine if a pseudo-polyphonic repeat chain (Section~\ref{sec:notation}) is present and whether the current note should be assigned to it or to a voice supported by the chain.

\begin{table}[!ht]
  \centering
  \small
  {\tabulinesep=1.5mm
  \begin{tabu}{l p{6cm} | c}
    Feature & Short Description & Formula \\ 
    \hline

    $*\phi_{59}$ & Does the pitch space of $n$ equal the pitch space of $l(v)$ & $ps(n) = ps(l(v))$ \\

    $*\phi_{77}$ & Does the beat duration of $n$ equal the beat duration of $l(v)$ & $bd(n) = bd(l(v))$ \\

    \stepcounter{fi} $\phi_{\thefi}$ & Are $n$ and $l(v)$ separated by two times the duration of $v(l)$ in beats & $on(n) = \mathit{off}(l(v)) + bd(l(v))$ \\

    \stepcounter{fi} $\phi_{\thefi}$ & Does $n$ belong to a pseudo-polyphonic repeat chain & $pp(n)$ \\

    \stepcounter{fi} $\phi_{\thefi}$ & Does $l(v)$ belong to a pseudo-polyphonic repeat chain& $pp(l(v))$ \\

    \stepcounter{fi} $\phi_{\thefi}$ & Does any note at onset $on(n) - bd(n)$ belong to a pseudo-polyphonic repeat chain & $pp(on(n) - bd(n))$ \\

    \stepcounter{fi} $\phi_{\thefi}$ & True if $l(v)$ is adjacent to $n$ in a pseudo-polyphonic repeat chain and no note at onset $on(n) - bd(n)$ belongs to a pseudo-polyphonic repeat chain& $\phi_{59} \land \phi_{77} \land \phi_{94} \land \phi_{95} \land \neg\phi_{97}$ \\

    \stepcounter{fi} $\phi_{\thefi}$ & True if neither $n$ nor $l(v)$ belongs to a pseudo-polyphonic repeat chain and there exists a note at onset $on(n) - bd(n)$ that belongs to a pseudo-polyphonic repeat chain & $\neg\phi_{95} \land \neg\phi_{96} \land \phi_{97}$ \\

    \stepcounter{fi} $\phi_{\thefi}$ & True if $n$ belongs to a pseudo-polyphonic repeat chain and $l(v)$ does not  & $\phi_{95} \land \neg\phi_{96}$ \\

    \stepcounter{fi} $\phi_{\thefi}$ & True if $n$ does not belong to a pseudo-polyphonic repeat chain and $v(l)$ does & $\neg\phi_{95} \land \phi_{96}$ \\
  \end{tabu}}
  \caption{
    Pair-level, pseudo-polyphony features in $\Phi(n, v)$.
  }
  \label{tab:pair-ppolphony-features}
\end{table}

\begin{figure}[H]
  \centering
  \includegraphics[width=0.4\columnwidth]{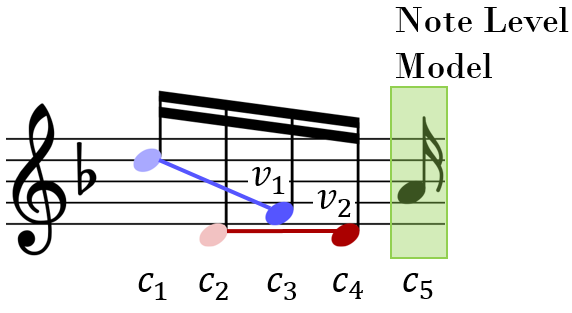}
  \caption{Pairing $n$ with the prominent melody in the pseudo-polyphonic voices in measure 82 from ``Forest".}
  \label{fig:pseudopoly_melody}
\end{figure}

Figure~\ref{fig:pseudopoly_melody} shows an example where the current note should be assigned to the melody supported by the repeat chain. Feature $\phi_{98}$ will be false and $\phi_{99}$ will be true when the system measures the compatibility between the note in $c_5$ and $v_1$. Thus, it should be able to learn a preference towards pairing the two. While measuring the compatibility between the note at $c_5$ and $v_2$, features $\phi_{98} - \phi_{100}$ will be false and $\phi_{101}$ will be true, which should help the system avoid pairing the two.

\begin{figure}[H]
  \centering
  \includegraphics[width=0.4\columnwidth]{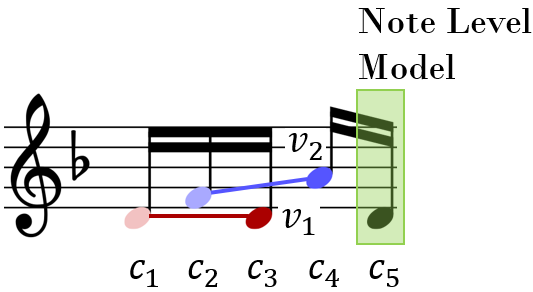}
  \caption{Pairing $n$ with the repeated chain of notes in the pseudo-polyphonic voices in measure 82 from ``Forest".}
  \label{fig:pseudopoly_chain}
\end{figure}

Figure~\ref{fig:pseudopoly_chain} shows an example where the current note should be assigned to the repeat chain. The note in $c_5$ is now part of a pseudo-polyphonic repeat chain. $\phi_{98}$ will be true and and $\phi_{99}$ will be false when the system measures the compatibility between the note in $c_5$ and $v_1$. Thus, based on the behavior learned from the data, it should be able to pair the two. While measuring the compatibility between the note in $c_5$ and $v_2$, features $\phi_{98}$, $\phi_{99}$ and $\phi_{101}$ will be false while $\phi_{100}$ will be true, which should discourage the system from pairing the two.

\begin{figure}[H]
  \centering
  \includegraphics[width=0.3\columnwidth]{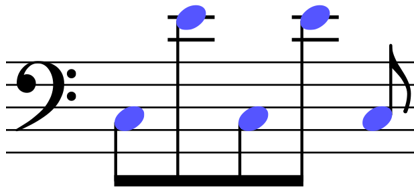}
  \caption{Simultaneous pseudo-polyphonic repeat chains in measure 1 from ``Count on Me".}
  \label{fig:pseudopoly_simultaneous}
\end{figure}

If two pseudo-polyphonic repeat chains are playing simultaneously, as in Figure~\ref{fig:pseudopoly_simultaneous}, then neither chain results into a salient voice. 
% has any melody to bring to focus for the listeners. This ultimately causes the notes of the two chains to blend together not into a discernable melody but instead into an oscillating beat. 
Hence the importance of  feature $\phi_{98}$ that asks whether or not a note at onset $on(n) - bd(n)$ belongs to a pseudo-polyphonic repeat chain.

\subsection{Empty Voice Feature}

Finally, we introduce the empty voice feature $\phi_{102}$, which is set to 1 only for the empty voice, i.e. $\phi_{102}(v)$ = 1, $\forall  v = \epsilon$. All the remaining features in any feature vector for an empty voice $\Phi(n, \epsilon)$ are set to zero. This allows the empty voice to activate a bias parameter $w_{102}$, which is equivalent to learning a threshold $-w_{102}$ that the weighted combination of the remaining features must exceed in order for the note to be assigned to an existing, non-empty active voice. Otherwise, the note $n$ will be assigned to the empty voice, meaning it will start a new voice.

\begin{table}[H]
  \centering
  \small
  {\tabulinesep=1.5mm
  \begin{tabu}{l l | c}
    Feature & Short Description & Formula \\ 
    \hline

    \stepcounter{fi} $\phi_{\thefi}$ & Is $v$ an empty voice & $v = \epsilon$ \\

  \end{tabu}}
  \caption{
    Empty voice feature in $\Phi(n, v)$.
  }
  \label{tab:empty-feature}
\end{table}

\pagebreak

\section{Voice Separation Features for the Chord-Level Model}
\label{sec:appendix-chord-level}

\subsection{Basic Concepts and Notation}
\label{sec:chord-notation}

\begin{table}[!ht]
  \small
  \centering
  \begin{tabular}{|c|p{13cm}|}
    \hline
    Symbol & \multicolumn{1}{c|}{Description} \\
    \hline
    \hline

    \multicolumn{2}{|c|}{\bf{Variables}} \\
    \hline
    \hline

    $n$ & Unassigned note in the current chord $c_t$. \\
    \hline
    $c_{m}$ & Chord of note $m$ with notes sorted in descending order of pitch. \\
    \hline
    $\vec{n}$ & Ordered vector of notes upon which $v$ diverges for the divergence level features. Notes in $\vec{n}$ are arranged in descending order of pitch. \\
    \hline
    $n_i$ & $i^{th}$ note in $\vec{n}$. \\
    \hline
    $n_{-i}$ & $i^{th}$ note from the end in $\vec{n}$. \\
    \hline
    \hline

    \multicolumn{2}{|c|}{\bf{Pitch Functions}} \\

    \hline
    \hline
    $ps(m)$ & Absolute pitch space of note $m$. Pitch values are taken from a pitch space in which C$_4$ has value 60 and a difference of 1 corresponds to one half step, e.g. C$_5$ has value 72. \\
    \hline
    \hline

    \multicolumn{2}{|c|}{\bf{Temporal Functions}} \\

    \hline
    \hline

    $on(m)$ & Onset of note $m$ measured in beats relative to the first note in the score. \\
    \hline
    $\mathit{off}(m)$ & Offset of note $m$ measured in beats relative to the first note in the score. The offset is defined as the time at which a note ends. \\
    \hline
    \hline

    \multicolumn{2}{|c|}{\bf{Positional Functions}} \\

    \hline
    \hline

    $cp(m)$ & Position of note $m$ in the ordered chord $c_m$. Position is relative to the topmost note in $c_m$. \\
    \hline
    \hline
  \end{tabular}
  \caption{
    Note and chord variable and function symbols used in the definitions of chord-level assignment model features.
  }
\label{tab:note-symbol-table2}
\end{table}

\begin{table}[!ht]
  \small
  \centering
  \begin{tabular}{|c|p{13cm}|}
    \hline

    Symbol & \multicolumn{1}{c|}{Description} \\

    \hline
    \hline

    \multicolumn{2}{|c|}{\bf{Variables}} \\
    \hline
    $v$ & Active voice. $v$ may either be a complete or paritial active voice. Furthermore, in our feature definitions, $v$ is comprised only of notes that lie within the beat horizon of $l(v)$. \\
    \hline
    $l(u)$ & The last note in the active voice $u$. \\
    \hline
    $\mathcal{V}$ & The ordered sequence  of active voices at the current chord $c_t$. \\
    \hline
    $\vec{v}$ & Ordered vector of active voices converging upon $n$ for the convergence level features. Voices maintain the same relative positions established in $\mathcal{V}$. \\
    \hline
    $v_i$ & $i^{th}$ active voice in $\vec{v}$. \\
    \hline
    $v_{- i}$ & $i^{th}$ active voice from the end in $\vec{v}$. \\
    \hline
    \hline

    \multicolumn{2}{|c|}{\bf{Positional Functions}} \\

    \hline
    \hline
    $vp(u)$ & Position of active voice $u$ in the ordered active voice sequence $\mathcal{V}$. Position is relative to the topmost voice in $\mathcal{V}$. \\
    \hline
    $cvp(u)$ & Position of active voice $u$ relative to the complete active voices in the ordered active voice sequence $\mathcal{V}$. \\
    \hline
    $dp(m)$ & Depth of the voice tree diverging from note $m$ \\
    \hline
    $div(u, w)$ & Shared divergent note between voices $u$ and $w$. \\
    \hline
    $zip(u, w)$ & List of note pairs where the containing Shared divergent note between voices $u$ and $w$. \\
    \hline
    $unq(u, w)$ & Number of unique onsets among the notes in voices $u$ and $w$. \\
    \hline
    $x(m, u)$ & Boolean function that returns true if the pairing of note $m$ and voice $u$ results in a cross of another voice. \\
    \hline
  \end{tabular}
  \caption{
    Voice variable and function symbols used in the definitions of our chord-level assignment model features.
  }
  \label{tab:voice-symbol-table2}
\end{table}
The basic concepts and notation that are used to define the input features specific to the chord-level assignment model are listed in Table~\ref{tab:note-symbol-table2} (for notes and chords) and Table~\ref{tab:voice-symbol-table2} (for voices).

\subsection{Chord-Level Model: Convergence Features}
\label{sec:convergence-features}

\subsubsection{Pitch Features (Table~\ref{tab_convergence_pitch_features})}

\begin{table}[h]
  \centering

  {\tabulinesep=1.5mm
  \begin{tabu}{l p{6cm} | c}

    Feature & Short Description & Formula \\ 
    \hline

    \stepcounter{fi} $\phi_{\thefi}$ & Absolute pitch distance between $n$ and the average pitch of the $l(v)$ in $\vec{v}$ & $\Bigl|ps(n) - \frac{1}{|\vec{v}|}\displaystyle\sum_{v \in \vec{v}} ps(l(v))\Bigr|$ \\

    \stepcounter{fi} $\vec{\phi}_{\thefi}$ & Direction of the pitch space of $n$ relative to the average pitch of the $l(v)$ in $\vec{v}$ & $ps(n) >=< \frac{1}{|\vec{v}|}\displaystyle\sum_{v \in \vec{v}} ps(l(v))$ \\

    \stepcounter{fi} $\vec{\phi}_{\thefi}$ & Absolute average pitch of the $l(v)$ in $\vec{v}$ & $\frac{1}{|\vec{v}|}\displaystyle\sum_{v \in \vec{v}} ps(l(v))$ \\

    \stepcounter{fi} $\phi_{\thefi}$ & Absolute pitch distance between $n$ and $l(v_0)$ & $|ps(n) - ps(l(v_0))|$ \\

    \stepcounter{fi} $\vec{\phi}_{\thefi}$ & Direction of the pitch space of $n$ relative to the pitch of $l(v_0)$ & $ps(n) >=< ps(l(v_0))$ \\

    \stepcounter{fi} $\phi_{\thefi}$ & Absolute pitch distance between $n$ and $l(v_{-1})$ & $|ps(n) - ps(l(v_{-1}))|$ \\

    \stepcounter{fi} $\vec{\phi}_{\thefi}$ & Direction of the pitch space of $n$ relative to the pitch of $l(v_{-1})$ & $ps(n) >=< ps(l(v_{-1}))$ \\

    \stepcounter{fi} $\phi_{\thefi}$ & Absolute pitch distance between $l(v_{0})$ and $l(v_{-1})$ & $|ps(l(v_{0})) - ps(l(v_{-1}))|$ \\

    \stepcounter{fi} $\phi_{\thefi}$ & Absolute pitch distance between $l(v_0)$ and $l(v_1)$ & $|ps(l(v_0)) - ps(l(v_1))|$ \\

    \stepcounter{fi} $\phi_{\thefi}$ & Absolute pitch distance between $l(v_{-1})$ and $l(v_{-2})$ & $|ps(l(v_{-1})) - ps(l(v_{-2}))|$ \\

    \stepcounter{fi} $\phi_{\thefi}$ & Absolute pitch distance between $n$ and the shared divergent note between $v_{0}$ and $v_{1}$ & $|ps(n) - ps(div(v_{0}, v_{1}))|$ \\

    \stepcounter{fi} $\phi_{\thefi}$ & Absolute pitch distance between $n$ and the shared divergent note between $v_{-1}$ and $v_{-2}$ & $|ps(n) - ps(div(v_{-1}, v_{-2}))|$ \\

  \end{tabu}}

  \caption{
    Convergence-level, pitch features in $\Phi(n, \vec{v})$.
  }

  \label{tab_convergence_pitch_features}
\end{table}

% JUSTIFY

\subsubsection{Temporal Features (Table~\ref{tab_convergence_temporal_features})}

\begin{table}[h]
  \centering

  {\tabulinesep=1.5mm
  \begin{tabu}{l p{6cm} | c}

    Feature & Short Description & Formula \\ 
    \hline

    \stepcounter{fi} $\phi_{\thefi}$ & Inter-onset distance between $n$ and $l(v_{0})$ & $|on(n) - on(l(v_{0}))|$ \\

    \stepcounter{fi} $\phi_{\thefi}$ & Onset-offset distance between $n$ and $l(v_{0})$ & $max(on(n) - \mathit{off}(l(v_{0})), 0)$ \\

    \stepcounter{fi} $\vec{\phi}_{\thefi}$ & Direction of the onset of $n$ relative to the offset of $l(v_{0})$ & $on(n) >=< \mathit{off}(l(v_{0}))$ \\

    \stepcounter{fi} $\phi_{\thefi}$ & Inter-onset distance between $n$ and $l(v_{-1})$ & $|on(n) - on(l(v_{-1}))|$ \\

    \stepcounter{fi} $\phi_{\thefi}$ & Onset-offset distance between $n$ and $l(v_{-1})$ & $max(on(n) - \mathit{off}(l(v_{-1})), 0)$ \\

    \stepcounter{fi} $\vec{\phi}_{\thefi}$ & Direction of the onset of $n$ relative to the offset of $l(v_{-1})$ & $on(n) >=< \mathit{off}(l(v_{-1}))$ \\

    \stepcounter{fi} $\vec{\phi}_{\thefi}$ & Direction of the onset of $l(v_{0})$ relative to the onset of $l(v_{1})$ & $on(l(v_{0})) >=< on(l(v_{1}))$ \\

    \stepcounter{fi} $\vec{\phi}_{\thefi}$ & Direction of the onset of $l(v_{-1})$ relative to the onset of $l(v_{-2})$ & $on(l(v_{-1})) >=< on(l(v_{-2}))$ \\

  \end{tabu}}

  \caption{
    Convergence-level, temporal features in $\Phi(n, \vec{v})$.
  }

  \label{tab_convergence_temporal_features}
\end{table}

\subsubsection{Positional Features (Table~\ref{tab_convergence_positional_features})}

\begin{table}[h]
  \centering

  {\tabulinesep=1.5mm
  \begin{tabu}{l p{8cm} | c}

    Feature & Short Description & Formula \\ 
    \hline

    \stepcounter{fi} $\phi_{\thefi}$ & Is $v_{0}$ a complete active voice & $dp(v_{0}) = 0$ \\

    \stepcounter{fi} $\phi_{\thefi}$ & Is $v_{-1}$ a complete active voice & $dp(v_{-1}) = 0$ \\

    \stepcounter{fi} $\phi_{\thefi}$ & Absolute difference between the chord positions of $l(v_{0})$ and $l(v_{1})$ & $|cp(l(v_{0})) - cp(l(v_{1}))|$ \\

    \stepcounter{fi} $\vec{\phi}_{\thefi}$ & Direction of the chord positions of $l(v_{0})$ relative to the chord position of $l(v_{1})$ & $cp(l(v_{0})) >=< cp(l(v_{1}))$ \\

    \stepcounter{fi} $\phi_{\thefi}$ & Absolute difference between the chord positions of $l(v_{-1})$ and $l(v_{-2})$ & $|cp(l(v_{-1})) - cp(l(v_{-2}))|$ \\

    \stepcounter{fi} $\vec{\phi}_{\thefi}$ & Direction of the chord positions of $l(v_{-1})$ relative to the chord position of $l(v_{-2})$ & $cp(l(v_{-1})) >=< cp(l(v_{-2}))$ \\

    \stepcounter{fi} $\phi_{\thefi}$ & Difference between the complete active voice positions of $v_{0}$ and $v_{1}$ & $cvp(v_{1}) - cvp(v_{0})$ \\

    \stepcounter{fi} $\phi_{\thefi}$ & Difference between the complete active voice positions of $v_{-1}$ and $v_{-2}$ & $cvp(v_{-1}) - cvp(v_{-2})$ \\

    \stepcounter{fi} $\vec{\phi}_{\thefi}$ & Absolute average difference in chord position between the notes in $v_{0}$ that synchronize with the notes in $v_{1}$ & $\frac{\displaystyle\sum_{m,p \in zip(v_{0}, v_{1})} |cp(m) - cp(p)|}{|zip(v_{0}, v_{1})|}$ \\

    \stepcounter{fi} $\vec{\phi}_{\thefi}$ & Absolute average difference in chord position between the notes in $v_{-1}$ that synchronize with the notes in $v_{-2}$ & $\frac{\displaystyle\sum_{m,p \in zip(v_{-1}, v_{-2})} |cp(m) - cp(p)|}{|zip(v_{-1}, v_{-2})|}$ \\

    \stepcounter{fi} $\vec{\phi}_{\thefi}$ & Number of unique onsets among the notes in $v_{0}$ and $v_{1}$ & $unq(v_{0}, v_{1})$ \\

    \stepcounter{fi} $\vec{\phi}_{\thefi}$ & Absolute difference between $\phi_{\thefi-1}$ and the number of notes in $v_{0}$ that synchronize with the notes in $v_{1}$ & $|\phi_{\thefi-1} - zip(v_{0}, v_{1})|$ \\

    \stepcounter{fi} $\vec{\phi}_{\thefi}$ & Number of unique onsets among the notes in $v_{-1}$ and $v_{-2}$ & $unq(v_{-1}, v_{-2})$ \\

    \stepcounter{fi} $\vec{\phi}_{\thefi}$ & Absolute difference between $\phi_{\thefi-1}$ and the number of notes in $v_{-1}$ that synchronize with the notes in $v_{-2}$ & $|\phi_{\thefi-1} - zip(v_{-1}, v_{-2})|$ \\

    \stepcounter{fi} $\vec{\phi}_{\thefi}$ & Length of $\vec{v}$ & $|\vec{v}|$ \\

    \stepcounter{fi} $\vec{\phi}_{\thefi}$ & Number of complete active voices in $\vec{v}$ & $\displaystyle\sum_{v \in \vec{v}} dp(v_0) = 0$ \\

    \stepcounter{fi} $\vec{\phi}_{\thefi}$ & Number of partial active voices in $\vec{v}$ & $\displaystyle\sum_{v \in \vec{v}} dp(v_0) \neq 0$ \\

  \end{tabu}}

  \caption{
    Convergence-level, positional features in $\Phi(n, \vec{v})$.
  }

  \label{tab_convergence_positional_features}
\end{table}

\subsubsection{Empty Convergence Set Feature (Table~\ref{tab_convergence_empty_feature})}

\begin{table}[!ht]
  \centering

  {\tabulinesep=1.5mm
  \begin{tabu}{l p{7.5cm} | c}
    Feature & Short Description & Formula \\ 
    \hline

    \stepcounter{fi} $\phi_{\thefi}$ & Does the joint assignment $j$ contain at least one convergence pairing & $\exists \vec{v} \in j.$ $|\vec{v}| > 1 $ \\

  \end{tabu}}

  \caption{
    Empty convergence set feature in $\Phi(n, \vec{v})$.
  }

  \label{tab_convergence_empty_feature}
\end{table}

\subsection{Chord-Level Model: Divergence Features}
\label{sec:divergence-features}

\subsubsection{Pitch Features (Table~\ref{tab_divergence_pitch_features})}

\begin{table}[!ht]
  \centering

  {\tabulinesep=1.5mm
  \begin{tabu}{l p{6cm} | c}

    Feature & Short Description & Formula \\ 
    \hline

    \stepcounter{fi} $\phi_{\thefi}$ & Absolute pitch distance between $l(v)$ and the average pitch of the $n$ in $\vec{n}$ & $\Bigl|ps(l(v)) - \frac{1}{|\vec{n}|}\displaystyle\sum_{n \in \vec{n}} ps(n)\Bigr|$ \\

    \stepcounter{fi} $\vec{\phi}_{\thefi}$ & Direction of the pitch space of $l(v)$ relative to the average pitch of the $n$ in $\vec{n}$ & $ps(l(v)) >=< \frac{1}{|\vec{n}|}\displaystyle\sum_{n \in \vec{n}} ps(n)$ \\

    \stepcounter{fi} $\vec{\phi}_{\thefi}$ & Absolute average pitch of the $n$ in $\vec{n}$ & $\frac{1}{|\vec{n}|}\displaystyle\sum_{n \in \vec{n}} ps(n)$ \\

    \stepcounter{fi} $\phi_{\thefi}$ & Absolute pitch distance between $l(v)$ and $n_0$ & $|ps(l(v)) - ps(n_0)|$ \\

    \stepcounter{fi} $\vec{\phi}_{\thefi}$ & Direction of the pitch space of $l(v)$ relative to the pitch of $n_0$ & $ps(l(v)) >=< ps(n_0)$ \\

    \stepcounter{fi} $\phi_{\thefi}$ & Absolute pitch distance between $l(v)$ and $n_{-1}$ & $|ps(l(v)) - ps(n_{-1})|$ \\

    \stepcounter{fi} $\vec{\phi}_{\thefi}$ & Direction of the pitch space of $l(v)$ relative to the pitch of $n_{-1}$ & $ps(l(v)) >=< ps(n_{-1})$ \\

    \stepcounter{fi} $\phi_{\thefi}$ & Absolute pitch distance between $n_{0}$ and $n_{-1}$ & $|ps(n_{0}) - ps(n_{-1})|$ \\

    \stepcounter{fi} $\phi_{\thefi}$ & Absolute pitch distance between $n_0$ and $n_1$ & $|ps(n_0) - ps(n_1)|$ \\

    \stepcounter{fi} $\phi_{\thefi}$ & Absolute pitch distance between $n_{-1}$ and $n_{-2}$ & $|ps(n_{-1}) - ps(n_{-2})|$ \\

    \stepcounter{fi} $\phi_{\thefi}$ & Absolute pitch distance between $n_0$ and the topmost convergent note of $v$ & $|ps(n_0) - ps(conv(v)_0)|$ \\

    \stepcounter{fi} $\phi_{\thefi}$ & Absolute pitch distance between $n_0$ and the bottommost divergent note of $v$ & $|ps(n_0) - ps(conv(v)_{-1})|$ \\

    \stepcounter{fi} $\phi_{\thefi}$ & Absolute pitch distance between $n_{-1}$ and the topmost convergent note of $v$ & $|ps(n_{-1}) - ps(conv(v)_0)|$ \\

    \stepcounter{fi} $\phi_{\thefi}$ & Absolute pitch distance between $n_{-1}$ and the bottommost divergent note of $v$ & $|ps(n_{-1}) - ps(conv(v)_{-1})|$ \\

  \end{tabu}}

  \caption{
    Divergence-level, pitch features in $\Phi(\vec{n}, v)$.
  }

  \label{tab_divergence_pitch_features}
\end{table}

% JUSTIFY

\subsubsection{Temporal Features (Table~\ref{tab_divergence_temporal_features})}

\begin{table}[!ht]
  \centering

  {\tabulinesep=1.5mm
  \begin{tabu}{l p{6cm} | c}

    Feature & Short Description & Formula \\ 
    \hline

    \stepcounter{fi} $\phi_{\thefi}$ & Inter-onset distance between $n_{0}$ and $l(v)$ & $|on(n_0) - on(l(v))|$ \\

    \stepcounter{fi} $\phi_{\thefi}$ & Onset-offset distance between $n_{0}$ and $l(v)$ & $max(on(n_0) - \mathit{off}(l(v)), 0)$ \\

    \stepcounter{fi} $\vec{\phi}_{\thefi}$ & Direction of the onset of $n_0$ relative to the offset of $l(v)$ & $on(n_0) >=< \mathit{off}(l(v))$ \\

    \stepcounter{fi} $\phi_{\thefi}$ & Inter-onset distance between $n_{-1}$ and $l(v)$ & $|on(n_{-1}) - on(l(v))|$ \\

    \stepcounter{fi} $\phi_{\thefi}$ & Onset-offset distance between $n_{-1}$ and $l(v)$ & $max(on(n_{-1}) - \mathit{off}(l(v)), 0)$ \\

    \stepcounter{fi} $\vec{\phi}_{\thefi}$ & Direction of the onset of $n_{-1}$ relative to the offset of $l(v)$ & $on(n_{-1}) >=< \mathit{off}(l(v))$ \\

    \stepcounter{fi} $\vec{\phi}_{\thefi}$ & Direction of the onset of $n_{0}$ relative to the onset of $n_{1}$ & $on(n_{0}) >=< on(n_{1})$ \\

    \stepcounter{fi} $\vec{\phi}_{\thefi}$ & Direction of the onset of $n_{-1}$ relative to the onset of $n_{-2}$ & $on(n_{-1}) >=< on(n_{-2})$ \\

  \end{tabu}}

  \caption{
    Divergence-level, temporal features in $\Phi(\vec{n}, v)$.
  }

  \label{tab_divergence_temporal_features}
\end{table}

\subsubsection{Positional Features (Table~\ref{tab_divergence_pitch_features})}

\begin{table}[!ht]
  \centering

  {\tabulinesep=1.5mm
  \begin{tabu}{l p{7.5cm} | c}

    Feature & Short Description & Formula \\ 
    \hline

    \stepcounter{fi} $\phi_{\thefi}$ & Absolute difference between the chord positions of $n_{0}$ and $n_{1}$ & $|cp(n_{0}) - cp(n_{1})|$ \\

    \stepcounter{fi} $\vec{\phi}_{\thefi}$ & Direction of the chord positions of $n_{0}$ relative to the chord position of $n_{1}$ & $cp(n_{0}) >=< cp(n_{1})$ \\

    \stepcounter{fi} $\phi_{\thefi}$ & Absolute difference between the chord positions of $n_{-1}$ and $n_{-2}$ & $|cp(n_{-1}) - cp(n_{-2})|$ \\

    \stepcounter{fi} $\vec{\phi}_{\thefi}$ & Direction of the chord positions of $n_{-1}$ relative to the chord position of $n_{-2}$ & $cp(n_{-1}) >=< cp(n_{-2})$ \\

    \stepcounter{fi} $\vec{\phi}_{\thefi}$ & Length of $\vec{n}$ & $|\vec{n}|$ \\

    \stepcounter{fi} $\vec{\phi}_{\thefi}$ & Number of notes in $\vec{n}$ that are currently paired with $v$ & $\displaystyle\sum_{n \in \vec{n}} l(v) \in lt(n)$ \\

  \end{tabu}}

  \caption{
    Divergence-level, positional features in $\Phi(\vec{n}, v)$.
  }

  \label{tab_divergence_positional_features}
\end{table}

\subsubsection{Empty Divergence Set Feature (Table~\ref{tab_divergence_empty_feature})}

\begin{table}[!ht]
  \centering

  {\tabulinesep=1.5mm
  \begin{tabu}{l p{7.5cm} | c}
    Feature & Short Description & Formula \\ 
    \hline

    \stepcounter{fi} $\phi_{\thefi}$ & Does the joint assignment $j$ contain at least one divergence pairing & $\exists \vec{n} \in j.$ $|\vec{n}| > 1 $ \\

  \end{tabu}}

  \caption{
    Empty divergence set feature in $\Phi(\vec{n}, v)$.
  }

  \label{tab_divergence_empty_feature}
\end{table}

\subsection{Chord-Level Model: Assignment Features}
\label{sec:assignment-features}

\subsubsection{Pitch Features (Table~\ref{tab_assignment_pitch_features})}

\begin{table}[H]
  \centering

  {\tabulinesep=1.5mm
  \begin{tabu}{l p{6cm} | c}

    Feature & Short Description & Formula \\ 
    \hline

    \stepcounter{fi} $\phi_{\thefi}$ & Average pitch distance between the notes in $c_t$ and the last notes of their pairing voices in $j$ & $\frac{1}{|flat(j)|}\displaystyle\sum_{(n, v) \in flat(j)} |ps(n) - ps(l(v))|$ \\

  \end{tabu}}

  \caption{
    Assignment-level, pitch feature in $\Phi(c_t, j)$.
  }

  \label{tab_assignment_pitch_features}
\end{table}

\subsubsection{Positional Features (Table~\ref{tab_assignment_positional_features})}

\begin{table}[H]
  \centering

  {\tabulinesep=1.5mm
  \begin{tabu}{l p{6cm} | c}

    Feature & Short Description & Formula \\ 
    \hline

    \stepcounter{fi} $\phi_{\thefi}$ & Is there a crossing pair in the joint assignment $j$ & $\exists (n,v) \in flat(j).\ x(n, v)$ \\

    \stepcounter{fi} $\vec{\phi}_{\thefi}$ & Absolute difference between the number of non-empty voices in $j$ and the number of notes in $c_t$ & $|c_t| - \displaystyle\sum_{(\_, v) \in flat(j)} v \neq \epsilon$ \\

    \stepcounter{fi} $\vec{\phi}_{\thefi}$ & Number of non-empty voices in $j$ & $\displaystyle\sum_{(\_, v) \in flat(j)} v \neq \epsilon$ \\

    \stepcounter{fi} $\vec{\phi}_{\thefi}$ & Number of empty voices in $j$ & $\displaystyle\sum_{(\_, v) \in flat(j)} v = \epsilon$ \\

    \stepcounter{fi} $\vec{\phi}_{\thefi}$ & Number of convergence pairings in $j$ & $\displaystyle\sum_{(\_, \vec{v}) \in j} |\vec{v}| > 1$ \\

    \stepcounter{fi} $\vec{\phi}_{\thefi}$ & Number of divergence pairings in $j$ & $\displaystyle\sum_{(\vec{n},\_) \in rev(j)} |\vec{n}| > 1$ \\

  \end{tabu}}

  \caption{
    Assignment-level, positional feature in $\Phi(c_t, j)$.
  }

  \label{tab_assignment_positional_features}
\end{table}

\end{document}